\documentclass[superscriptaddress,twocolumn]{revtex4}
\usepackage{doi}
\usepackage{hyperref}
\hypersetup{
  colorlinks=true,        
  linkcolor=blue,         
  citecolor=cyan,         
}

\usepackage[latin1]{inputenc}
\usepackage{graphicx}
\usepackage{dcolumn}
\usepackage{bm}
\usepackage{color}
\usepackage{enumitem}
\usepackage{amsmath}
\usepackage{amssymb}

\begin{document}

\title{Shadows and gravitational weak lensing by the Schwarzschild black hole in the string cloud background with quintessential field}

\author{G. Mustafa}
\email{gmustafa3828@gmail.com}
\affiliation{Department of Physics, Zhejiang Normal University,
Jinhua 321004, China}

\author{Farruh Atamurotov}
\email{atamurotov@yahoo.com}

\affiliation{Inha University in Tashkent, Ziyolilar 9, Tashkent 100170, Uzbekistan}
\affiliation{Akfa University, Milliy Bog Street 264, Tashkent 111221, Uzbekistan}
\affiliation{National University of Uzbekistan, Tashkent 100174, Uzbekistan} 
\affiliation{Tashkent State Technical University, Tashkent 100095, Uzbekistan}

\author{Ibrar Hussain}
\email{ibrar.hussain@seecs.nust.edu.pk}
\affiliation{School of Electrical Engineering and Computer Science,
National University of Sciences and Technology, H-12, Islamabad, Pakistan}

\author{Sanjar Shaymatov}\email{sanjar@astrin.uz}
\affiliation{Institute for Theoretical Physics and Cosmology, Zheijiang University of Technology, Hangzhou 310023, China}
\affiliation{Akfa University, Milliy Bog Street 264, Tashkent 111221, Uzbekistan}
\affiliation{Ulugh Beg Astronomical Institute, Astronomicheskaya 33, Tashkent 100052, Uzbekistan}
\affiliation{Institute of Fundamental and Applied Research, National Research University TIIAME, Kori Niyoziy 39, Tashkent 100000, Uzbekistan}
\affiliation{National University of Uzbekistan, Tashkent 100174, Uzbekistan} 
\affiliation{Tashkent State Technical University, Tashkent 100095, Uzbekistan}

\author{Ali \"Ovg\"un}
\email{ali.ovgun@emu.edu.tr}
\homepage{https://www.aovgun.com}
\affiliation{Physics Department, Eastern Mediterranean University, Famagusta, 99628 North Cyprus via Mersin 10, Turkey}


\begin{abstract}
In this work, we observe that in the presence of the string cloud parameter $a$ and the quintessence parameter $\gamma$, with the equation of state parameter $\omega_q={-2}/{3}$ the radius of the shadow of the Schwarzschild black hole increases as compared with the pure Schwarzschild black hole case. The existence of both quintessential dark energy and cloud of strings magnify the shadow size and hence the strength of the gravitational field around the Schwarzschild black hole increases. Using the data collected by the Event Horizon Telescope (EHT) collaboration for the M87* and Sgr A*, we obtain upper bounds on the values of the parameters $a$ and $\gamma$. Further, we see the effects of the parameters $a$ and $\gamma$ on the rate of emission energy for the Schwarzschild black hole. We notice that the rate of emission energy is higher in the presence of clouds of string and quintessence. Moreover, we study the weak deflection angle using the Gauss-Bonnet theorem. We show the influence of the cloud of string parameter $a$ and the quintessential parameter $\gamma$ on the weak deflection angle. We notice that both the parameters $a$ and $\gamma$ increase the deflection angle $\alpha$.
\end{abstract}
\maketitle


\section{Introduction}
\label{introduction}

Theory of General Relativity (GR) is so far the most accepted theory of gravity. It has been verified by observing the gravitational waves by LIGO in the recent past \cite{Abbott16a}. The observation of the images of supermassive black hole shadows at the centre of the near by galaxy M87 and also at the centre of our own galaxy by the Event Horizon Telescope (ETH) Collaboration \cite{Akiyama19L1,Akiyama19L6}, further confirms the GR. However, there are still some unresolved issues like the accelerated expansion of our Universe, which cannot be completely explained by the theory of GR. The presence of the cosmological constant in the field equations of GR may be a candidate to explain the accelerated expansion of the Universe~\cite{Cruz05,Stuchlik11,Grenon10,
Rezzolla03a,Arraut15,Faraoni15,Shaymatov18a}. 
However, alternatives have so far been proposed to explain the behavior of the dark energy.
Of them the quintessence matter field is a well-accepted one as an alternative form of the dark energy~\cite{Peebles03,Wetterich88,Caldwell09}. For example, Kiselev proposed a black hole solution with the quintessence represented by the equation of state $p = \omega_{q}\rho$, where $\omega_q$ is given in the range $(-1;-1/3)$ \cite{Kiselev2003aa} and $(-1;-2/3)$ \cite{Hellerman2001JHEP}. In contrast to the quintessence, the case with $\omega_{q}=-1$ corresponds to the vacuum energy represented by the cosmological constant $\Lambda$. Besides, the effect of string clouds, which are assumed as a collection of strings formed in the early era after the big-bang in the structuring of the Universe, due to the asymmetry breaking, has also been considered in the study of black hole spacetime \cite{Toledoa}. The dynamics of test particles and photons in the vicinity of these stringy static black holes have been analysed \cite{Batool16,Mustafa2021,Fathi2022}.

Recent experiments associated with BlackHoleCam and the EHT have provided the shadows of the supermassive black hole that exists at the center of the M87 galaxy and the milky way galaxy  ~\cite{Akiyama19L1,Akiyama19L6}.  Note that the black hole shadow reflects the gravitational lensing, and thus black hole shadows and gravitational lensing effects are of primary importance to test the GR and understand more deeply the nature of the background geometry very near the black hole horizon. Further the study of black hole shadow and gravitational lensing may be helpful in putting constraints on the parameters of different alternative theories of gravity. There exists a vast literature on these lines (see for  example~\cite{Hioki09,Atamurotov13,Atamurotov13b,Abdujabbarov13aa,Atamurotov2016a,Papnoi2015,Abdujabbarov15a,Rahul:2020a,Cunha20a,Cunha17a,Atamurotov21b,Jafarzade21a,Afrin21a,Ghasemi2020a,Bambi:2019tjh,Vagnozzi:2019apd,Khodadi:2020jij}). 
Gravitational lensing is currently the main astrophysical test in GR to provide information on the geometry of the compact objects. In this regard, gravitational lensing in both weak and strong field regime has also become very important in understanding unexplored properties of gravitational compact objects. The strong gravitational lensing was proposed and considered by Virbhadra and Ellis~\cite{Virbha:2000a}. Following that, an extensive body of work has been done in a large variety of contexts~\cite[see, e.g.][]{Bozza:2001a,Bozza:2002b,Zhao:2017a,Vazquez04,Eiroa:2002b,Eiroa:2004a,Chak:2017a,Perlick04,Abu:2017a,Islam20egb,Virbha:2002a,Kumar2020a,Bakhtiyor2021a}. From an astrophysical point of view, all photons mostly go through a plasma medium. On the  other hand, the plasma medium can affect angular positions of an equivalent image, thus giving various wavelengths in observations. Therefore, this is the most intriguing and important reason why one needs to consider the plasma medium in the analysis of gravitational lensing. Thus a large amount of work has addressed the impact of plasma medium on the gravitational lensing effects in the weak field regime ~\cite{Bin:2010a,Babar2021a,Abu:2013a,Babar2021b,Hakimov2016a,Rog:2015a,Far:2021a,Car:2018a,Atamurotov215,Atamurotov216,Atamurotov22epjp,Atamurotov2022EPJP3,AtamurotovGhosh2022,AtamurotovAllo2022}. 

In this paper, we use the method by Gibbons and Werner who proposed new methodology to calculate weak deflection angle using the Gauss-Bonnet theorem (GBT) of asymptotically flat spacetime \cite{Gibbons:2008rj}. Werner has applied it to stationary spacetimes using the Finsler-Randers type geometry \cite{Werner_2012}. Then Ishihara et al. have extended the method of Gibbons and Werner to non-asymptotically flat spacetimes using the finite distance corrections of the source and the receiver \cite{Ishihara_2016,Ishihara:2016sfv}. Next T. Ono et al. have showed that it is also working for the axisymmetric spacetimes \cite{Ono:2017pie}. Li et al. have studied the finite distance method using the massive particles and Jacobi-Maupertuis Randers-Finsler metric within the framework of the GBT \cite{Li:2020dln,Li:2020wvn}. Crisnejo and Gallo have calculated the weak deflection angle in a plasma medium using GBT \cite{Crisnejo:2018uyn}. The study of the gravitational lensing for black holes has also been carried out in the literature \cite{Ovgun:2018fnk,Ovgun:2019wej,Ovgun:2018oxk,Javed:2019kon,Javed:2019rrg,Javed:2019ynm,Javed:2020lsg,Javed:2019qyg,Ovgun:2018fte,Javed:2019jag,Pantig:2020odu,Pantig2022,Pantig2022a,Okyay:2021nnh,Arakida:2017hrm,Arakida:2020xil,Zhang:2021ygh,Li:2021xhy,DCarvalho:2021zpf,Ali:2021psk,Fu:2021akc,Kumar:2019pjp,Jusufi:2018jof,21}. In this paper, we study the shadow and gravitational lensing in the weak field limit in the Schwarzschild black hole spacetime with the string cloud and quintessence background. We use the data released by the EHT collaboration for the M87* and Sgr A* to restrict the string cloud parameter $a$ and the quintessence parameter $\gamma$.  

The work is arranged as follows: In Sec.~\ref{Sec:metric} we discuss the spacetime metric of the Schwarzschild black hole in the presence of string cloud in quintessential background. In Sec.~\ref{Sec:shadow} we study the effects of the string clouds and quintessence on the shadow of the Schwarzschild black hole. In the same Section we also investigate the effects of string clouds and quintessence on the energy emission rate for the Schwarzschild black hole. We analyse the weak deflection angle of photon beam by the Schwarzschild black hole in string clouds with quintessential field in Sec.~\ref{Sec:GBT}. Finally, in Sec.~\ref{Sec:conclusion} we give a conclusion of our work.

\section{\label{Sec:metric}
Schwarzschild black hole metric in the string cloud background with quintessence}

Here in this section we derive the Kiselev black hole in the background of string clouds. The detailed solution has already been derived in the literature Ref. \cite{34,SC1}. 

Here the action can be represented as
\begin{equation}\label{S}
S=\frac{1}{2}\int{dx^4\sqrt{-g}}(R - L_{m}),
\end{equation}
where $g$ stands for the determinant of the metric tensor $g_{\mu\nu}$, $R$ represents the scalar curvature and $L_{m}$ is responsible for the matter parts of the action. It should be mentioned that the matter part is further consists of two parts namely the string clouds and the quintessence i.e $L_m=L_s+L_q$, and are defined below.

The Lagrangian density for the string clouds is given by \cite{SC1}
\begin{equation}\label{S1}
 L_s=k\big(-\frac{1}{2}\Sigma^{\mu\nu}\Sigma_{\mu\nu} \big),
\end{equation}
where the constant $k$ is related to the tension of the string and the bivector
\begin{equation}\label{S3}
\Sigma^{\mu\nu}=\epsilon^{a\beta}\frac{\partial x^\mu}{\partial \lambda^{a}}\frac{\partial x^\nu}{\partial \lambda^{\beta}}.
\end{equation}
In the last equation $\epsilon^{a \beta}$ is the two-dimensional Levi-Civita tensor and $\lambda^{a} (\lambda^{a}=\lambda^0, \lambda^1)$, is used for the parameterization of the world sheet that is described by the string with the induced metric  \cite{SC1}
\begin{equation}\label{S4}
h_{a \beta}=g_{\mu\nu}\frac{\partial x^\mu}{\partial \lambda^{a}}\frac{\partial x^\nu}{\partial \lambda^{\beta}}.
\end{equation}
The $\Sigma^{\mu \nu}$ describes the following identities \cite{SC1,SC2} 
\begin{equation}\label{S5}
\Sigma^{\mu [a}\Sigma^{\beta \sigma]}=0,\quad \nabla_\mu\Sigma^{\mu [a}\Sigma^{\beta \sigma]}=0, \quad \Sigma^{\mu a}\Sigma_{a\sigma}\Sigma^{\sigma \nu}={\bf h} \Sigma^{\nu\mu},
\end{equation}
where ${\bf h}$ denotes the determinant of $h_{a \beta}$. Varying the Lagrangian density with respect to $g_{\mu\nu}$, one obtain \cite{SC2} 
\begin{equation}\label{S6}
T^{\mu\nu}=\rho_s\frac{\Sigma^{\mu a}\Sigma^{\nu}_{a}}{\sqrt{-\bf h}},
\end{equation}
where $\rho_s$ stands for the string cloud density. Using the three identities given above in (\ref{S5}), one obtain $\partial_\mu(\sqrt{-g}\Sigma^{\mu a})=0$. Therefore, for the spherically symmetric static spacetime the non-zero components of the stress-energy-momentum tensor in the background of string clouds are  \cite{Vilenkin:2000jqa,Barriola:1989hx}
\begin{eqnarray}\label{11}
T_{0}^0=T_{1}^1=-\frac{a}{r^2}\, , \\
T_{2}^2=T_{3}^3=0\, , 
\end{eqnarray}
where the cloud of strings correspond to the constant $a$. The line element is then given by \cite{SC0}
\begin{eqnarray}\label{S7}
ds^2=-(1-a-\frac{R_s}{r})dt^2+(1-a-\frac{R_s}{r})^{-1}dr^2 + r^2 d\Omega^2. \end{eqnarray}
Throughout we use $d\Omega^2=d\theta^2 + sin^2\theta d\phi^2$ and $R_s=2M$, where $M$ refers to the black hole mass. In the above metric, one can find the black hole horizon as follows:
\begin{eqnarray}\label{hor22}
r_{H}=\frac{2M}{1-a}\, .
\end{eqnarray}
From the above expression (\ref{hor22}), one can see that the horizon radius increases/enlarges in the case when $a<1$. 
For our analysis here we focus on the case $a<1$ since it acts as an attractive gravitational charge. It is here worth noting that the string cloud model has been proposed to explain the field theory that stems from distance interactions existing between particles. Accordingly these interactions correspond to a particular behaviour of gravitational field. Hence, it is assumed that this gravitational field can be produced by the elements of strings. With this in mind, the above mentioned string cloud parameter, $a$, can be proposed to reveal a reasonable behaviour of such field theory. Thus, it is potentially important to understand more deeply the nature of string cloud parameter $a$ to bring out its effect on the astrophysical phenomena, such as black hole shadow and deflection angle of light.

For the quintessence part we write \cite{Q1} 
\begin{equation}\label{q}
L_q=-\frac{1}{2}g^{\mu\nu}\partial_\mu\varphi \partial_\nu\varphi-V(\varphi),
\end{equation}
where $\varphi$ represents the quintessence field and $V(\varphi)$ stands for the potential term. The non-zero components of the stress-energy-momentum tensor of the matter fluid for the Kiselev black hole solution are given as \cite{Toledoa,SC1}
\begin{eqnarray}\label{1}
T_{0}^0=T_{1}^1=\rho_{q}\, ,  
\end{eqnarray}
\begin{eqnarray}\label{2}
T_{2}^2=T_{3}^3=-\frac{\rho_{q}}{2}(3\omega_q+1)\, ,\end{eqnarray}
where $p_q$ and $\rho_q$ denotes respectively the pressure and the density of the quintessence and $\omega_q$ represents the equation of state parameter for the quintessence. The Kiselev black hole line element is 
\begin{eqnarray}\label{88}
ds^2&=&-(1-\frac{R_s}{r}-\frac{\gamma}{r^{3\omega_q+1}})dt^2 \\ \nonumber
&&+(1-\frac{R_s}{r}-\frac{\gamma}{r^{3\omega_q+1}})^{-1}dr^2 + r^2 d\Omega^2\, ,  
\end{eqnarray}
where the density
\begin{eqnarray} \label{9}
\rho_{q}=-\frac{\gamma}{2}\frac{3\omega_q}{r^{3\omega_q+1}}\, ,
\end{eqnarray}
and $\gamma$ is used for the quintessence parameter.

The Kiselev black hole was then analysed in the background of the string clouds \cite{Toledoa}. It was assumed that the quintessence and the string clouds are not interacting and the surviving components of the total stress-energy-momentum tensor for the two matter-energy contents were obtained as \cite{Toledoa}
\begin{eqnarray}\label{13}
T_{0}^0=T_{1}^1=\rho_{q}+\frac{a}{r^2}\, , \\  
T_{2}^2=T_{3}^3=-\frac{\rho_{q}}{2}(3\omega_q+1)\, . 
\end{eqnarray}
The Kiselev black hole with string clouds is given as \cite{34,Toledoa} 
\begin{eqnarray}\label{20}
ds^2&=&-(1-a-\frac{R_s}{r}-\frac{\gamma}{r^{3\omega_q+1}})dt^2\\ \nonumber
&&+(1-a-\frac{R_s}{r}-\frac{\gamma}{r^{3\omega_q+1}})^{-1}dr^2 + r^2 d\Omega^2\, .
\end{eqnarray}

Note that $\gamma$ and $\omega_q$ respectively represent the quintessential field parameter and the equation of state parameter. The quintessence equation of state is given as $p_q=\omega_{q}\rho_q$, with $\omega_q\in(-1;-1/3)$. It is worth noting here that $\omega_q=-1$ refers to the matter field with the vacuum energy defied by the cosmological constant $\Lambda$, while $\omega_q=-1/3$ represents another matter field corresponds to the frustrated network of cosmic strings (for detail see for example \cite{Liaqat2022,Miranda2014}). With this in mind, we further restrict ourselves to the case $\omega_q=-2/3$  that implicitly describes the pure quintessential field. The above metric (\ref{20}) reduces to the Schwarzschild metric in the case of $a=0$ and $\gamma=0$. For the  structure of the horizon of the metric represented by (\ref{20}); see Ref.~\cite{Toledoa}. 

\begin{figure*}
\centering
 \includegraphics[width=0.5\textwidth]{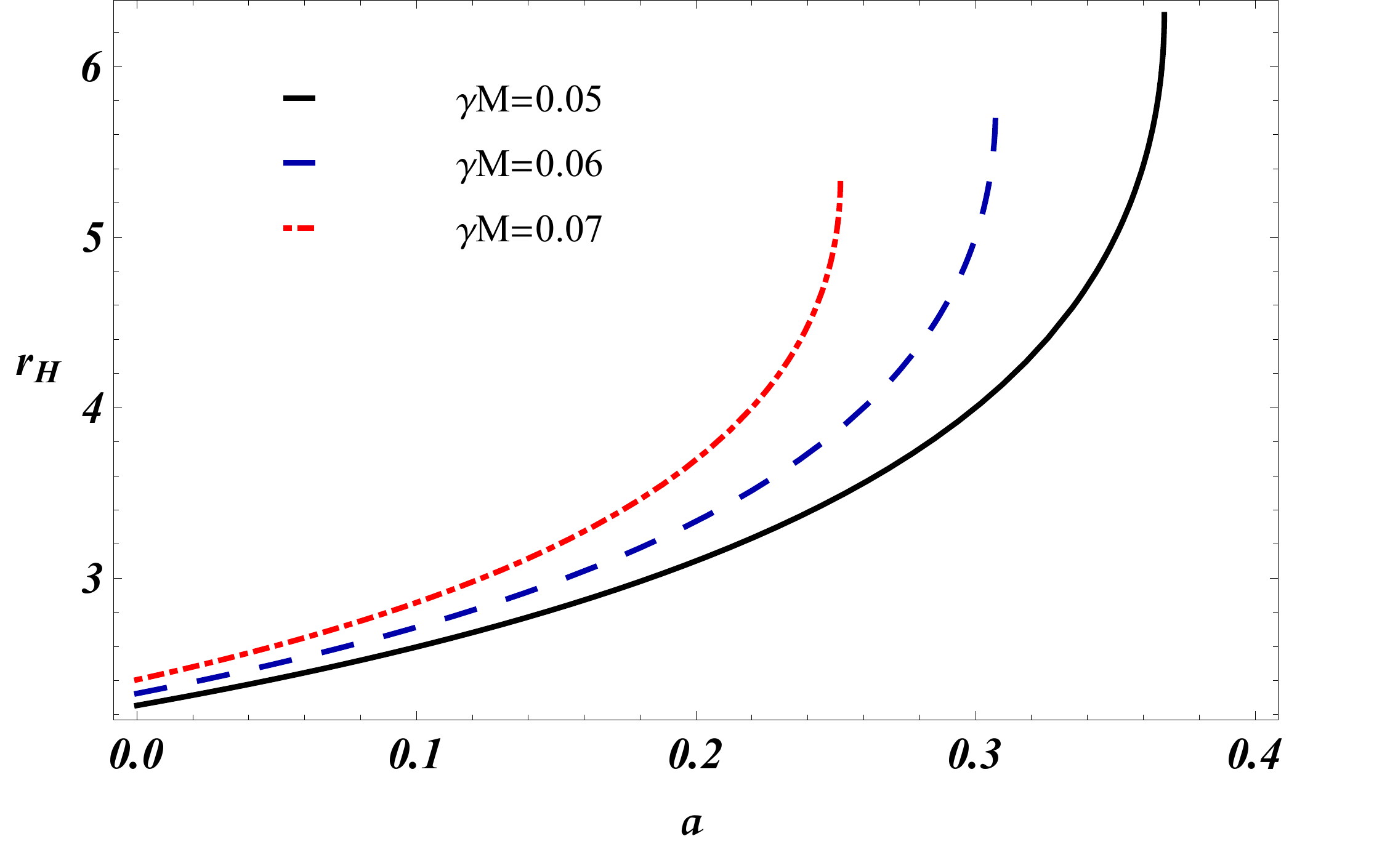}%
 \includegraphics[width=0.5\textwidth]{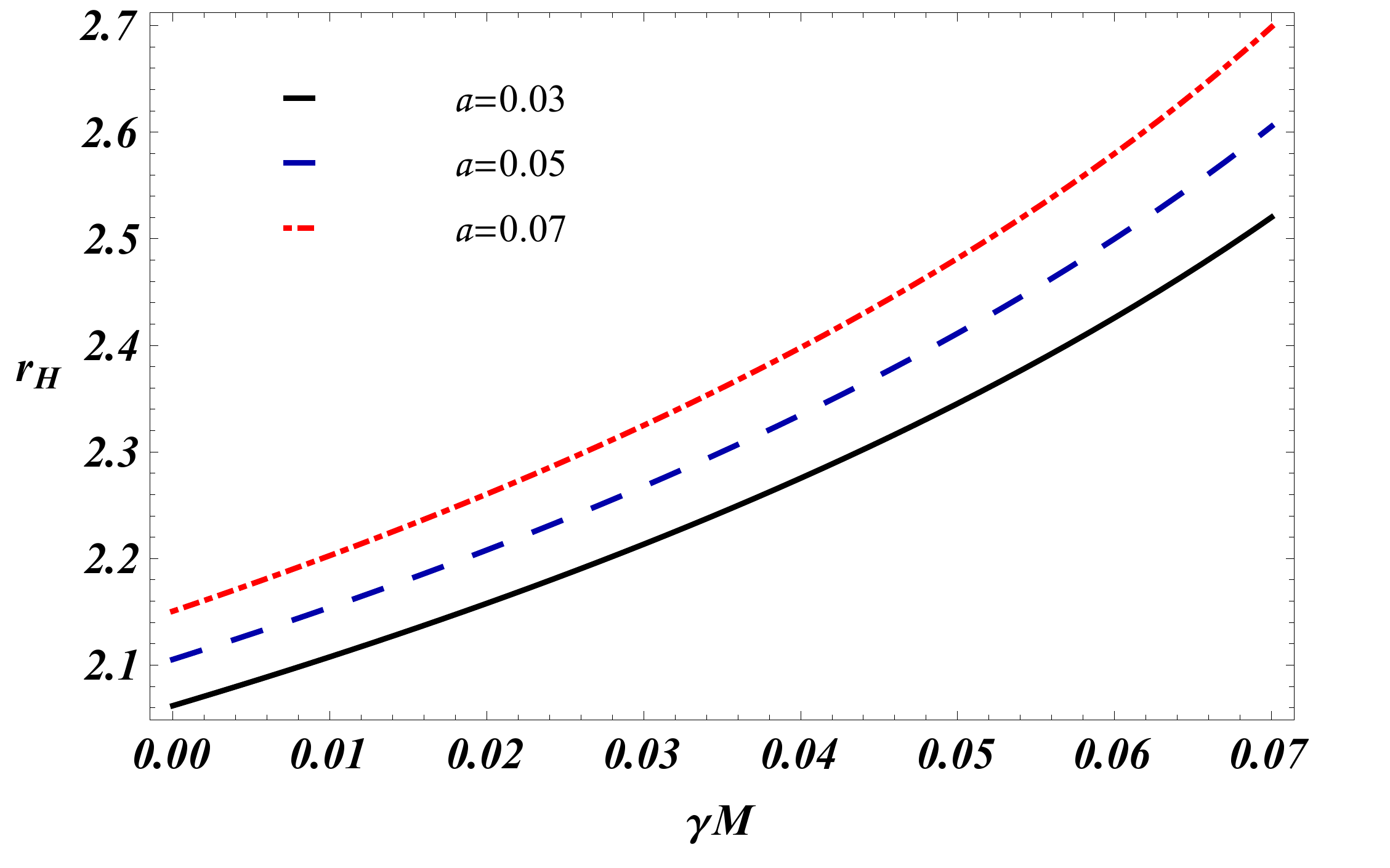}
\caption{Shows the graphic behavior of $r_{H}$ for the string cloud parameter $a$ (left panel) and for the quintessence parameter $\gamma$ (right panel), for $\omega_q=-\frac{2}{3}$. \label{H1}}
\end{figure*}

Let us then introduce the black hole horizon. The horizon is located at the root of $\gamma r^2+r(a-1)+2M=0$, that solves to give 
\begin{eqnarray}
r_{H}&&=\frac{1-a-\sqrt{a^2-2 a-4R_s \gamma+1}}{\gamma}\, ,\\ 
r_{q}&&=\frac{1-a+\sqrt{a^2-2 a-4R_s \gamma+1}}{\gamma}\, ,
\end{eqnarray}
where $r_h$ and $r_q$ refer to the black hole horizon and the quintessential cosmological horizon, respectively. Interestingly, it turns out that the cosmological horizon never vanishes regardless of the fact that there exists no source. We note that the cosmological horizon is located at a distance far way from the black hole. In Fig.~\ref{H1}, we show the black hole horizon as a function of the string cloud parameter $a$ and the quintessence parameter $\gamma$. From Fig.~\ref{H1}, we see that as the parameters $a$ and $\gamma$ increase, the event horizon also increases.


\section{\label{Sec:shadow}
Shadows of the Schwarzschild black holes surrounded by clouds of strings and quintessence}

\subsection{Geodesic equations}
The Lagrangian for the spacetime metric (\ref{20}) with $\omega_q={-2}/{3}$ is
\begin{eqnarray}\label{lag}
\mathcal{L}=m\bigg(-f(r)\frac{\dot{t}^{2}}{2}+\frac{\dot{r}^{2}}{2f(r)}+\frac{r^{2}}{2}(\dot{\theta}^{2}+sin^{2}\theta \dot{\phi}^{2})\bigg),
\end{eqnarray}
with $f(r)=1-a-\frac{R_s}{r}-\gamma r$. For (\ref{lag}) there are two conserved quantities, the specific angular momentum $\mathfrak{L}$ and the specific energy $\mathcal{E}$, given as \cite{Mustafa2021}
\begin{eqnarray}\label{lag1}
\dot{t}=-\mathcal{E}\left(f(r)\right)^{-1},\;\;\;\;\dot{\phi}=\frac{\mathfrak{L}}{r^{2}sin^{2}\theta}
\end{eqnarray}
Restricting ourselves to the equatorial plane, we take $\theta=\frac{\pi}{2}$, and (\ref{lag}) becomes
\begin{equation}\label{lag2}
-f(r)\dot{t}^{2}+\frac{\dot{r}^{2}}{f(r)}+r^{2}\dot{\phi}^{2}=-\lambda,
\end{equation}

\begin{figure*}
\centering
 \includegraphics[width=0.5\textwidth]{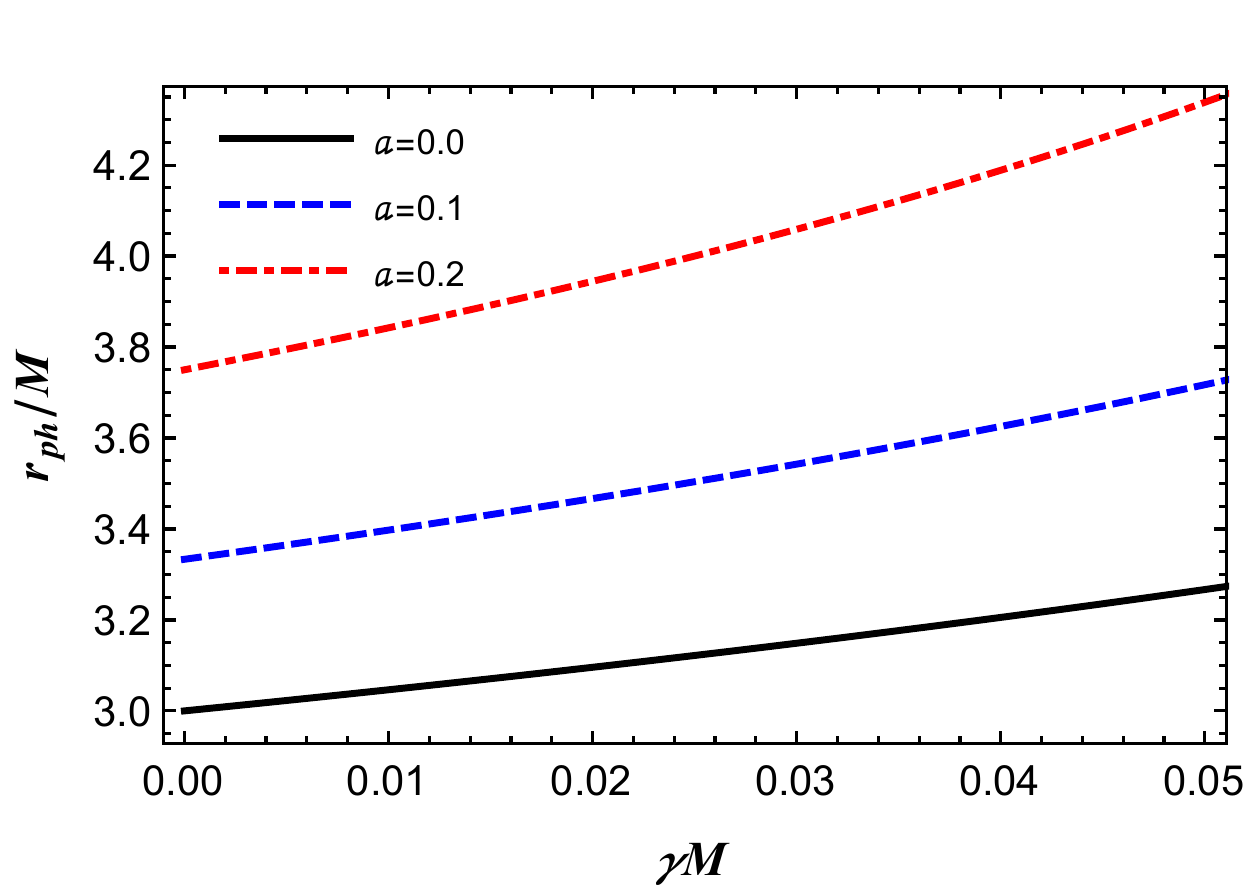}%
 \includegraphics[width=0.5\textwidth]{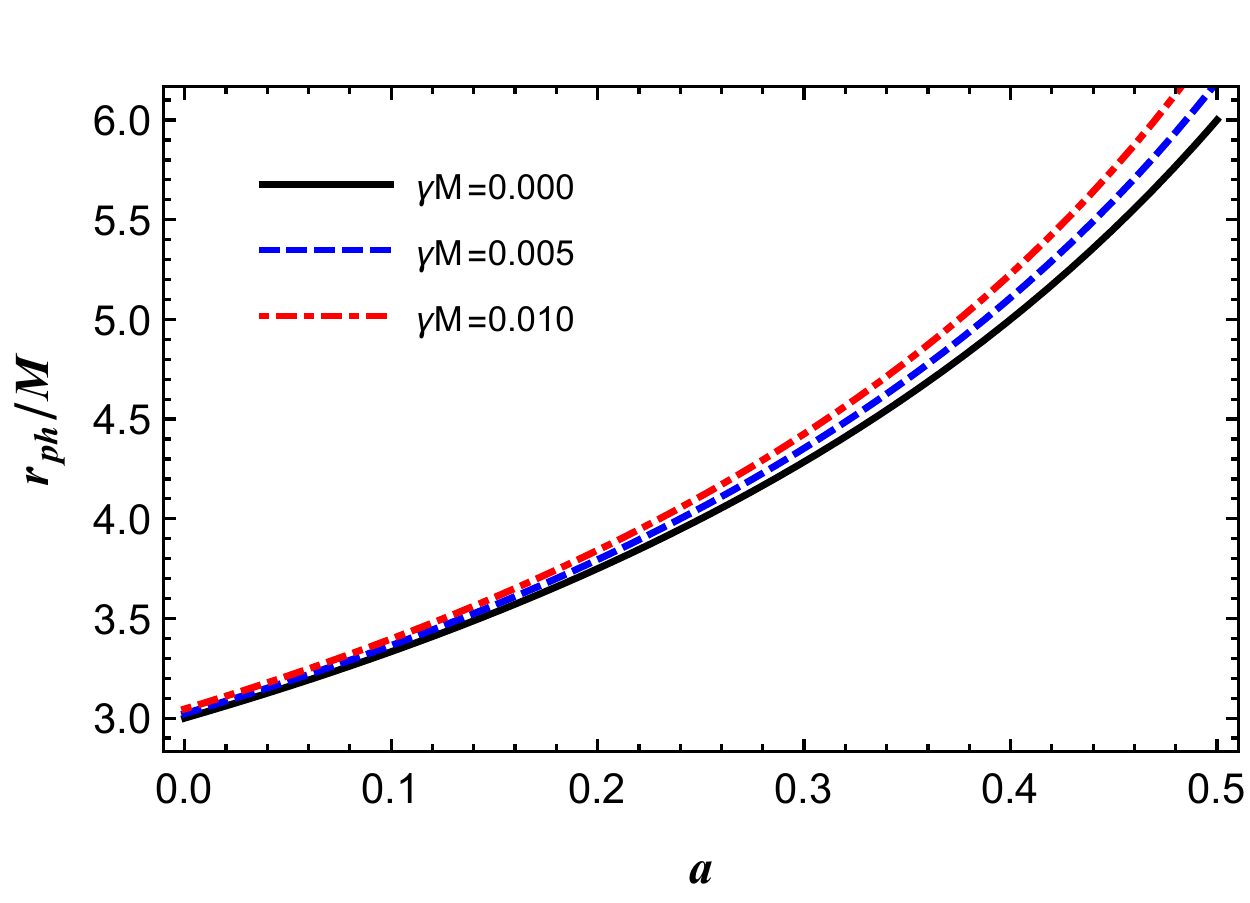}
\caption{The dependence of the photon orbits $r_{ph}$ on the quintessence parameter $\gamma$ (left panel) and on the string cloud parameter $a$ (right panel). \label{photonorbits}}
\end{figure*}

where $\lambda=0$ and $\lambda=1$ corresponds to the null and timelike geodesics respectively. By using the normalization condition, the equation of motion becomes
\begin{equation}\label{lag3}
\bigg(\frac{dr}{d\chi}\bigg)^{2}+ V_{eff}(r)=\mathcal{E}^{2},
\end{equation}
where $V_{eff}(r)$ denotes the effective potential and expressed as:
\begin{equation}\label{lag4}
V_{eff}(r)= f(r)\left(\lambda+\frac{{\mathfrak{L}}^2}{r^2}\right),
\end{equation}
For the current analysis $\mathcal{E}^2$ and $\mathfrak{L}^2$ are expressed as \cite{Mustafa2021}
\begin{eqnarray}
\mathcal{E}^{2}&&=\frac{2 r}{r^2 \left(-\frac{2}{r (a+\gamma r-1)+R_s}-\gamma\right)+R_s},\label{lag5}\\
\mathfrak{L}^{2}&&=\frac{r^4 \left(\gamma - \frac{R_s}{r^2}\right)}{r^2 \left(-\frac{2}{r (a+\gamma r-1)+R_s}-\gamma\right)+R_s}. \label{lag6}
\end{eqnarray}

We move then to consider the photon orbit around the Schwarzschild black hole surrounded by quintessential field in the string cloud background. The photon orbit $r_{ph}$ can be derived from the standard condition ${V}_{eff}'=0$ with $\lambda=0$ and we can write the orbit $r_{ph}$ as 
\begin{eqnarray}
r_{ph}&&=\frac{1-a-\sqrt{a^2-2 a-3R_s \gamma+1}}{\gamma}\, . \label{lag7}
\end{eqnarray}
From the above equation it is easily seen that the photon orbit $r_{ph}$ reduces to the one for the Schwarzschild case when $a\to0$ and $\gamma\to 0$. In Fig.~\ref{photonorbits}, we show the radius of the photon sphere with respect to the quintessence parameter $\gamma$ and string cloud parameter $a$. In Fig.~\ref{photonorbits}, left panel represents the effect of the string cloud parameter $a$ on the radius of the photon sphere. While the right panel shows the impact of the quintessence parameter $\gamma$ on the radius of the photon sphere. As can be seen from Fig.~\ref{photonorbits}, the radius of the photon sphere shifts upward as a consequence of an increase in the values of both the parameters $a$ and $\gamma$, thus resulting in an increase in the radius of the photon sphere.
\subsection{Black hole shadow}

In string theory the one-dimensional strings are considered to be the fundamental building blocks of nature instead of elementary particles. In gravity the one-dimensional analogue of cloud of dust is taken as cloud of string \cite{Toledoa}, to investigate the possible measurable effects of these clouds on the strong gravitational fields of black holes. Letelier was the first to generalize the Schwarzschild black hole solution in the presence of spherically symmetric static cloud of string and have obtained some interesting features of the resulting black hole spacetime \cite{SC0}. The Letelier generalization of the Schwarzschild black hole was in the sense that the metric of the black hole spacetime with clouds of strings corresponds locally to the geometry of the Schwarzschild spacetime with a solid deficit angle. from the metric of the Schwarzschild black hole with clouds of string and quintessence, given above, one can infer that the string cloud parameter $a$ is responsible for the solid deficit angle. The quintessence dark energy has its own role in the theory of gravity and especially in the surrounding of a black hole as discussed in the first section above. Consequently, in this subsection we explore the shadow of the Schwarzschild black hole in the string cloud background with quintessential dark energy. 
For the angular radius of the black hole shadow we consider \cite{Perlick2022rev,Atamurotov2022papnoi}
\begin{eqnarray}\label{shadow nonrotating1}
\sin^2\alpha_{sh}=\frac{h(r_{ph})^2}{h(r_{obs})^2},
\end{eqnarray}
where
\begin{eqnarray}
h(r)^2=\frac{g_{22}}{g_{00}}=\frac{r^2}{f(r)},\label{eq:h}
\end{eqnarray}
$\alpha_{sh}$ is the angular radius of the black hole shadow and $r_{obs}$ is the observe position. We assume that distant observer $r_{obs}$ is located at the cosmological radius $r_q$ that is located at the distance far away from the black hole. It is worth noting that in an astrophysical realistic scenario the quintessential field parameter $\gamma$ is supposed to be constant in the Universe and extremely small, and thus the cosmological horizon is supposed to be located at large distances.  The quantity $r_{ph}$ is the radius of the photon sphere as it was mentioned previously.

Now we combine Eqs.~(\ref{shadow nonrotating1}) and ~(\ref{eq:h}), and for an observer the Eq. (\ref{shadow nonrotating1}) takes the following form

\begin{eqnarray}\label{shadow nonrotating2}
\sin^2 \alpha_{sh}=\frac{r_{ph}^2}{f(r_{ph})}\frac{f(r_{obs})}{r^2_{obs}}.
\end{eqnarray}

One can find the radius of black hole shadow for observer at large distance using Eq. (\ref{shadow nonrotating2}) as \cite{Perlick2022rev}

\begin{eqnarray}\label{shadow nonrotating3}
R_{sh}=r_{obs} \sin \alpha_{sh} = \frac{r_{ph}}{\sqrt{f(r_{ph})}}{\sqrt{f(r_{obs})}}.
\end{eqnarray}

\begin{figure*}
\centering
 \includegraphics[width=0.31\textwidth]{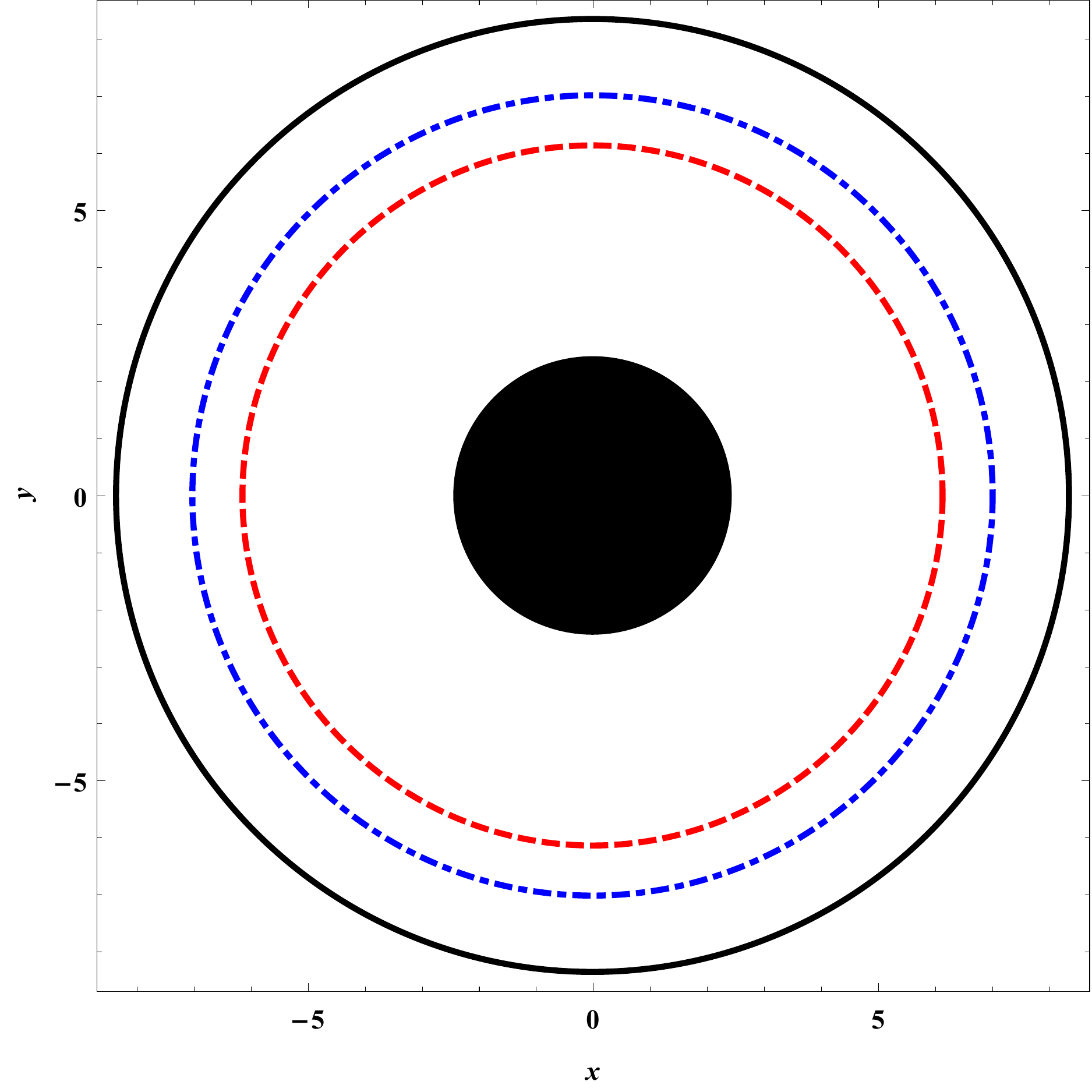}%
 \includegraphics[width=0.31\textwidth]{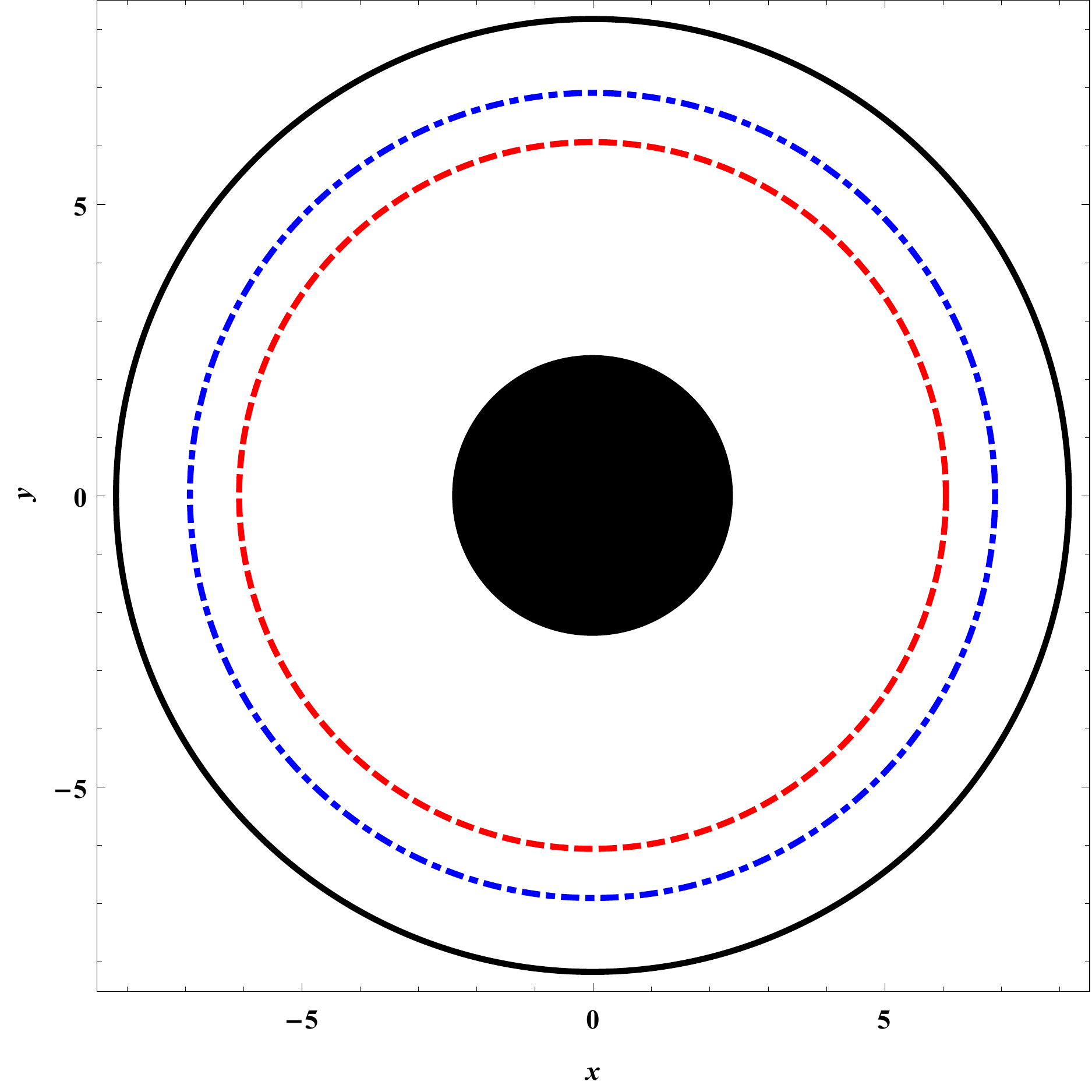}
 \includegraphics[width=0.31\textwidth]{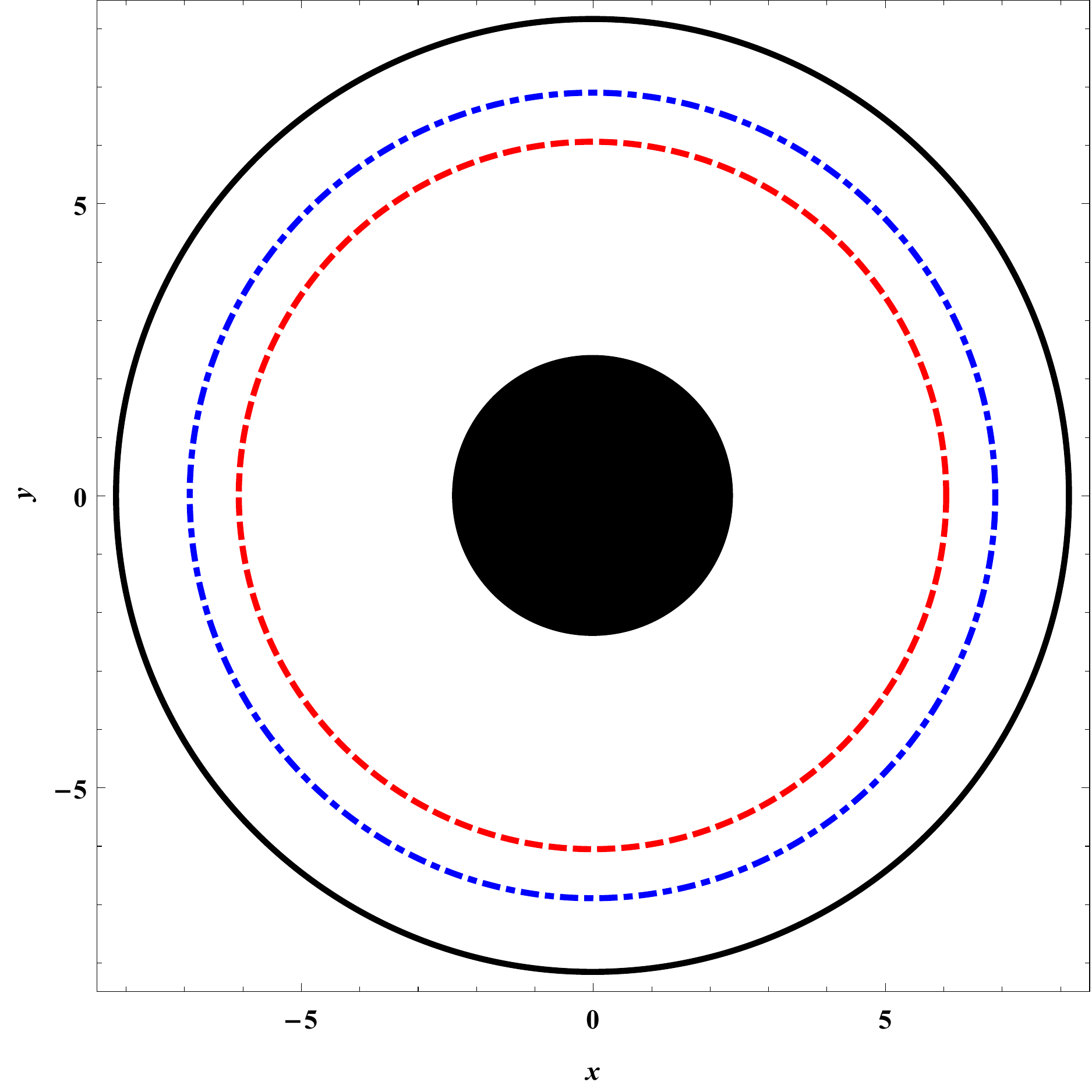}
\caption{Shows the shadows of black hole with $a=0.05$ (left panel), $a=0.03$ (middle panel), and $a=0.01$ (right panel) for three different values of quintessence parameter $\gamma M=0.05$, $\gamma M=0.03$, and $\gamma M=0.01$ outer circle to inner circle respectively. \label{F1}}
\end{figure*}

\begin{figure*}
\centering
 \includegraphics[width=0.31\textwidth]{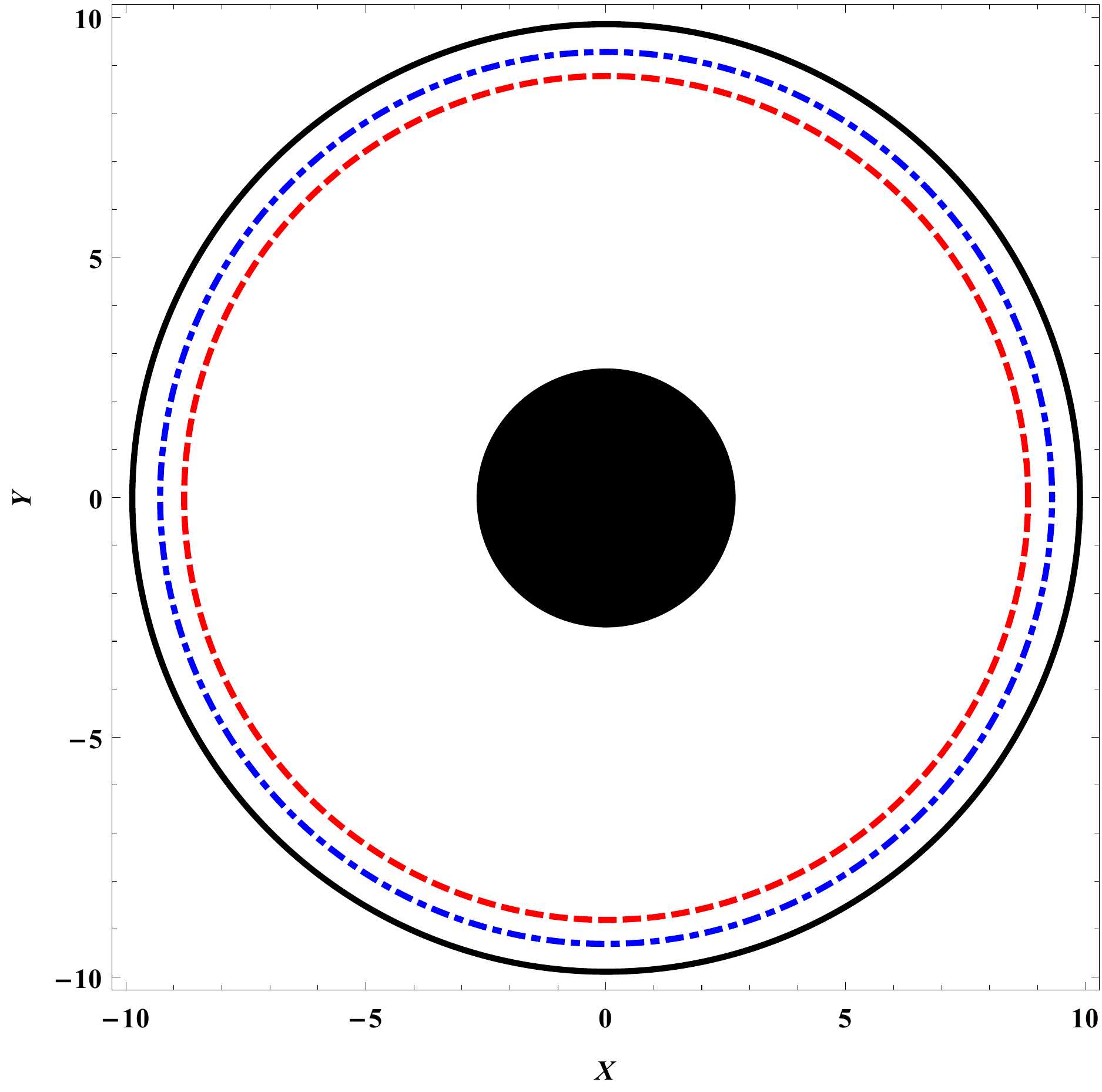}%
 \includegraphics[width=0.31\textwidth]{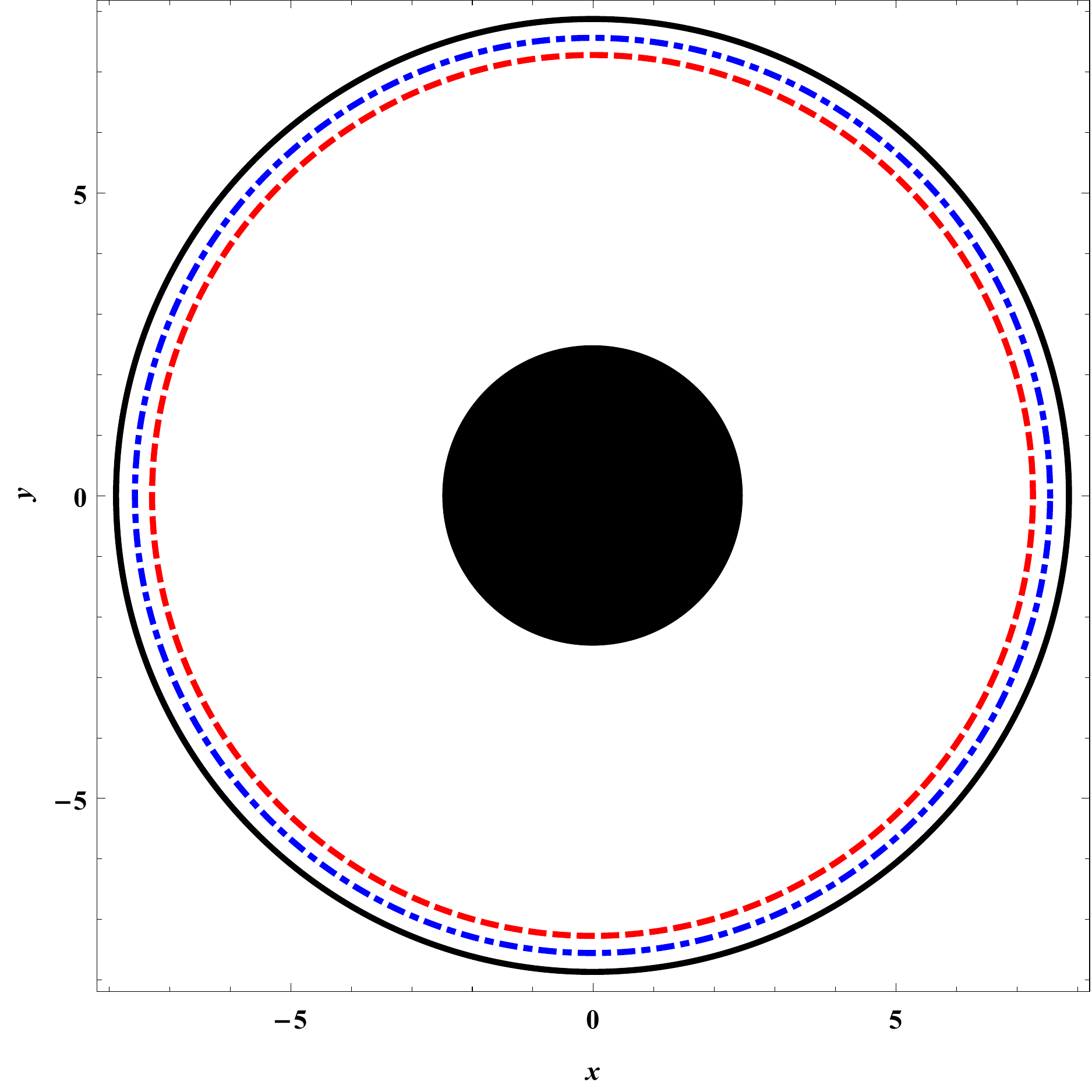}
 \includegraphics[width=0.31\textwidth]{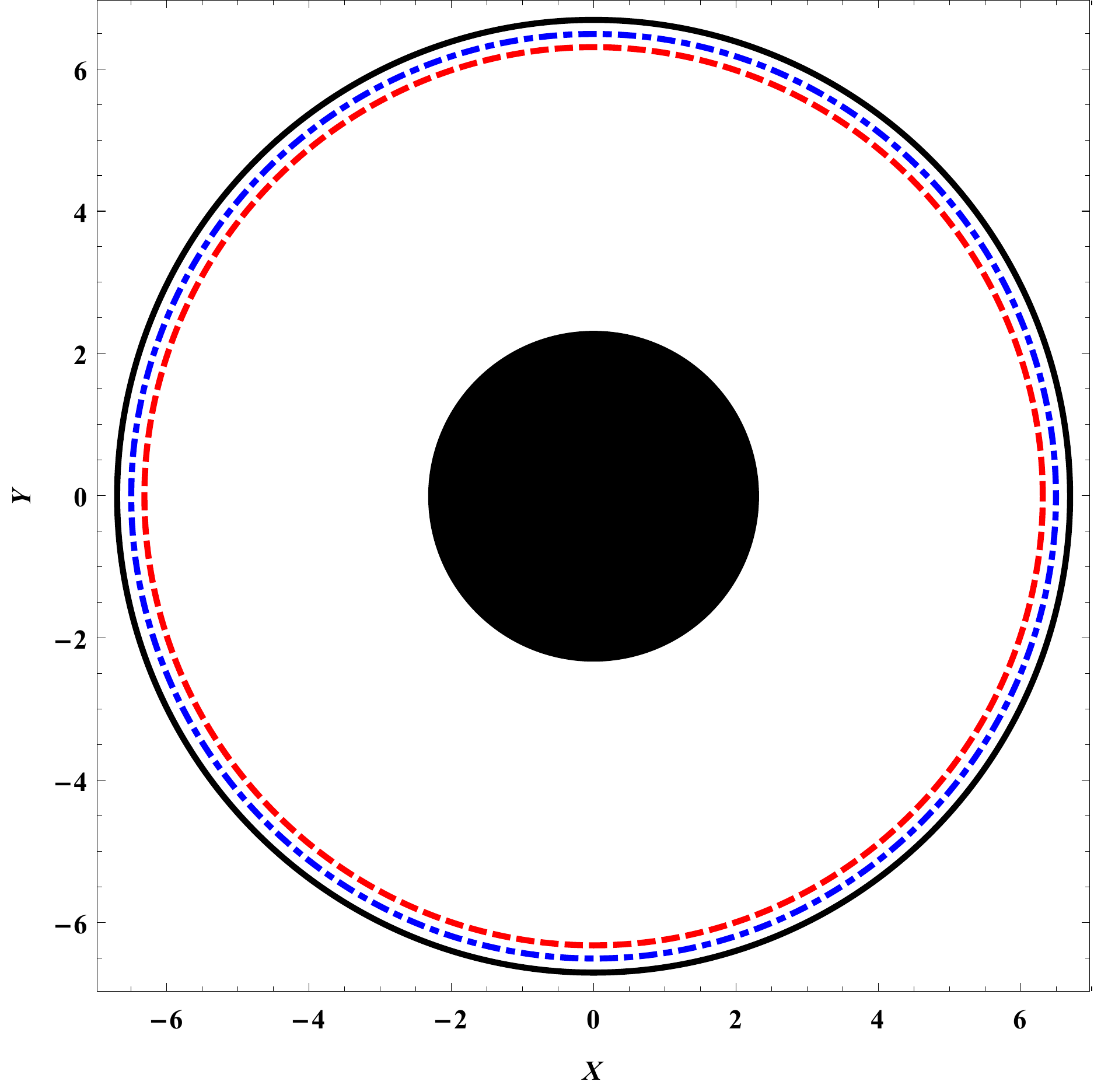}
\caption{Shows the shadows of black hole with $\gamma M=0.05$ (left panel), $\gamma M=0.03$ (middle panel), and $\gamma M=0.01$ (right panel) for three different values of string cloud parameter $a=0.05$, $a=0.03$, and $a=0.01$ outer circle to inner circle respectively. \label{F2}}
\end{figure*}

\begin{figure*}
\centering
 \includegraphics[width=0.31\textwidth]{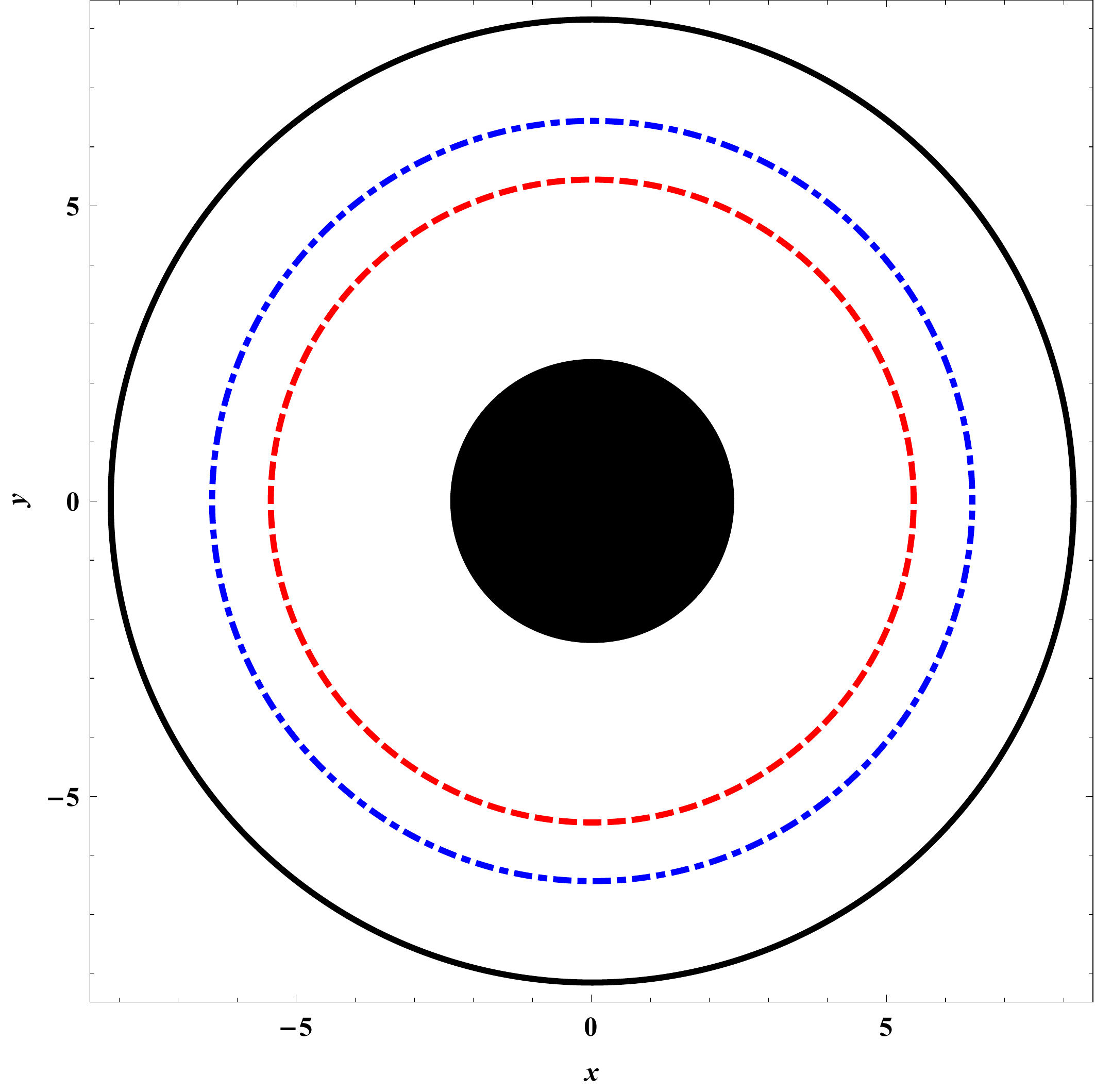}%
 \includegraphics[width=0.31\textwidth]{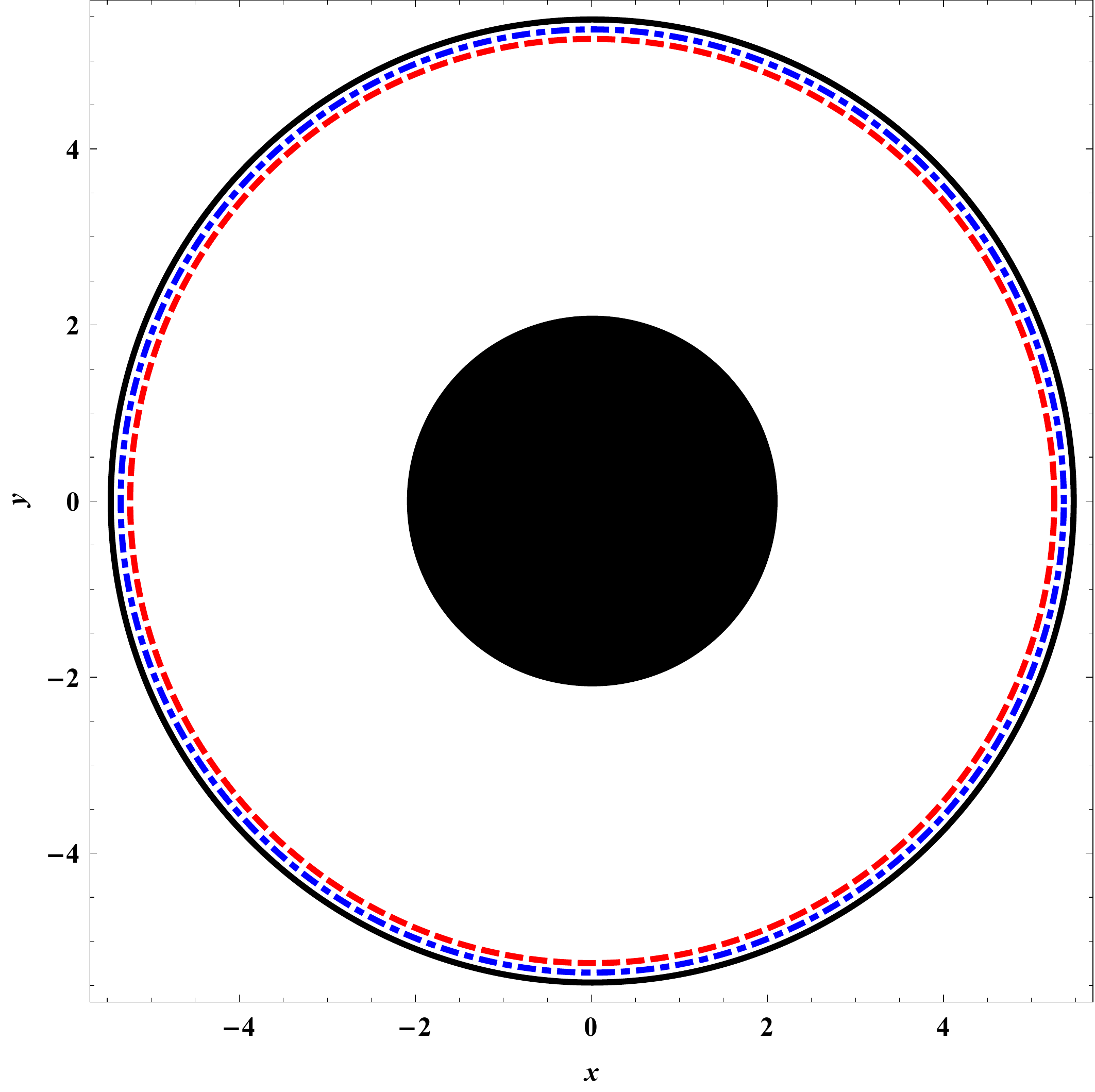}
 \includegraphics[width=0.31\textwidth]{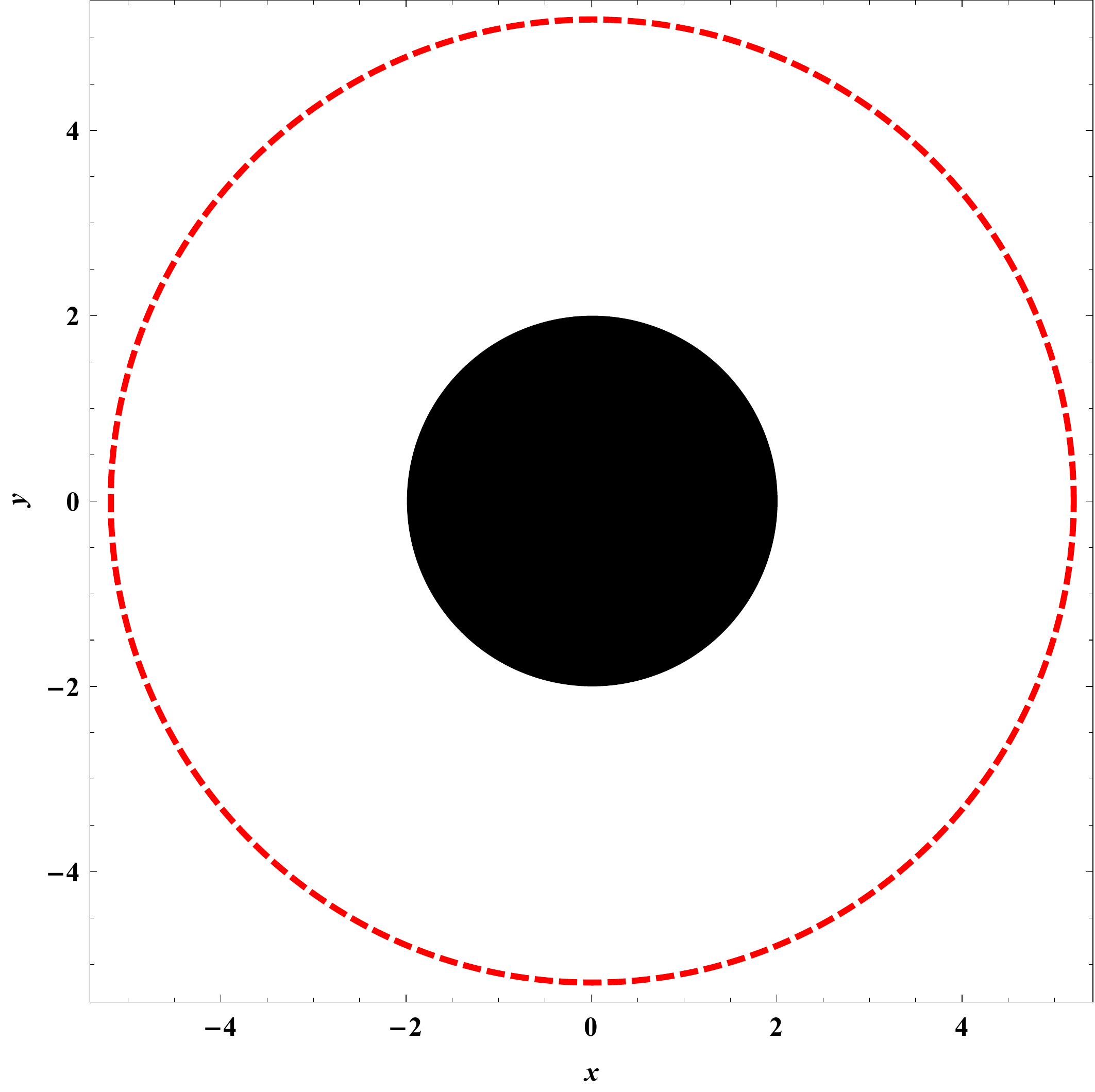}
\caption{Shows the shadows of black hole with $a=0.0$ (left panel) for $\gamma M=0.05$, $\gamma M=0.03$, and $\gamma M=0.01$.   Shadows of black hole with $\gamma M=0.0$ (right panel) for three different values of $a=0.05$, $a=0.03$, and $a=0.01$ outer circle to inner circle respectively. The last panel shows the Schwarzschild black hole shadow.\label{F3}}
\end{figure*}

\begin{figure*}
\centering
 \includegraphics[width=0.5\textwidth]{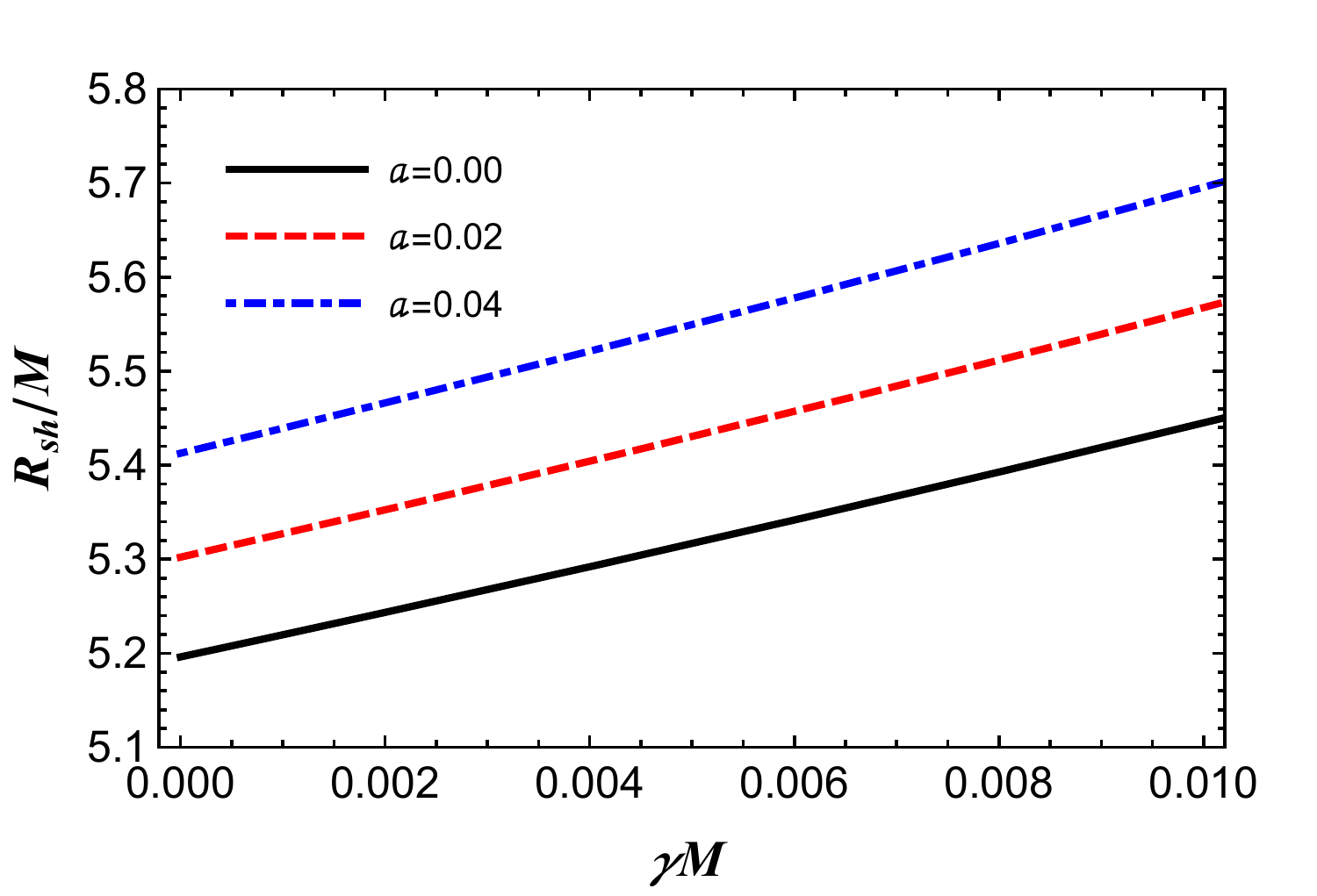}%
 \includegraphics[width=0.5\textwidth]{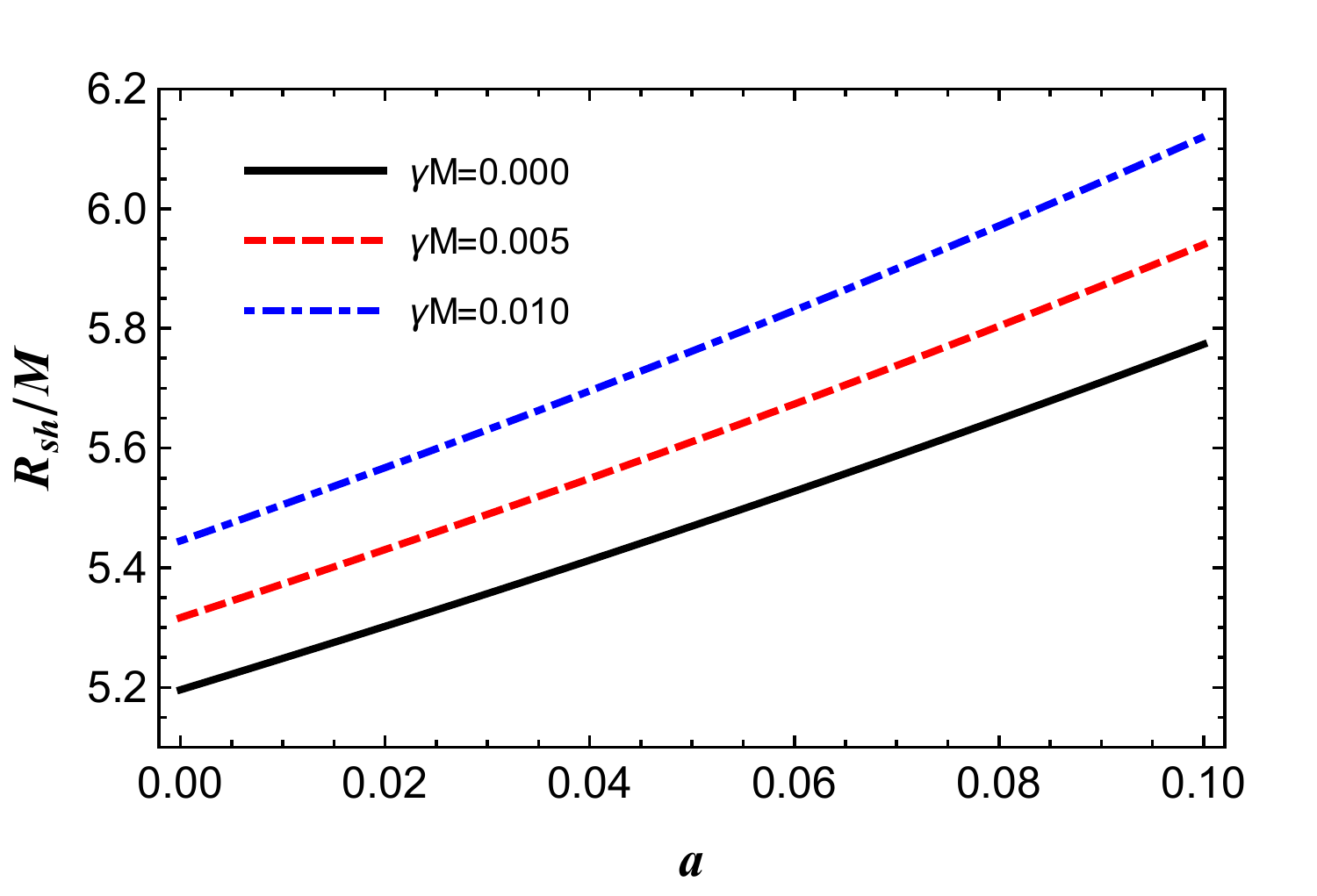}
\caption{The dependence of the black hole shadow radius on the quintessential field parameter $\gamma M$ and the string cloud parameter $a$. Left panel: the shadow radius is plotted for various combinations of $a$. Right panel: the shadow radius is plotted for various combinations of $\gamma M$.}    \label{plot:bhshadowlast}
\end{figure*}

Finally Eq. (\ref{shadow nonrotating3}) can explain the shadow of a static black hole. In order to discuss the size and the shape of the shadow of the Schwarzschild black holes surrounded by the clouds of string and quintessence, we may show black hole shadow plots using two celestial coordinates for the observer\cite{Jusufi2020obs,Hioki09}, namely, $X$ and $Y$ (where $R_{sh}=\sqrt{X^2+Y^2}$), using above equation black hole shadows are represented in Figs.~\ref{F1}, \ref{F2}, \ref{F3} or \ref{plot:bhshadowlast}.


\begin{figure*}
\centering
 \includegraphics[width=0.5\textwidth]{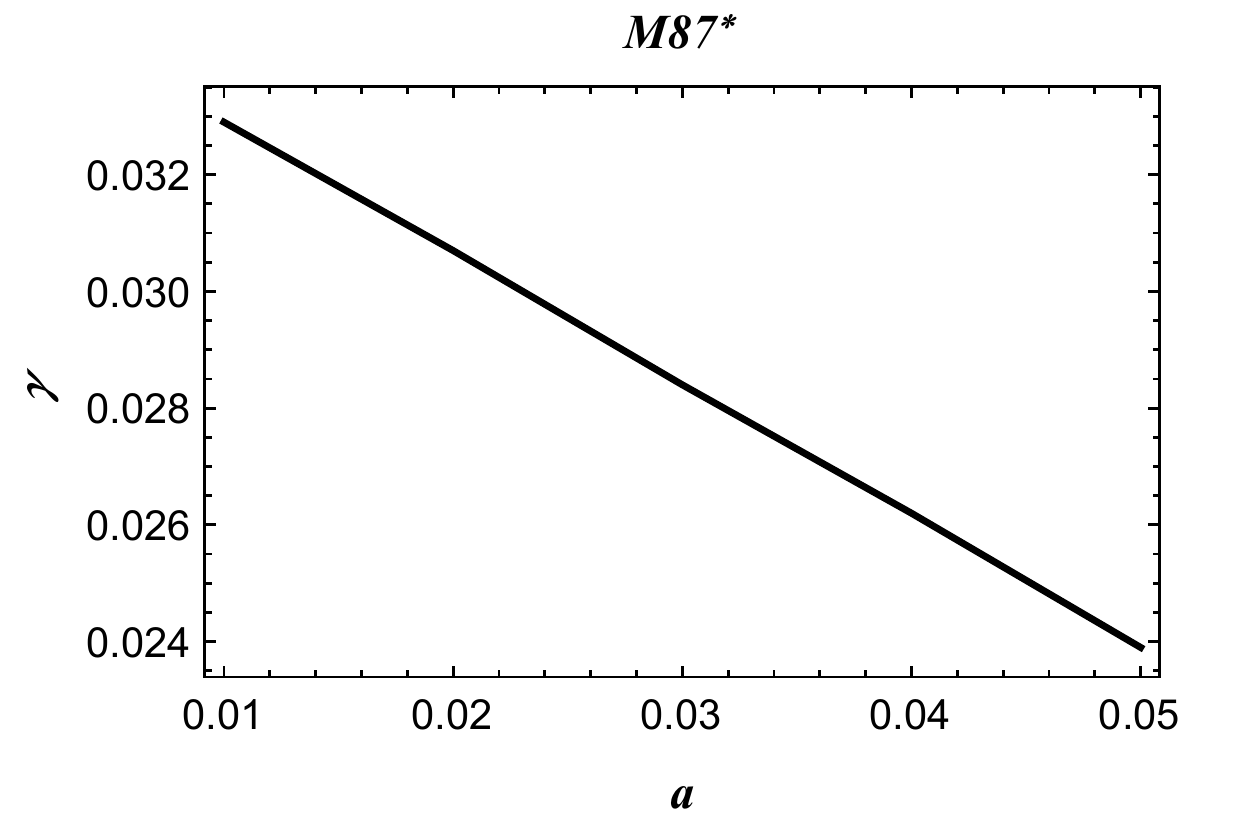}%
 \includegraphics[width=0.5\textwidth]{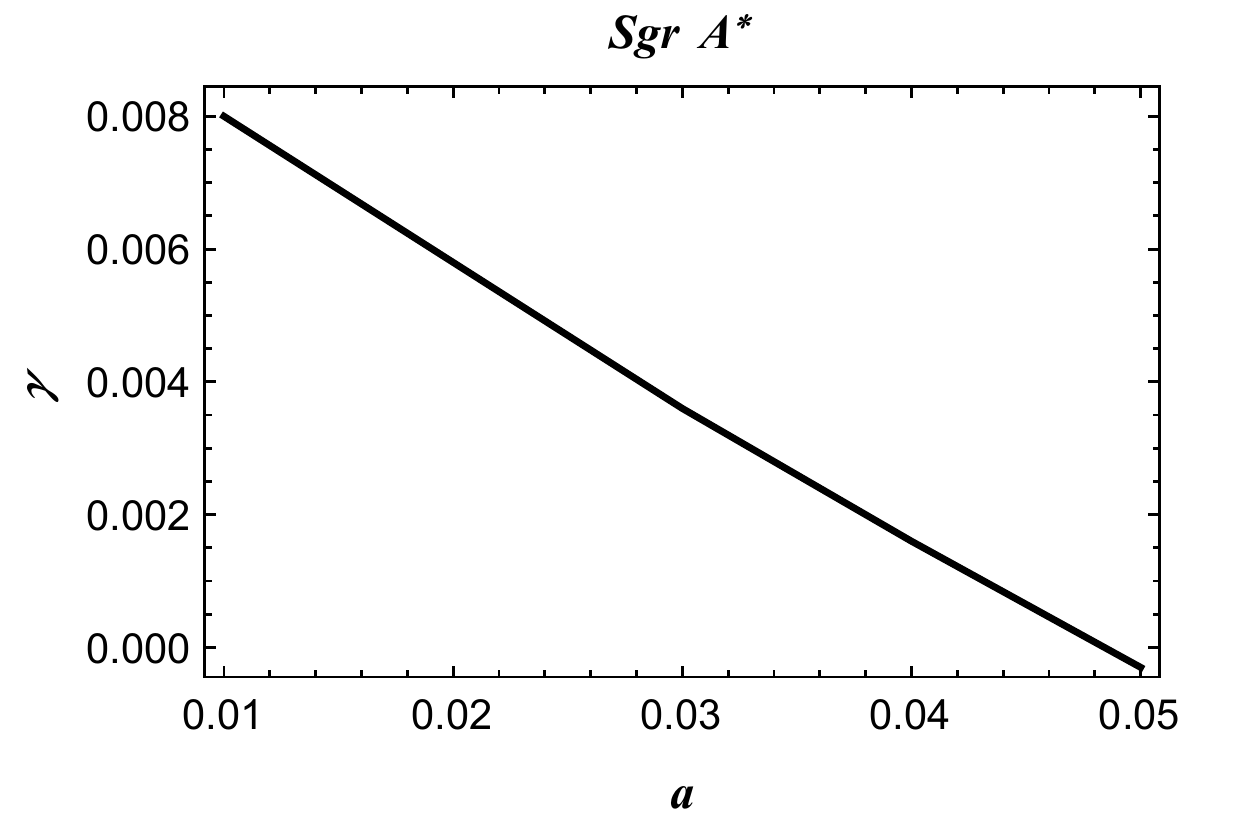}
\caption{Shows the quintessential field parameter as the function of string cloud parameter $a$ for the supermassive black holes in the galaxy M87 and the Sgr A$^{\star}$. Here, we have set $M=1$. \label{plot:con}}
\end{figure*}

From the Fig.~\ref{F1} above, we can see that for a fixed value of the quintessence parameter $\gamma$, the effect of the string cloud parameter $a$ on the radius of the black hole shadow is negligible. In the same figure we can observe that the radius of the black hole shadow gets bigger when the parameter $\gamma$ increases. This suggest that the gravitational field becomes stronger in the presence of the quintessence. This also confirms the repulsive nature of the quintessence dark energy. In the Fig.~\ref{F2}, for increasing $\gamma$ we see an increase in the radius of the shadow of the black hole. In the same figure we observe that the effect of the parameter $a$ on the radius of the black hole shadow is just similar to that of the parameter $\gamma$. Here we see that the radius of the shadow increases as the value of the parameter $a$ increases. This can be interpreted as that the existence of the string clouds may strengthen the gravitational field.  In the Fig.~\ref{F3}, we notice that if the parameter $a=0$, i.e there are no string clouds, then for the parameter $\gamma$ getting smaller we observe the radius of the shadow gets shrink. Similarly if there is no quintessence i.e $\gamma=0$, then for the parameter $a$ getting smaller, we have smaller radius of the shadow. In the right panel of the Fig.~\ref{F3}, we have the radius of the pure Schwarzschild black hole shadow. Here we see in the absence of the string clouds and quintessence, i.e when both the parameters $a=0$ and $\gamma=0$, we have smaller radius of the shadow. Hence the presence of both the string clouds and the quintessence push the black hole shadow outwards and therefore the combined effect of both the quintessences and cloud of strings is an attractive in nature and increases the strength of the gravitational field. The effect of both parameters is clearly shown in Fig.~\ref{plot:bhshadowlast}. This explicitly shows that $R_{sh}$ increases with increasing $a$ and $\gamma$. 
Further we noticed that the size of the Schwarzschild black hole shadow is more sensitive to the quintessence parameter than the string cloud parameter. 

For being a bit quantitative we try to explore theoretically the upper limits of parameters $\gamma$ and $a$. For that we constrain these two parameters using the observational data provided by the EHT collaboration for M87* and Sgr A* as a consequence of their shadows. For the M87* it is well known \cite{Akiyama19L1} that, the angular diameter of the shadow, the distance from the Earth and its mass have been reported as $\theta_\text{M87*} = 42 \pm 3 \:\mu$as,  $D = 16.8$ Mpc and $M_\text{M87*} = 6.5 \pm 0.90$x$10^9 \: M_\odot$, respectively. For the Sgr A* the data has been provided in recent EHT collaboration paper \cite{Akiyama2022sgr}. The above mentioned parameters for the M87*, in the case of the  Sgr A* are $\theta_\text{Sgr A*} = 48.7 \pm 7 \:\mu$as (EHT), $D = 8277\pm33$ pc and $M_\text{Sgr A*} = 4.3 \pm 0.013$x$10^6 \: M_\odot$ (VLTI) \cite{Akiyama2022sgr}. Based on the data, we are able to evaluate the diameter of the shadow size per unit mass using the following expression \cite{Bambi:2019tjh},
\begin{equation}
    d_\text{sh} = \frac{D \theta}{M}\,.
\end{equation}
One can then therefore be allowed to obtain the shadow diameter theoretically via $d_\text{sh}^\text{theo} = 2R_\text{sh}$. Following the literature \cite{Akiyama19L1,Bambi:2019tjh,Akiyama2022sgr}, it is then straightforward to obtain the diameter of the shadow image as follows: $d^\text{M87*}_\text{sh} = (11 \pm 1.5)M$ for M87* and  $d^\text{Sgr A*}_\text{sh} = (9.5 \pm 1.4)M$ for Sgr A*. Following the data results, we find the upper values of $\gamma$ and $a$ for the supermassive black holes in the galaxy M87 and the Sgr A$^{\star}$ and show their upper values in the Tables~\ref{table1}. We interestingly observe that the upper limit of the quintessential parameter $\gamma$ decreases once the string cloud parameter $a$ grows. For that what we can expect is that the effect that seems from the string cloud would be a bit larger on the geometry as compered to the one for the quintessential field. The behaviour of Table \ref{table1} is also demonstrated in Fig.~\ref{plot:con}. As can be seen from Fig.~\ref{plot:con} the upper threshold value of $\gamma$ would be larger for the supermassive black hole in the galaxy M87 as compered to the Sgr A$^{\star}$.

\begin{table}
\caption{The upper values of $\gamma$ and $a$ are tabulated for the supermassive black holes M87* and the Sgr A$^{\star}$. Note that we set $M=1$.} \label{table1}
\centering
\begin{tabular}{l l l l l l}
\hline\hline\noalign{\smallskip}
$a_{M87*}$& $0.01$   & $0.02$   & $0.03$ & $0.04$ & $0.05$   \\
$\gamma_{M87*}$   & 0.0329   &0.0307  &  0.0284 & 0.0262 & 0.0239  \\
\\ \hline\noalign{\smallskip}
$a_{Sgr A*}$& $0.01$   & $0.02$ & $0.03$ & $0.04$ & $0.046$   \\
$\gamma_{Sgr A*}$   & 0.0080 &0.0058  &  0.0036 & 0.0015 & 0.0000  \\
\\ \hline\hline\noalign{\smallskip}
\end{tabular}
\end{table}

\subsection{Rate of emission energy} \label{Sec:emission}
Due to the quantum fluctuations in a black hole spacetime the creation and annihilation of pairs of particles take place in the vicinity of the horizon of the black hole. During this process particles having positive energy escape from the black hole through the quantum tunneling. In the region where the Hawking-radiation takes place, the black hole evaporates in a definite period of time. Here in this subsection we consider the associated rate of the energy emission. Near a limiting constant value $\sigma_{lim}$, at a high energy the cross section of absorption of a black hole modulates slightly. As a consequence the shadow cast by the black hole causes the high energy cross section of absorption by the black hole for the observer located at finite distance $r_0$. The limiting constant value $\sigma_{lim}$, which is related to the radius of the photon sphere is given as \cite{Atamurotov2022papnoi}
\begin{equation}\label{sigmalast}
\sigma_{lim} \approx \pi R_{sh}^2,
\end{equation}
where $R_{sh}$ denotes the radius of the black hole shadow. The equation for the rate of the energy emission of a black hole is \cite{Atamurotov2022papnoi}
\begin{equation}\label{emissionenergyeq}
\frac{d^2 {\cal E}}{d\omega dt}= \frac{2 \pi^2 \sigma_{lim}}{\exp[{\omega/T}]-1}\omega^3,
\end{equation}
where $T=\kappa/2 \pi$ is the expression for the Hawking temperature and $\kappa$ is the notation used for the surface gravity. Combining Eq.(\ref{sigmalast}) with the Eq.(\ref{emissionenergyeq}), we arrive at an alternate form for the expression of emission energy rate as 

\begin{equation}\label{emissionenergyeqlast}
\frac{d^2 {\cal E}}{d\omega dt}= \frac{2\pi^3 R_{sh}^2}{e^{\omega/T}-1}\omega^3.
\end{equation}

\begin{figure*}
 \begin{center}
   \includegraphics[scale=0.65]{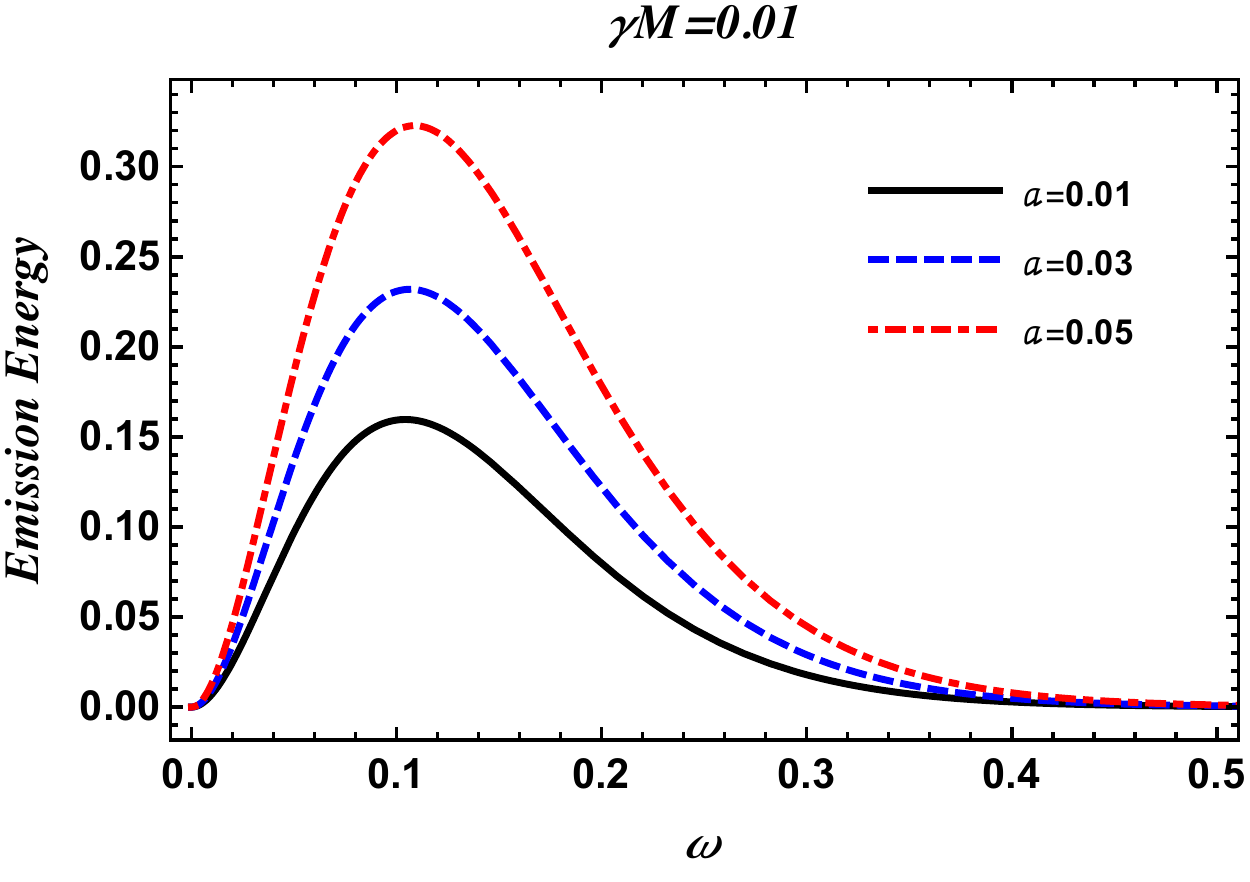}
   \includegraphics[scale=0.65]{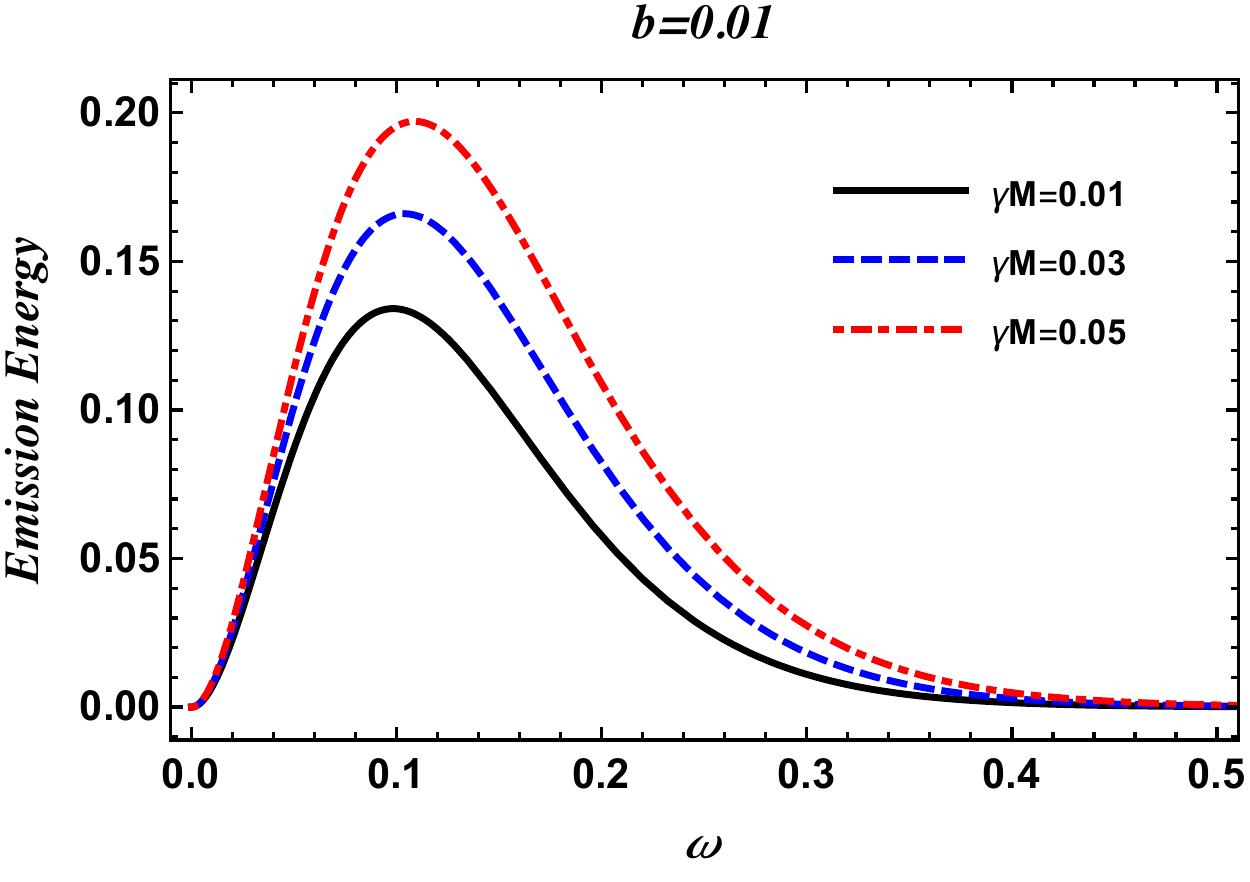}
  \end{center}
\caption{Plots showing the rate of energy emission varying with the frequency for different values of the cloud of string parameter $a$ (Left panel) and the quintessence parameter $\gamma$ (Right panel).}\label{plot:emission}
\end{figure*}

Variation of the energy emission rate with respect to the frequency of photon $\omega$, for different values of the parameters $a$ and $\gamma$ is represented in Fig.~\ref{plot:emission}. We see that with an increase in the values of the parameters $b$ and $\gamma$, the peak of the graph of the rate of the energy emission increases. This indicates that for the higher energy emission rate the evaporation process of the black hole is high.


\section{Weak Deflection Angle of the Schwarzschild black hole in the string cloud background with quintessential field using GBT}\label{Sec:GBT}

In this section, we study the effect of a string cloud on the weak deflection angle by the Schwarzschild black hole with quintessential energy. For this purpose, we will use the notion of the GBT.



First, we obtain the corresponding optical metric for the Schwarzschild black hole in the string cloud background with quintessential field given in \ref{20} as follows:

\begin{equation}
d \sigma ^ { 2 } = g _ { i j } ^ { \mathrm { opt } } d x ^ { i } d x ^ { j } =\frac{1}{f(r)}\left(\frac{d r^{2}}{f(r)}+r^2 d \phi^{2}\right),
\end{equation}

where $f(r)=(1-a-\frac{2 M}{r}-\frac{\gamma}{r^{3\omega_q+1}})$.

We obtain the Gaussian curvature for the above optical metric as follows: 

\begin{eqnarray}
\mathcal{K}\approx
\frac{9 a \gamma \omega _q}{2 r^3}-\frac{3 a \gamma \omega _q \log (r)}{r^3}+\frac{a \gamma}{r^3}+\frac{2 a M}{r^3}+\frac{6 \gamma M \omega _q}{r^4}\notag \\-\frac{9 \gamma M \omega _q \log (r)}{r^4}+\frac{3 \gamma M}{r^4}-\frac{9 \gamma \omega _q}{2 r^3}\notag \\+\frac{3 \gamma \omega _q \log (r)}{r^3}-\frac{\gamma}{r^3}-\frac{2 M}{r^3},
\end{eqnarray}

Note that there is a contribution from the cloud of strings on the Gaussian curvature. Then we have also
\begin{equation}
\frac{d\sigma}{d\phi}\bigg|_{C_{R}}=
\left( \frac {r^2 } { f ( R ) } \right) ^ { 1 / 2 },
\end{equation}

which has the following limit:
\begin{equation}
\lim_{R\to\infty} \kappa_g\frac{d\sigma}{d\phi}\bigg|_{C_R}\approx 1\,.
\end{equation}

At spatial infinity, $R\to\infty$, and by using the straight light approximation $r=b/\sin\phi$, the GBT reduces to \cite{Gibbons:2008rj}:

\begin{equation}
 \int^{\pi+\alpha}_0 \left[\kappa_g\frac{d\sigma}{d\phi}\right]\bigg|_{C_R}d\phi =\pi-\lim_{R\to\infty}\int^\pi_0\int^{\infty}_{\frac{b}{\sin\phi}}\mathcal{K} dS.
\end{equation}

We calculate the weak deflection angle in the weak limit approximation as follows:

\begin{eqnarray}\label{eq:perf}
\alpha \approx
\frac{4 M}{b}+\frac{2 \gamma}{b}+\frac{a \gamma}{b}+\frac{2 a M}{b}+\frac{9 \gamma \omega _q}{b}+\frac{27 a^2 \gamma \omega _q}{8 b} \notag \\+\frac{3 a^2 \gamma}{4 b}+\frac{3 a^2 M}{2 b}+\frac{9 a \gamma \omega _q}{2 b}.
\end{eqnarray}

\begin{figure}[hpt]
 \centering

   \includegraphics[scale=0.60]{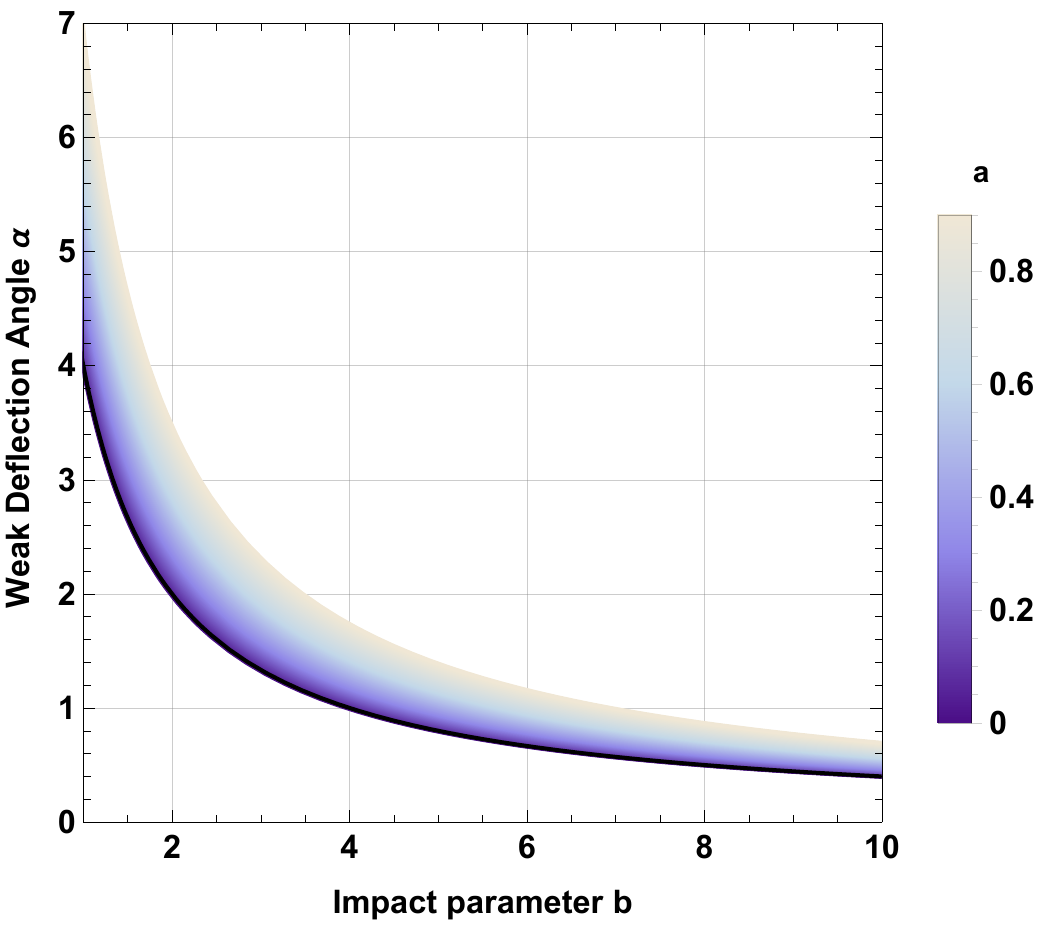}
    \includegraphics[scale=0.60]{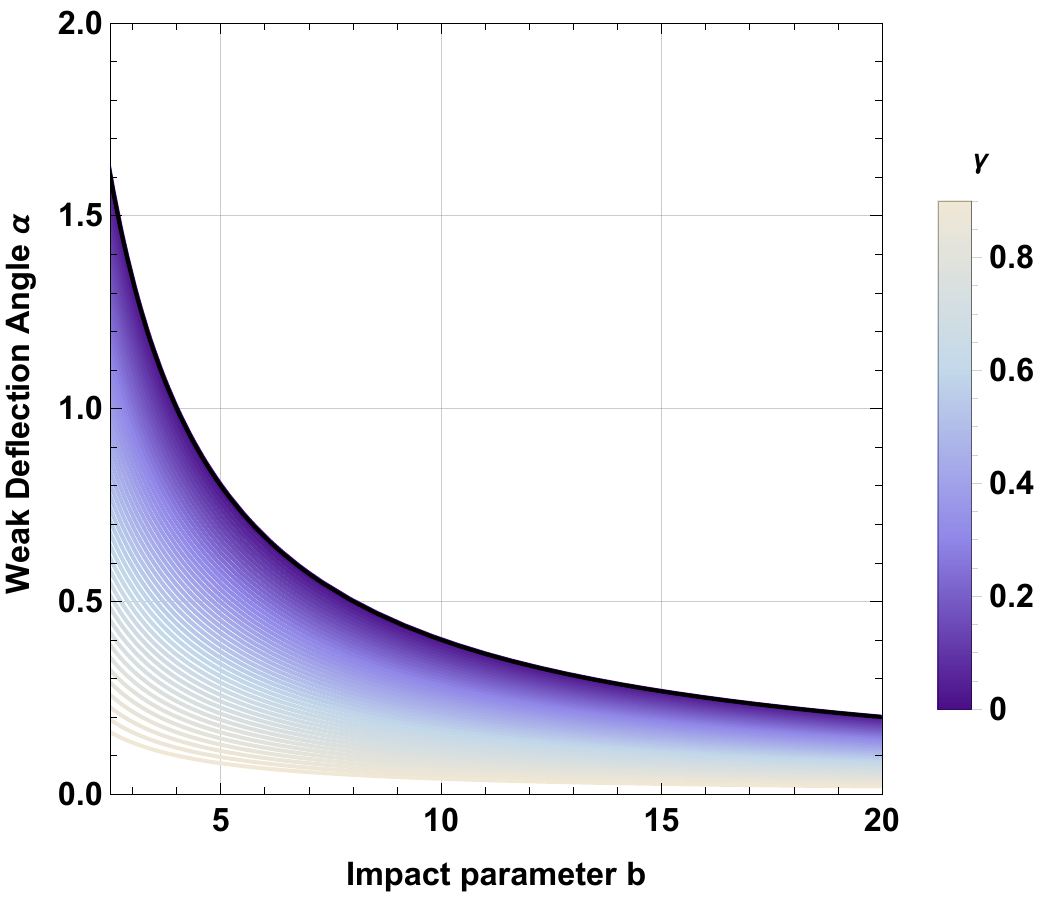}
\caption{ The first panel shows the dependence of $\alpha$ on string cloud parameter $a$, for fixed  $M=1$, $\omega_q=-2/3$ and $\gamma=0.01$, where it is compared with the Schwarzschild black hole case (plot with a black solid line). The second panel shows the dependence of $\alpha$ on the parameter $\gamma$,  for fixed $a = 0.01$,  $M=1$, $\omega_q=-2/3$, where it is compared with the Schwarzschild black hole case (plot with a black solid line).}\label{plot:deflectionangle1}

  \end{figure}

Hence, we see the effect of the string cloud on the deflection angle in weak field limits using the GBT.  As a result, the cloud of string parameter $a$ and the quintessence parameter $\gamma$ increase the deflection angle $\alpha$, and can be seen from the Eq. \ref{eq:perf} for positive values of $\omega _q$, (similarly in the paper \cite{Sharif:2015oua}). However for negative values of $\omega _q$  deflection angle $\alpha$ decreases as can be seen in Fig. \ref{plot:deflectionangle1}.

\section{Conclusions}
\label{Sec:conclusion}

In this work we have considered the Schwarzschild black hole in the background of string clouds and quintessence field to look at the shadows and the gravitational deflections angle of photons. In the Fig. \ref{H1} the black hole horizon radius is plotted against the string cloud parameter $a$ and the quintessence parameter $\gamma$. We have seen that with both the parameters $a$ and $\gamma$ the horizon of the Schwarzschild black hole increases. From the Fig. \ref{photonorbits} We have concluded that the radius of the photon sphere for $\omega_q=-2/3$, increases if the parameters $a$ and $\gamma$ get bigger. We have observed the impact of the string cloud parameter $a$ and the quintessence parameter $\gamma$ on both the shadow and the deflection angle in the Schwarzschild spacetime in the presence of the cloud of strings and quintessence dark energy. In the Figs. \ref{F1}, \ref{F2}, \ref{F3} and \ref{plot:bhshadowlast} the dependencies of the shadows of the Schwarzschild black hole with quintessence and cloud of strings on the parameters $a$ and $\gamma$ are presented. In these graphs we have noticed that with increase in the values of the parameters $a$ and $\gamma$ the radius of the shadow cast by the Schwarzschild black hole with quintessence and string clouds increases. We have shown that if both the parameters $a=0$ and $\gamma=0$, the black hole shadow is smaller as compared with the one obtained in the case of nonzero values of these parameters $a$ and $\gamma$. This fact is evident from the right panel of the Fig. \ref{F3}. From this it may be speculated that the black hole shadow can be bigger in the presence of string clouds and quintessence for spherically symmetric static black holes, and the effect of both the quintessence and string cloud is repulsive in this scenario. As the size of the shadow of the black hoe gets bigger, this suggest that the intensity of the gravitational field of the Schwarzschild black hole is increased in the presence of the string clouds and quintessence dark energy. Another observation is that the radius of the shadow of the black hole is more sensitive to the quintessence parameter $\gamma$ as compared to the string cloud parameter $a$. We have used the observational data provided by the EHT collaboration for the shadows of the black holes at the centre of the Messier 87 and the Milky way galaxies, to obtain upper limits for the values of the parameters $a$ and $\gamma$. These values are presented in the Table~\ref{table1} given in the Sec. III above. 

The evaporation of a black hole by the Hawking radiation can be explained in terms of the emission rate of energy. The emission rate of energy of a black hole is therefore attracts researchers and has linked with the radius of the black hole shadow. Therefore, in this study we have also considered the emission rate of energy for the Schwarzschild black hole with string clouds and quintessence. in the Fig.~\ref{plot:emission}, we have seen that with increase in the the string cloud parameters $a$ and quintessence parameter $\gamma$, the peak of the graph for the rate of the energy emission increases. This going up of the peak of the graph indicates that for the higher rate of the emission energy, the rate of evaporation of the black hole is also high. 

Last we have also calculated the weak deflection angle using the GBT to probe the effect of the cloud of string parameter $a$, and quintessence parameter $\gamma$ on the weak deflection angle $\alpha$. We have shown that the deflection angle obtained in Eq. \eqref{eq:perf} increases with the cloud of string parameter $a$, but decreases with the the quintessence parameter $\gamma$ for negative values of $\omega _q$, shown in Fig. \ref{plot:deflectionangle1}.

\section*{Data Availability Statement}
This manuscript has no associated data, or the data will not be deposited. (There are no observational data related to this
article. The necessary calculations and graphic discussion can be made available on request.)

\section*{Acknowledgments}
G. Mustafa is very thankful to Prof. Gao Xianlong from the Department of Physics, Zhejiang Normal University, for his kind support and help during this research. Further, G. Mustafa acknowledges the Grant No. ZC304022919 to support his Postdoctoral Fellowship at Zhejiang Normal University. F.A. acknowledges the support of Inha University in Tashkent and research work has been supported by the Visitor Research Fellowship at Zhejiang Normal University. This research is partly supported by Research Grant FZ-20200929344; F-FA-2021-510 and F-FA-2021-432 of the Uzbekistan Ministry for Innovative Development.

\bibliographystyle{apsrev4-1}  
\bibliography{gravreferences,Lensing}

\begin{thebibliography}{110}%
\makeatletter
\providecommand \@ifxundefined [1]{%
 \@ifx{#1\undefined}
}%
\providecommand \@ifnum [1]{%
 \ifnum #1\expandafter \@firstoftwo
 \else \expandafter \@secondoftwo
 \fi
}%
\providecommand \@ifx [1]{%
 \ifx #1\expandafter \@firstoftwo
 \else \expandafter \@secondoftwo
 \fi
}%
\providecommand \natexlab [1]{#1}%
\providecommand \enquote  [1]{``#1''}%
\providecommand \bibnamefont  [1]{#1}%
\providecommand \bibfnamefont [1]{#1}%
\providecommand \citenamefont [1]{#1}%
\providecommand \href@noop [0]{\@secondoftwo}%
\providecommand \href [0]{\begingroup \@sanitize@url \@href}%
\providecommand \@href[1]{\@@startlink{#1}\@@href}%
\providecommand \@@href[1]{\endgroup#1\@@endlink}%
\providecommand \@sanitize@url [0]{\catcode `\\12\catcode `\$12\catcode
  `\&12\catcode `\#12\catcode `\^12\catcode `\_12\catcode `\%12\relax}%
\providecommand \@@startlink[1]{}%
\providecommand \@@endlink[0]{}%
\providecommand \url  [0]{\begingroup\@sanitize@url \@url }%
\providecommand \@url [1]{\endgroup\@href {#1}{\urlprefix }}%
\providecommand \urlprefix  [0]{URL }%
\providecommand \Eprint [0]{\href }%
\providecommand \doibase [0]{http://dx.doi.org/}%
\providecommand \selectlanguage [0]{\@gobble}%
\providecommand \bibinfo  [0]{\@secondoftwo}%
\providecommand \bibfield  [0]{\@secondoftwo}%
\providecommand \translation [1]{[#1]}%
\providecommand \BibitemOpen [0]{}%
\providecommand \bibitemStop [0]{}%
\providecommand \bibitemNoStop [0]{.\EOS\space}%
\providecommand \EOS [0]{\spacefactor3000\relax}%
\providecommand \BibitemShut  [1]{\csname bibitem#1\endcsname}%
\let\auto@bib@innerbib\@empty
\bibitem [{\citenamefont {{Abbott}}\ and\ \citenamefont {et~al. {(Virgo and
  LIGO Scientific Collaborations)}}(2016)}]{Abbott16a}%
  \BibitemOpen
  \bibfield  {author} {\bibinfo {author} {\bibfnamefont {B.~P.}\ \bibnamefont
  {{Abbott}}}\ and\ \bibinfo {author} {\bibnamefont {et~al. {(Virgo and LIGO
  Scientific Collaborations)}}},\ }\href {\doibase
  10.1103/PhysRevLett.116.061102} {\bibfield  {journal} {\bibinfo  {journal}
  {Phys. Rev. Lett.}\ }\textbf {\bibinfo {volume} {116}},\ \bibinfo {eid}
  {061102} (\bibinfo {year} {2016})},\ \Eprint
  {http://arxiv.org/abs/1602.03837} {arXiv:1602.03837 [gr-qc]} \BibitemShut
  {NoStop}%
\bibitem [{\citenamefont {{Akiyama}}\ and\ \citenamefont {et~al. {(Event
  Horizon Telescope Collaboration)}}(2019{\natexlab{a}})}]{Akiyama19L1}%
  \BibitemOpen
  \bibfield  {author} {\bibinfo {author} {\bibfnamefont {K.}~\bibnamefont
  {{Akiyama}}}\ and\ \bibinfo {author} {\bibnamefont {et~al. {(Event Horizon
  Telescope Collaboration)}}},\ }\href {\doibase 10.3847/2041-8213/ab0ec7}
  {\bibfield  {journal} {\bibinfo  {journal} {Astrophys. J.}\ }\textbf
  {\bibinfo {volume} {875}},\ \bibinfo {eid} {L1} (\bibinfo {year}
  {2019}{\natexlab{a}})},\ \Eprint {http://arxiv.org/abs/1906.11238}
  {arXiv:1906.11238 [astro-ph.GA]} \BibitemShut {NoStop}%
\bibitem [{\citenamefont {{Akiyama}}\ and\ \citenamefont {et~al. {(Event
  Horizon Telescope Collaboration)}}(2019{\natexlab{b}})}]{Akiyama19L6}%
  \BibitemOpen
  \bibfield  {author} {\bibinfo {author} {\bibfnamefont {K.}~\bibnamefont
  {{Akiyama}}}\ and\ \bibinfo {author} {\bibnamefont {et~al. {(Event Horizon
  Telescope Collaboration)}}},\ }\href {\doibase 10.3847/2041-8213/ab1141}
  {\bibfield  {journal} {\bibinfo  {journal} {Astrophys. J.}\ }\textbf
  {\bibinfo {volume} {875}},\ \bibinfo {eid} {L6} (\bibinfo {year}
  {2019}{\natexlab{b}})},\ \Eprint {http://arxiv.org/abs/1906.11243}
  {arXiv:1906.11243 [astro-ph.GA]} \BibitemShut {NoStop}%
\bibitem [{\citenamefont {{Cruz}}\ \emph {et~al.}(2005)\citenamefont {{Cruz}},
  \citenamefont {{Olivares}},\ and\ \citenamefont {{Villanueva}}}]{Cruz05}%
  \BibitemOpen
  \bibfield  {author} {\bibinfo {author} {\bibfnamefont {N.}~\bibnamefont
  {{Cruz}}}, \bibinfo {author} {\bibfnamefont {M.}~\bibnamefont {{Olivares}}},
  \ and\ \bibinfo {author} {\bibfnamefont {J.~R.}\ \bibnamefont
  {{Villanueva}}},\ }\href {\doibase 10.1088/0264-9381/22/6/016} {\bibfield
  {journal} {\bibinfo  {journal} {Class. Quantum Grav.}\ }\textbf {\bibinfo
  {volume} {22}},\ \bibinfo {pages} {1167} (\bibinfo {year} {2005})},\ \Eprint
  {http://arxiv.org/abs/gr-qc/0408016} {gr-qc/0408016} \BibitemShut {NoStop}%
\bibitem [{\citenamefont {{Stuchl{\'{\i}}k}}\ and\ \citenamefont
  {{Schee}}(2011)}]{Stuchlik11}%
  \BibitemOpen
  \bibfield  {author} {\bibinfo {author} {\bibfnamefont {Z.}~\bibnamefont
  {{Stuchl{\'{\i}}k}}}\ and\ \bibinfo {author} {\bibfnamefont {J.}~\bibnamefont
  {{Schee}}},\ }\href {\doibase 10.1088/1475-7516/2011/09/018} {\bibfield
  {journal} {\bibinfo  {journal} {Journal of Cosmology and Astroparticles}\
  }\textbf {\bibinfo {volume} {9}},\ \bibinfo {eid} {018} (\bibinfo {year}
  {2011})}\BibitemShut {NoStop}%
\bibitem [{\citenamefont {{Grenon}}\ and\ \citenamefont
  {{Lake}}(2010)}]{Grenon10}%
  \BibitemOpen
  \bibfield  {author} {\bibinfo {author} {\bibfnamefont {C.}~\bibnamefont
  {{Grenon}}}\ and\ \bibinfo {author} {\bibfnamefont {K.}~\bibnamefont
  {{Lake}}},\ }\href {\doibase 10.1103/PhysRevD.81.023501} {\bibfield
  {journal} {\bibinfo  {journal} {Phys. Rev. D}\ }\textbf {\bibinfo {volume}
  {81}},\ \bibinfo {eid} {023501} (\bibinfo {year} {2010})},\ \Eprint
  {http://arxiv.org/abs/0910.0241} {arXiv:0910.0241 [astro-ph.CO]} \BibitemShut
  {NoStop}%
\bibitem [{\citenamefont {{Rezzolla}}\ \emph {et~al.}(2003)\citenamefont
  {{Rezzolla}}, \citenamefont {{Zanotti}},\ and\ \citenamefont
  {{Font}}}]{Rezzolla03a}%
  \BibitemOpen
  \bibfield  {author} {\bibinfo {author} {\bibfnamefont {L.}~\bibnamefont
  {{Rezzolla}}}, \bibinfo {author} {\bibfnamefont {O.}~\bibnamefont
  {{Zanotti}}}, \ and\ \bibinfo {author} {\bibfnamefont {J.~A.}\ \bibnamefont
  {{Font}}},\ }\href {\doibase 10.1051/0004-6361:20031457} {\bibfield
  {journal} {\bibinfo  {journal} {Astronomy and Astrophysics}\ }\textbf
  {\bibinfo {volume} {412}},\ \bibinfo {pages} {603} (\bibinfo {year}
  {2003})},\ \Eprint {http://arxiv.org/abs/gr-qc/0310045} {gr-qc/0310045}
  \BibitemShut {NoStop}%
\bibitem [{\citenamefont {{Arraut}}(2015)}]{Arraut15}%
  \BibitemOpen
  \bibfield  {author} {\bibinfo {author} {\bibfnamefont {I.}~\bibnamefont
  {{Arraut}}},\ }\href {\doibase 10.1142/S0218271815500224} {\bibfield
  {journal} {\bibinfo  {journal} {Int. J. Mod. Phys. D}\ }\textbf {\bibinfo
  {volume} {24}},\ \bibinfo {eid} {1550022} (\bibinfo {year}
  {2015})}\BibitemShut {NoStop}%
\bibitem [{\citenamefont {{Faraoni}}(2015)}]{Faraoni15}%
  \BibitemOpen
  \bibinfo {editor} {\bibfnamefont {V.}~\bibnamefont {{Faraoni}}},\ ed.,\ \href
  {\doibase 10.1007/978-3-319-19240-6} {\emph {\bibinfo {title} {Lecture Notes
  in Physics, Berlin Springer Verlag}}},\ \bibinfo {series} {Lecture Notes in
  Physics, Berlin Springer Verlag}, Vol.\ \bibinfo {volume} {907}\ (\bibinfo
  {year} {2015})\BibitemShut {NoStop}%
\bibitem [{\citenamefont {{Shaymatov}}\ \emph {et~al.}(2018)\citenamefont
  {{Shaymatov}}, \citenamefont {{Ahmedov}}, \citenamefont {{Stuchl{\'\i}k}},\
  and\ \citenamefont {{Abdujabbarov}}}]{Shaymatov18a}%
  \BibitemOpen
  \bibfield  {author} {\bibinfo {author} {\bibfnamefont {S.}~\bibnamefont
  {{Shaymatov}}}, \bibinfo {author} {\bibfnamefont {B.}~\bibnamefont
  {{Ahmedov}}}, \bibinfo {author} {\bibfnamefont {Z.}~\bibnamefont
  {{Stuchl{\'\i}k}}}, \ and\ \bibinfo {author} {\bibfnamefont {A.}~\bibnamefont
  {{Abdujabbarov}}},\ }\href {\doibase 10.1142/S0218271818500888} {\bibfield
  {journal} {\bibinfo  {journal} {International Journal of Modern Physics D}\
  }\textbf {\bibinfo {volume} {27}},\ \bibinfo {eid} {1850088} (\bibinfo {year}
  {2018})}\BibitemShut {NoStop}%
\bibitem [{\citenamefont {{Peebles}}\ and\ \citenamefont
  {{Ratra}}(2003)}]{Peebles03}%
  \BibitemOpen
  \bibfield  {author} {\bibinfo {author} {\bibfnamefont {P.~J.}\ \bibnamefont
  {{Peebles}}}\ and\ \bibinfo {author} {\bibfnamefont {B.}~\bibnamefont
  {{Ratra}}},\ }\href {\doibase 10.1103/RevModPhys.75.559} {\bibfield
  {journal} {\bibinfo  {journal} {Reviews of Modern Physics}\ }\textbf
  {\bibinfo {volume} {75}},\ \bibinfo {pages} {559} (\bibinfo {year} {2003})},\
  \Eprint {http://arxiv.org/abs/astro-ph/0207347} {astro-ph/0207347}
  \BibitemShut {NoStop}%
\bibitem [{\citenamefont {{Wetterich}}(1988)}]{Wetterich88}%
  \BibitemOpen
  \bibfield  {author} {\bibinfo {author} {\bibfnamefont {C.}~\bibnamefont
  {{Wetterich}}},\ }\href {\doibase 10.1016/0550-3213(88)90193-9} {\bibfield
  {journal} {\bibinfo  {journal} {Nuclear Physics B}\ }\textbf {\bibinfo
  {volume} {302}},\ \bibinfo {pages} {668} (\bibinfo {year}
  {1988})}\BibitemShut {NoStop}%
\bibitem [{\citenamefont {{Caldwell}}\ and\ \citenamefont
  {{Kamionkowski}}(2009)}]{Caldwell09}%
  \BibitemOpen
  \bibfield  {author} {\bibinfo {author} {\bibfnamefont {R.}~\bibnamefont
  {{Caldwell}}}\ and\ \bibinfo {author} {\bibfnamefont {M.}~\bibnamefont
  {{Kamionkowski}}},\ }\href {\doibase 10.1038/458587a} {\bibfield  {journal}
  {\bibinfo  {journal} {Nature}\ }\textbf {\bibinfo {volume} {458}},\ \bibinfo
  {pages} {587} (\bibinfo {year} {2009})}\BibitemShut {NoStop}%
\bibitem [{\citenamefont {{Kiselev}}(2003)}]{Kiselev2003aa}%
  \BibitemOpen
  \bibfield  {author} {\bibinfo {author} {\bibfnamefont {V.~V.}\ \bibnamefont
  {{Kiselev}}},\ }\href {\doibase 10.1088/0264-9381/20/6/310} {\bibfield
  {journal} {\bibinfo  {journal} {Classical and Quantum Gravity}\ }\textbf
  {\bibinfo {volume} {20}},\ \bibinfo {pages} {1187} (\bibinfo {year}
  {2003})},\ \Eprint {http://arxiv.org/abs/gr-qc/0210040} {arXiv:gr-qc/0210040
  [gr-qc]} \BibitemShut {NoStop}%
\bibitem [{\citenamefont {{Hellerman}}\ \emph {et~al.}(2001)\citenamefont
  {{Hellerman}}, \citenamefont {{Kaloper}},\ and\ \citenamefont
  {{Susskind}}}]{Hellerman2001JHEP}%
  \BibitemOpen
  \bibfield  {author} {\bibinfo {author} {\bibfnamefont {S.}~\bibnamefont
  {{Hellerman}}}, \bibinfo {author} {\bibfnamefont {N.}~\bibnamefont
  {{Kaloper}}}, \ and\ \bibinfo {author} {\bibfnamefont {L.}~\bibnamefont
  {{Susskind}}},\ }\href {\doibase 10.1088/1126-6708/2001/06/003} {\bibfield
  {journal} {\bibinfo  {journal} {Journal of High Energy Physics}\ }\textbf
  {\bibinfo {volume} {2001}},\ \bibinfo {eid} {003} (\bibinfo {year} {2001})},\
  \Eprint {http://arxiv.org/abs/hep-th/0104180} {arXiv:hep-th/0104180 [hep-th]}
  \BibitemShut {NoStop}%
\bibitem [{\citenamefont {{Toledo}}\ and\ \citenamefont
  {{Bezerra}}(2018)}]{Toledoa}%
  \BibitemOpen
  \bibfield  {author} {\bibinfo {author} {\bibfnamefont {J.~d.~M.}\
  \bibnamefont {{Toledo}}}\ and\ \bibinfo {author} {\bibfnamefont {V.~B.}\
  \bibnamefont {{Bezerra}}},\ }\href {\doibase 10.1140/epjc/s10052-018-6001-z}
  {\bibfield  {journal} {\bibinfo  {journal} {European Physical Journal C}\
  }\textbf {\bibinfo {volume} {78}},\ \bibinfo {eid} {534} (\bibinfo {year}
  {2018})}\BibitemShut {NoStop}%
\bibitem [{\citenamefont {{Batool}}\ and\ \citenamefont
  {{Hussain}}(2017)}]{Batool16}%
  \BibitemOpen
  \bibfield  {author} {\bibinfo {author} {\bibfnamefont {M.}~\bibnamefont
  {{Batool}}}\ and\ \bibinfo {author} {\bibfnamefont {I.}~\bibnamefont
  {{Hussain}}},\ }\href {\doibase 10.1142/S021827181741005X} {\bibfield
  {journal} {\bibinfo  {journal} {International Journal of Modern Physics D}\
  }\textbf {\bibinfo {volume} {26}},\ \bibinfo {eid} {1741005} (\bibinfo {year}
  {2017})}\BibitemShut {NoStop}%
\bibitem [{\citenamefont {{Mustafa}}\ and\ \citenamefont
  {{Hussain}}(2021)}]{Mustafa2021}%
  \BibitemOpen
  \bibfield  {author} {\bibinfo {author} {\bibfnamefont {G.}~\bibnamefont
  {{Mustafa}}}\ and\ \bibinfo {author} {\bibfnamefont {I.}~\bibnamefont
  {{Hussain}}},\ }\href {\doibase 10.1140/epjc/s10052-021-09195-5} {\bibfield
  {journal} {\bibinfo  {journal} {European Physical Journal C}\ }\textbf
  {\bibinfo {volume} {81}},\ \bibinfo {eid} {419} (\bibinfo {year}
  {2021})}\BibitemShut {NoStop}%
\bibitem [{\citenamefont {{Fathi}}\ \emph {et~al.}(2022)\citenamefont
  {{Fathi}}, \citenamefont {{Olivares}},\ and\ \citenamefont
  {{Villanueva}}}]{Fathi2022}%
  \BibitemOpen
  \bibfield  {author} {\bibinfo {author} {\bibfnamefont {M.}~\bibnamefont
  {{Fathi}}}, \bibinfo {author} {\bibfnamefont {M.}~\bibnamefont {{Olivares}}},
  \ and\ \bibinfo {author} {\bibfnamefont {J.~R.}\ \bibnamefont
  {{Villanueva}}},\ }\href@noop {} {\bibfield  {journal} {\bibinfo  {journal}
  {arXiv e-prints}\ ,\ \bibinfo {eid} {arXiv:2205.13261}} (\bibinfo {year}
  {2022})},\ \Eprint {http://arxiv.org/abs/2205.13261} {arXiv:2205.13261
  [gr-qc]} \BibitemShut {NoStop}%
\bibitem [{\citenamefont {{Hioki}}\ and\ \citenamefont
  {{Maeda}}(2009)}]{Hioki09}%
  \BibitemOpen
  \bibfield  {author} {\bibinfo {author} {\bibfnamefont {K.}~\bibnamefont
  {{Hioki}}}\ and\ \bibinfo {author} {\bibfnamefont {K.-I.}\ \bibnamefont
  {{Maeda}}},\ }\href {\doibase 10.1103/PhysRevD.80.024042} {\bibfield
  {journal} {\bibinfo  {journal} {Phys. Rev. D}\ }\textbf {\bibinfo {volume}
  {80}},\ \bibinfo {eid} {024042} (\bibinfo {year} {2009})}\BibitemShut
  {NoStop}%
\bibitem [{\citenamefont {{Atamurotov}}\ \emph
  {et~al.}(2013{\natexlab{a}})\citenamefont {{Atamurotov}}, \citenamefont
  {{Abdujabbarov}},\ and\ \citenamefont {{Ahmedov}}}]{Atamurotov13}%
  \BibitemOpen
  \bibfield  {author} {\bibinfo {author} {\bibfnamefont {F.}~\bibnamefont
  {{Atamurotov}}}, \bibinfo {author} {\bibfnamefont {A.}~\bibnamefont
  {{Abdujabbarov}}}, \ and\ \bibinfo {author} {\bibfnamefont {B.}~\bibnamefont
  {{Ahmedov}}},\ }\href {\doibase 10.1007/s10509-013-1548-5} {\bibfield
  {journal} {\bibinfo  {journal} {Astrophys Space Sci}\ }\textbf {\bibinfo
  {volume} {348}},\ \bibinfo {pages} {179} (\bibinfo {year}
  {2013}{\natexlab{a}})}\BibitemShut {NoStop}%
\bibitem [{\citenamefont {{Atamurotov}}\ \emph
  {et~al.}(2013{\natexlab{b}})\citenamefont {{Atamurotov}}, \citenamefont
  {{Abdujabbarov}},\ and\ \citenamefont {{Ahmedov}}}]{Atamurotov13b}%
  \BibitemOpen
  \bibfield  {author} {\bibinfo {author} {\bibfnamefont {F.}~\bibnamefont
  {{Atamurotov}}}, \bibinfo {author} {\bibfnamefont {A.}~\bibnamefont
  {{Abdujabbarov}}}, \ and\ \bibinfo {author} {\bibfnamefont {B.}~\bibnamefont
  {{Ahmedov}}},\ }\href {\doibase 10.1103/PhysRevD.88.064004} {\bibfield
  {journal} {\bibinfo  {journal} {Phys. Rev. D}\ }\textbf {\bibinfo {volume}
  {88}},\ \bibinfo {eid} {064004} (\bibinfo {year}
  {2013}{\natexlab{b}})}\BibitemShut {NoStop}%
\bibitem [{\citenamefont {{Abdujabbarov}}\ \emph {et~al.}(2013)\citenamefont
  {{Abdujabbarov}}, \citenamefont {{Atamurotov}}, \citenamefont {{Kucukakca}},
  \citenamefont {{Ahmedov}},\ and\ \citenamefont {{Camci}}}]{Abdujabbarov13aa}%
  \BibitemOpen
  \bibfield  {author} {\bibinfo {author} {\bibfnamefont {A.}~\bibnamefont
  {{Abdujabbarov}}}, \bibinfo {author} {\bibfnamefont {F.}~\bibnamefont
  {{Atamurotov}}}, \bibinfo {author} {\bibfnamefont {Y.}~\bibnamefont
  {{Kucukakca}}}, \bibinfo {author} {\bibfnamefont {B.}~\bibnamefont
  {{Ahmedov}}}, \ and\ \bibinfo {author} {\bibfnamefont {U.}~\bibnamefont
  {{Camci}}},\ }\href {\doibase 10.1007/s10509-012-1337-6} {\bibfield
  {journal} {\bibinfo  {journal} {Astrophys. Space. Sci.}\ }\textbf {\bibinfo
  {volume} {344}},\ \bibinfo {pages} {429} (\bibinfo {year} {2013})},\ \Eprint
  {http://arxiv.org/abs/1212.4949} {arXiv:1212.4949 [physics.gen-ph]}
  \BibitemShut {NoStop}%
\bibitem [{\citenamefont {{Atamurotov}}\ \emph {et~al.}(2016)\citenamefont
  {{Atamurotov}}, \citenamefont {{Ghosh}},\ and\ \citenamefont
  {{Ahmedov}}}]{Atamurotov2016a}%
  \BibitemOpen
  \bibfield  {author} {\bibinfo {author} {\bibfnamefont {F.}~\bibnamefont
  {{Atamurotov}}}, \bibinfo {author} {\bibfnamefont {S.~G.}\ \bibnamefont
  {{Ghosh}}}, \ and\ \bibinfo {author} {\bibfnamefont {B.}~\bibnamefont
  {{Ahmedov}}},\ }\href {\doibase 10.1140/epjc/s10052-016-4122-9} {\bibfield
  {journal} {\bibinfo  {journal} {European Physical Journal C}\ }\textbf
  {\bibinfo {volume} {76}},\ \bibinfo {eid} {273} (\bibinfo {year} {2016})},\
  \Eprint {http://arxiv.org/abs/1506.03690} {arXiv:1506.03690 [gr-qc]}
  \BibitemShut {NoStop}%
\bibitem [{\citenamefont {{Papnoi}}\ \emph {et~al.}(2014)\citenamefont
  {{Papnoi}}, \citenamefont {{Atamurotov}}, \citenamefont {{Ghosh}},\ and\
  \citenamefont {{Ahmedov}}}]{Papnoi2015}%
  \BibitemOpen
  \bibfield  {author} {\bibinfo {author} {\bibfnamefont {U.}~\bibnamefont
  {{Papnoi}}}, \bibinfo {author} {\bibfnamefont {F.}~\bibnamefont
  {{Atamurotov}}}, \bibinfo {author} {\bibfnamefont {S.~G.}\ \bibnamefont
  {{Ghosh}}}, \ and\ \bibinfo {author} {\bibfnamefont {B.}~\bibnamefont
  {{Ahmedov}}},\ }\href {\doibase 10.1103/PhysRevD.90.024073} {\bibfield
  {journal} {\bibinfo  {journal} {Phys. Rev. D}\ }\textbf {\bibinfo {volume}
  {90}},\ \bibinfo {eid} {024073} (\bibinfo {year} {2014})},\ \Eprint
  {http://arxiv.org/abs/1407.0834} {arXiv:1407.0834 [gr-qc]} \BibitemShut
  {NoStop}%
\bibitem [{\citenamefont {{Abdujabbarov}}\ \emph {et~al.}(2015)\citenamefont
  {{Abdujabbarov}}, \citenamefont {{Atamurotov}}, \citenamefont {{Dadhich}},
  \citenamefont {{Ahmedov}},\ and\ \citenamefont
  {{Stuchl{\'{\i}}k}}}]{Abdujabbarov15a}%
  \BibitemOpen
  \bibfield  {author} {\bibinfo {author} {\bibfnamefont {A.}~\bibnamefont
  {{Abdujabbarov}}}, \bibinfo {author} {\bibfnamefont {F.}~\bibnamefont
  {{Atamurotov}}}, \bibinfo {author} {\bibfnamefont {N.}~\bibnamefont
  {{Dadhich}}}, \bibinfo {author} {\bibfnamefont {B.}~\bibnamefont
  {{Ahmedov}}}, \ and\ \bibinfo {author} {\bibfnamefont {Z.}~\bibnamefont
  {{Stuchl{\'{\i}}k}}},\ }\href {\doibase 10.1140/epjc/s10052-015-3604-5}
  {\bibfield  {journal} {\bibinfo  {journal} {Eur. Phys. J. C}\ }\textbf
  {\bibinfo {volume} {75}},\ \bibinfo {eid} {399} (\bibinfo {year} {2015})},\
  \Eprint {http://arxiv.org/abs/1508.00331} {arXiv:1508.00331 [gr-qc]}
  \BibitemShut {NoStop}%
\bibitem [{\citenamefont {Kumar}\ and\ \citenamefont
  {Ghosh}(2020)}]{Rahul:2020a}%
  \BibitemOpen
  \bibfield  {author} {\bibinfo {author} {\bibfnamefont {R.}~\bibnamefont
  {Kumar}}\ and\ \bibinfo {author} {\bibfnamefont {S.~G.}\ \bibnamefont
  {Ghosh}},\ }\href {\doibase 10.1088/1475-7516/2020/07/053} {\bibfield
  {journal} {\bibinfo  {journal} {J.~Cosmol.~A.~P}\ }\textbf {\bibinfo {volume}
  {2020}},\ \bibinfo {pages} {053} (\bibinfo {year} {2020})}\BibitemShut
  {NoStop}%
\bibitem [{\citenamefont {{Cunha}}\ \emph {et~al.}(2020)\citenamefont
  {{Cunha}}, \citenamefont {{Eir{\'o}}}, \citenamefont {{Herdeiro}},\ and\
  \citenamefont {{Lemos}}}]{Cunha20a}%
  \BibitemOpen
  \bibfield  {author} {\bibinfo {author} {\bibfnamefont {P.~V.~P.}\
  \bibnamefont {{Cunha}}}, \bibinfo {author} {\bibfnamefont {N.~A.}\
  \bibnamefont {{Eir{\'o}}}}, \bibinfo {author} {\bibfnamefont {C.~A.~R.}\
  \bibnamefont {{Herdeiro}}}, \ and\ \bibinfo {author} {\bibfnamefont
  {J.~P.~S.}\ \bibnamefont {{Lemos}}},\ }\href {\doibase
  10.1088/1475-7516/2020/03/035} {\bibfield  {journal} {\bibinfo  {journal}
  {J.~Cosmol.~A.~P}\ }\textbf {\bibinfo {volume} {2020}},\ \bibinfo {eid} {035}
  (\bibinfo {year} {2020})},\ \Eprint {http://arxiv.org/abs/1912.08833}
  {arXiv:1912.08833 [gr-qc]} \BibitemShut {NoStop}%
\bibitem [{\citenamefont {{Cunha}}\ \emph {et~al.}(2017)\citenamefont
  {{Cunha}}, \citenamefont {{Herdeiro}}, \citenamefont {{Kleihaus}},
  \citenamefont {{Kunz}},\ and\ \citenamefont {{Radu}}}]{Cunha17a}%
  \BibitemOpen
  \bibfield  {author} {\bibinfo {author} {\bibfnamefont {P.~V.~P.}\
  \bibnamefont {{Cunha}}}, \bibinfo {author} {\bibfnamefont {C.~A.~R.}\
  \bibnamefont {{Herdeiro}}}, \bibinfo {author} {\bibfnamefont
  {B.}~\bibnamefont {{Kleihaus}}}, \bibinfo {author} {\bibfnamefont
  {J.}~\bibnamefont {{Kunz}}}, \ and\ \bibinfo {author} {\bibfnamefont
  {E.}~\bibnamefont {{Radu}}},\ }\href {\doibase
  10.1016/j.physletb.2017.03.020} {\bibfield  {journal} {\bibinfo  {journal}
  {Physics Letters B}\ }\textbf {\bibinfo {volume} {768}},\ \bibinfo {pages}
  {373} (\bibinfo {year} {2017})},\ \Eprint {http://arxiv.org/abs/1701.00079}
  {arXiv:1701.00079 [gr-qc]} \BibitemShut {NoStop}%
\bibitem [{\citenamefont {{Atamurotov}}\ \emph
  {et~al.}(2022{\natexlab{a}})\citenamefont {{Atamurotov}}, \citenamefont
  {{Papnoi}},\ and\ \citenamefont {{Jusufi}}}]{Atamurotov21b}%
  \BibitemOpen
  \bibfield  {author} {\bibinfo {author} {\bibfnamefont {F.}~\bibnamefont
  {{Atamurotov}}}, \bibinfo {author} {\bibfnamefont {U.}~\bibnamefont
  {{Papnoi}}}, \ and\ \bibinfo {author} {\bibfnamefont {K.}~\bibnamefont
  {{Jusufi}}},\ }\href {\doibase 10.1088/1361-6382/ac3e76} {\bibfield
  {journal} {\bibinfo  {journal} {Class. Quan. Grav.}\ }\textbf {\bibinfo
  {volume} {39}},\ \bibinfo {eid} {025014} (\bibinfo {year}
  {2022}{\natexlab{a}})},\ \Eprint {http://arxiv.org/abs/2104.14898}
  {arXiv:2104.14898 [gr-qc]} \BibitemShut {NoStop}%
\bibitem [{\citenamefont {{Jafarzade}}\ \emph {et~al.}(2021)\citenamefont
  {{Jafarzade}}, \citenamefont {{Kord Zangeneh}},\ and\ \citenamefont
  {{Lobo}}}]{Jafarzade21a}%
  \BibitemOpen
  \bibfield  {author} {\bibinfo {author} {\bibfnamefont {K.}~\bibnamefont
  {{Jafarzade}}}, \bibinfo {author} {\bibfnamefont {M.}~\bibnamefont {{Kord
  Zangeneh}}}, \ and\ \bibinfo {author} {\bibfnamefont {F.~S.~N.}\ \bibnamefont
  {{Lobo}}},\ }\href {\doibase 10.1088/1475-7516/2021/04/008} {\bibfield
  {journal} {\bibinfo  {journal} {J. Cosmol. A. P.}\ }\textbf {\bibinfo
  {volume} {2021}},\ \bibinfo {eid} {008} (\bibinfo {year} {2021})},\ \Eprint
  {http://arxiv.org/abs/2010.05755} {arXiv:2010.05755 [gr-qc]} \BibitemShut
  {NoStop}%
\bibitem [{\citenamefont {{Afrin}}\ \emph {et~al.}(2021)\citenamefont
  {{Afrin}}, \citenamefont {{Kumar}},\ and\ \citenamefont
  {{Ghosh}}}]{Afrin21a}%
  \BibitemOpen
  \bibfield  {author} {\bibinfo {author} {\bibfnamefont {M.}~\bibnamefont
  {{Afrin}}}, \bibinfo {author} {\bibfnamefont {R.}~\bibnamefont {{Kumar}}}, \
  and\ \bibinfo {author} {\bibfnamefont {S.~G.}\ \bibnamefont {{Ghosh}}},\
  }\href {\doibase 10.1093/mnras/stab1260} {\bibfield  {journal} {\bibinfo
  {journal} {Mon. Not. R. Astron. Soc.}\ }\textbf {\bibinfo {volume} {504}},\
  \bibinfo {pages} {5927} (\bibinfo {year} {2021})},\ \Eprint
  {http://arxiv.org/abs/2103.11417} {arXiv:2103.11417 [gr-qc]} \BibitemShut
  {NoStop}%
\bibitem [{\citenamefont {{Ghasemi-Nodehi}}\ \emph {et~al.}(2020)\citenamefont
  {{Ghasemi-Nodehi}}, \citenamefont {{Azreg-A{\"\i}nou}}, \citenamefont
  {{Jusufi}},\ and\ \citenamefont {{Jamil}}}]{Ghasemi2020a}%
  \BibitemOpen
  \bibfield  {author} {\bibinfo {author} {\bibfnamefont {M.}~\bibnamefont
  {{Ghasemi-Nodehi}}}, \bibinfo {author} {\bibfnamefont {M.}~\bibnamefont
  {{Azreg-A{\"\i}nou}}}, \bibinfo {author} {\bibfnamefont {K.}~\bibnamefont
  {{Jusufi}}}, \ and\ \bibinfo {author} {\bibfnamefont {M.}~\bibnamefont
  {{Jamil}}},\ }\href {\doibase 10.1103/PhysRevD.102.104032} {\bibfield
  {journal} {\bibinfo  {journal} {Phys. Rev. D}\ }\textbf {\bibinfo {volume}
  {102}},\ \bibinfo {eid} {104032} (\bibinfo {year} {2020})},\ \Eprint
  {http://arxiv.org/abs/2011.02276} {arXiv:2011.02276 [gr-qc]} \BibitemShut
  {NoStop}%
\bibitem [{\citenamefont {Bambi}\ \emph {et~al.}(2019)\citenamefont {Bambi},
  \citenamefont {Freese}, \citenamefont {Vagnozzi},\ and\ \citenamefont
  {Visinelli}}]{Bambi:2019tjh}%
  \BibitemOpen
  \bibfield  {author} {\bibinfo {author} {\bibfnamefont {C.}~\bibnamefont
  {Bambi}}, \bibinfo {author} {\bibfnamefont {K.}~\bibnamefont {Freese}},
  \bibinfo {author} {\bibfnamefont {S.}~\bibnamefont {Vagnozzi}}, \ and\
  \bibinfo {author} {\bibfnamefont {L.}~\bibnamefont {Visinelli}},\ }\href
  {\doibase 10.1103/PhysRevD.100.044057} {\bibfield  {journal} {\bibinfo
  {journal} {Phys. Rev. D}\ }\textbf {\bibinfo {volume} {100}},\ \bibinfo
  {pages} {044057} (\bibinfo {year} {2019})},\ \Eprint
  {http://arxiv.org/abs/1904.12983} {arXiv:1904.12983 [gr-qc]} \BibitemShut
  {NoStop}%
\bibitem [{\citenamefont {Vagnozzi}\ and\ \citenamefont
  {Visinelli}(2019)}]{Vagnozzi:2019apd}%
  \BibitemOpen
  \bibfield  {author} {\bibinfo {author} {\bibfnamefont {S.}~\bibnamefont
  {Vagnozzi}}\ and\ \bibinfo {author} {\bibfnamefont {L.}~\bibnamefont
  {Visinelli}},\ }\href {\doibase 10.1103/PhysRevD.100.024020} {\bibfield
  {journal} {\bibinfo  {journal} {Phys. Rev. D}\ }\textbf {\bibinfo {volume}
  {100}},\ \bibinfo {pages} {024020} (\bibinfo {year} {2019})},\ \Eprint
  {http://arxiv.org/abs/1905.12421} {arXiv:1905.12421 [gr-qc]} \BibitemShut
  {NoStop}%
\bibitem [{\citenamefont {Khodadi}\ \emph {et~al.}(2020)\citenamefont
  {Khodadi}, \citenamefont {Allahyari}, \citenamefont {Vagnozzi},\ and\
  \citenamefont {Mota}}]{Khodadi:2020jij}%
  \BibitemOpen
  \bibfield  {author} {\bibinfo {author} {\bibfnamefont {M.}~\bibnamefont
  {Khodadi}}, \bibinfo {author} {\bibfnamefont {A.}~\bibnamefont {Allahyari}},
  \bibinfo {author} {\bibfnamefont {S.}~\bibnamefont {Vagnozzi}}, \ and\
  \bibinfo {author} {\bibfnamefont {D.~F.}\ \bibnamefont {Mota}},\ }\href
  {\doibase 10.1088/1475-7516/2020/09/026} {\bibfield  {journal} {\bibinfo
  {journal} {JCAP}\ }\textbf {\bibinfo {volume} {09}},\ \bibinfo {pages} {026}
  (\bibinfo {year} {2020})},\ \Eprint {http://arxiv.org/abs/2005.05992}
  {arXiv:2005.05992 [gr-qc]} \BibitemShut {NoStop}%
\bibitem [{\citenamefont {Virbhadra}\ and\ \citenamefont
  {Ellis}(2000)}]{Virbha:2000a}%
  \BibitemOpen
  \bibfield  {author} {\bibinfo {author} {\bibfnamefont {K.~S.}\ \bibnamefont
  {Virbhadra}}\ and\ \bibinfo {author} {\bibfnamefont {G.~F.~R.}\ \bibnamefont
  {Ellis}},\ }\href {\doibase 10.1103/PhysRevD.62.084003} {\bibfield  {journal}
  {\bibinfo  {journal} {Phys.~Rev.~D.}\ }\textbf {\bibinfo {volume} {62}},\
  \bibinfo {pages} {084003} (\bibinfo {year} {2000})}\BibitemShut {NoStop}%
\bibitem [{\citenamefont {Bozza}\ \emph {et~al.}(2001)\citenamefont {Bozza},
  \citenamefont {Capozziello}, \citenamefont {Iovane},\ and\ \citenamefont
  {Scarpetta}}]{Bozza:2001a}%
  \BibitemOpen
  \bibfield  {author} {\bibinfo {author} {\bibfnamefont {â.}~\bibnamefont
  {Bozza}}, \bibinfo {author} {\bibfnamefont {S.}~\bibnamefont {Capozziello}},
  \bibinfo {author} {\bibfnamefont {G.}~\bibnamefont {Iovane}}, \ and\ \bibinfo
  {author} {\bibfnamefont {G.}~\bibnamefont {Scarpetta}},\ }\href {\doibase
  10.1023/A:1012292927358} {\bibfield  {journal} {\bibinfo  {journal} {Gen.
  Rel. Grav.}\ }\textbf {\bibinfo {volume} {33}},\ \bibinfo {pages} {1535}
  (\bibinfo {year} {2001})}\BibitemShut {NoStop}%
\bibitem [{\citenamefont {Bozza}(2002)}]{Bozza:2002b}%
  \BibitemOpen
  \bibfield  {author} {\bibinfo {author} {\bibfnamefont {â.}~\bibnamefont
  {Bozza}},\ }\href {\doibase 10.1103/PhysRevD.66.103001} {\bibfield  {journal}
  {\bibinfo  {journal} {Phys.~Rev.~D.}\ }\textbf {\bibinfo {volume} {66}},\
  \bibinfo {pages} {103001} (\bibinfo {year} {2002})}\BibitemShut {NoStop}%
\bibitem [{\citenamefont {Zhao}\ and\ \citenamefont {Xie}(2017)}]{Zhao:2017a}%
  \BibitemOpen
  \bibfield  {author} {\bibinfo {author} {\bibfnamefont {S.-S.}\ \bibnamefont
  {Zhao}}\ and\ \bibinfo {author} {\bibfnamefont {Y.}~\bibnamefont {Xie}},\
  }\href {\doibase 10.1016/j.physletb.2017.09.090} {\bibfield  {journal}
  {\bibinfo  {journal} {Phys.~Lett.~B.}\ }\textbf {\bibinfo {volume} {774}},\
  \bibinfo {pages} {357} (\bibinfo {year} {2017})}\BibitemShut {NoStop}%
\bibitem [{\citenamefont {{V{\'a}zquez}}\ and\ \citenamefont
  {{Esteban}}(2004)}]{Vazquez04}%
  \BibitemOpen
  \bibfield  {author} {\bibinfo {author} {\bibfnamefont {S.~E.}\ \bibnamefont
  {{V{\'a}zquez}}}\ and\ \bibinfo {author} {\bibfnamefont {E.~P.}\ \bibnamefont
  {{Esteban}}},\ }\href {\doibase 10.1393/ncb/i2004-10121-y} {\bibfield
  {journal} {\bibinfo  {journal} {Nuovo Cim. B}\ }\textbf {\bibinfo {volume}
  {119}},\ \bibinfo {pages} {489} (\bibinfo {year} {2004})}\BibitemShut
  {NoStop}%
\bibitem [{\citenamefont {Eiroa}\ \emph {et~al.}(2002)\citenamefont {Eiroa},
  \citenamefont {Romero},\ and\ \citenamefont {Torres}}]{Eiroa:2002b}%
  \BibitemOpen
  \bibfield  {author} {\bibinfo {author} {\bibfnamefont {E.~F.}\ \bibnamefont
  {Eiroa}}, \bibinfo {author} {\bibfnamefont {G.~E.}\ \bibnamefont {Romero}}, \
  and\ \bibinfo {author} {\bibfnamefont {D.~F.}\ \bibnamefont {Torres}},\
  }\href {\doibase 10.1103/PhysRevD.66.024010} {\bibfield  {journal} {\bibinfo
  {journal} {Phys.~Rev.~D.}\ }\textbf {\bibinfo {volume} {66}},\ \bibinfo
  {pages} {024010} (\bibinfo {year} {2002})}\BibitemShut {NoStop}%
\bibitem [{\citenamefont {Eiroa}\ and\ \citenamefont
  {Torres}(2004)}]{Eiroa:2004a}%
  \BibitemOpen
  \bibfield  {author} {\bibinfo {author} {\bibfnamefont {E.~F.}\ \bibnamefont
  {Eiroa}}\ and\ \bibinfo {author} {\bibfnamefont {D.~F.}\ \bibnamefont
  {Torres}},\ }\href {\doibase 10.1103/PhysRevD.69.063004} {\bibfield
  {journal} {\bibinfo  {journal} {Phys.~Rev.~D.}\ }\textbf {\bibinfo {volume}
  {69}},\ \bibinfo {pages} {063004} (\bibinfo {year} {2004})}\BibitemShut
  {NoStop}%
\bibitem [{\citenamefont {{Chakraborty}}\ and\ \citenamefont
  {Soumitra}(2017)}]{Chak:2017a}%
  \BibitemOpen
  \bibfield  {author} {\bibinfo {author} {\bibfnamefont {S.}~\bibnamefont
  {{Chakraborty}}}\ and\ \bibinfo {author} {\bibfnamefont {S.}~\bibnamefont
  {Soumitra}},\ }\href {\doibase 10.1088/1475-7516/2017/07/045} {\bibfield
  {journal} {\bibinfo  {journal} {J.~Cosmol.~A.~P.}\ }\textbf {\bibinfo
  {volume} {07}},\ \bibinfo {pages} {045} (\bibinfo {year} {2017})}\BibitemShut
  {NoStop}%
\bibitem [{\citenamefont {{Perlick}}(2004)}]{Perlick04}%
  \BibitemOpen
  \bibfield  {author} {\bibinfo {author} {\bibfnamefont {V.}~\bibnamefont
  {{Perlick}}},\ }\href {\doibase 10.12942/lrr-2004-9} {\bibfield  {journal}
  {\bibinfo  {journal} {Living Reviews in Relativity}\ }\textbf {\bibinfo
  {volume} {7}},\ \bibinfo {pages} {9} (\bibinfo {year} {2004})}\BibitemShut
  {NoStop}%
\bibitem [{\citenamefont {Abdujabbarov}\ \emph {et~al.}(2017)\citenamefont
  {Abdujabbarov}, \citenamefont {Ahmedov}, \citenamefont {Dadhich},\ and\
  \citenamefont {Atamurotov}}]{Abu:2017a}%
  \BibitemOpen
  \bibfield  {author} {\bibinfo {author} {\bibfnamefont {A.}~\bibnamefont
  {Abdujabbarov}}, \bibinfo {author} {\bibfnamefont {B.}~\bibnamefont
  {Ahmedov}}, \bibinfo {author} {\bibfnamefont {N.}~\bibnamefont {Dadhich}}, \
  and\ \bibinfo {author} {\bibfnamefont {F.}~\bibnamefont {Atamurotov}},\
  }\href {\doibase 10.1103/PhysRevD.96.084017} {\bibfield  {journal} {\bibinfo
  {journal} {Phys.~Rev.~D.}\ }\textbf {\bibinfo {volume} {96}},\ \bibinfo
  {pages} {084017} (\bibinfo {year} {2017})}\BibitemShut {NoStop}%
\bibitem [{\citenamefont {{Islam}}\ \emph {et~al.}(2020)\citenamefont
  {{Islam}}, \citenamefont {{Kumar}},\ and\ \citenamefont
  {{Ghosh}}}]{Islam20egb}%
  \BibitemOpen
  \bibfield  {author} {\bibinfo {author} {\bibfnamefont {S.~U.}\ \bibnamefont
  {{Islam}}}, \bibinfo {author} {\bibfnamefont {R.}~\bibnamefont {{Kumar}}}, \
  and\ \bibinfo {author} {\bibfnamefont {S.~G.}\ \bibnamefont {{Ghosh}}},\
  }\href {\doibase 10.1088/1475-7516/2020/09/030} {\bibfield  {journal}
  {\bibinfo  {journal} {JCAP}\ }\textbf {\bibinfo {volume} {2020}},\ \bibinfo
  {eid} {030} (\bibinfo {year} {2020})},\ \Eprint
  {http://arxiv.org/abs/2004.01038} {arXiv:2004.01038 [gr-qc]} \BibitemShut
  {NoStop}%
\bibitem [{\citenamefont {Virbhadra}\ and\ \citenamefont
  {Ellis}(2002)}]{Virbha:2002a}%
  \BibitemOpen
  \bibfield  {author} {\bibinfo {author} {\bibfnamefont {K.~S.}\ \bibnamefont
  {Virbhadra}}\ and\ \bibinfo {author} {\bibfnamefont {G.~F.~R.}\ \bibnamefont
  {Ellis}},\ }\href {\doibase 10.1103/PhysRevD.65.103004} {\bibfield  {journal}
  {\bibinfo  {journal} {Phys.~Rev.~D.}\ }\textbf {\bibinfo {volume} {65}},\
  \bibinfo {pages} {103004} (\bibinfo {year} {2002})}\BibitemShut {NoStop}%
\bibitem [{\citenamefont {{Kumar}}\ \emph
  {et~al.}(2020{\natexlab{a}})\citenamefont {{Kumar}}, \citenamefont
  {{Islam}},\ and\ \citenamefont {{Ghosh}}}]{Kumar2020a}%
  \BibitemOpen
  \bibfield  {author} {\bibinfo {author} {\bibfnamefont {R.}~\bibnamefont
  {{Kumar}}}, \bibinfo {author} {\bibfnamefont {S.~U.}\ \bibnamefont
  {{Islam}}}, \ and\ \bibinfo {author} {\bibfnamefont {S.~G.}\ \bibnamefont
  {{Ghosh}}},\ }\href {\doibase 10.1140/epjc/s10052-020-08606-3} {\bibfield
  {journal} {\bibinfo  {journal} {European Physical Journal C}\ }\textbf
  {\bibinfo {volume} {80}},\ \bibinfo {eid} {1128} (\bibinfo {year}
  {2020}{\natexlab{a}})},\ \Eprint {http://arxiv.org/abs/2004.12970}
  {arXiv:2004.12970 [gr-qc]} \BibitemShut {NoStop}%
\bibitem [{\citenamefont {{Narzilloev}}\ \emph {et~al.}(2021)\citenamefont
  {{Narzilloev}}, \citenamefont {{Shaymatov}}, \citenamefont {{Hussain}},
  \citenamefont {{Abdujabbarov}}, \citenamefont {{Ahmedov}},\ and\
  \citenamefont {{Bambi}}}]{Bakhtiyor2021a}%
  \BibitemOpen
  \bibfield  {author} {\bibinfo {author} {\bibfnamefont {B.}~\bibnamefont
  {{Narzilloev}}}, \bibinfo {author} {\bibfnamefont {S.}~\bibnamefont
  {{Shaymatov}}}, \bibinfo {author} {\bibfnamefont {I.}~\bibnamefont
  {{Hussain}}}, \bibinfo {author} {\bibfnamefont {A.}~\bibnamefont
  {{Abdujabbarov}}}, \bibinfo {author} {\bibfnamefont {B.}~\bibnamefont
  {{Ahmedov}}}, \ and\ \bibinfo {author} {\bibfnamefont {C.}~\bibnamefont
  {{Bambi}}},\ }\href {\doibase 10.1140/epjc/s10052-021-09617-4} {\bibfield
  {journal} {\bibinfo  {journal} {European Physical Journal C}\ }\textbf
  {\bibinfo {volume} {81}},\ \bibinfo {eid} {849} (\bibinfo {year} {2021})},\
  \Eprint {http://arxiv.org/abs/2109.02816} {arXiv:2109.02816 [gr-qc]}
  \BibitemShut {NoStop}%
\bibitem [{\citenamefont {Bisnovatyi-Kogan}\ and\ \citenamefont
  {Tsupko}(2010)}]{Bin:2010a}%
  \BibitemOpen
  \bibfield  {author} {\bibinfo {author} {\bibfnamefont {G.~S.}\ \bibnamefont
  {Bisnovatyi-Kogan}}\ and\ \bibinfo {author} {\bibfnamefont {O.~Y.}\
  \bibnamefont {Tsupko}},\ }\href {\doibase 10.1111/j.1365-2966.2010.16290.x}
  {\bibfield  {journal} {\bibinfo  {journal} {Mon.~Not.~R.~Astron.~Soc.}\
  }\textbf {\bibinfo {volume} {404}},\ \bibinfo {pages} {1790} (\bibinfo {year}
  {2010})}\BibitemShut {NoStop}%
\bibitem [{\citenamefont {{Babar}}\ \emph {et~al.}(2021)\citenamefont
  {{Babar}}, \citenamefont {{Atamurotov}},\ and\ \citenamefont
  {{Babar}}}]{Babar2021a}%
  \BibitemOpen
  \bibfield  {author} {\bibinfo {author} {\bibfnamefont {G.~Z.}\ \bibnamefont
  {{Babar}}}, \bibinfo {author} {\bibfnamefont {F.}~\bibnamefont
  {{Atamurotov}}}, \ and\ \bibinfo {author} {\bibfnamefont {A.~Z.}\
  \bibnamefont {{Babar}}},\ }\href {\doibase 10.1016/j.dark.2021.100798}
  {\bibfield  {journal} {\bibinfo  {journal} {Physics of the Dark Universe}\
  }\textbf {\bibinfo {volume} {32}},\ \bibinfo {eid} {100798} (\bibinfo {year}
  {2021})}\BibitemShut {NoStop}%
\bibitem [{\citenamefont {Morozova}\ \emph {et~al.}(2013)\citenamefont
  {Morozova}, \citenamefont {Ahmedov},\ and\ \citenamefont
  {Tursunov}}]{Abu:2013a}%
  \BibitemOpen
  \bibfield  {author} {\bibinfo {author} {\bibfnamefont {V.}~\bibnamefont
  {Morozova}}, \bibinfo {author} {\bibfnamefont {B.}~\bibnamefont {Ahmedov}}, \
  and\ \bibinfo {author} {\bibfnamefont {A.}~\bibnamefont {Tursunov}},\ }\href
  {\doibase 10.1007/s10509-013-1458-6} {\bibfield  {journal} {\bibinfo
  {journal} {Astrophys.~Space.~Sci.}\ }\textbf {\bibinfo {volume} {346}},\
  \bibinfo {pages} {513} (\bibinfo {year} {2013})}\BibitemShut {NoStop}%
\bibitem [{\citenamefont {Babar}\ \emph {et~al.}(2021)\citenamefont {Babar},
  \citenamefont {Atamurotov}, \citenamefont {Ul~Islam},\ and\ \citenamefont
  {Ghosh}}]{Babar2021b}%
  \BibitemOpen
  \bibfield  {author} {\bibinfo {author} {\bibfnamefont {G.~Z.}\ \bibnamefont
  {Babar}}, \bibinfo {author} {\bibfnamefont {F.}~\bibnamefont {Atamurotov}},
  \bibinfo {author} {\bibfnamefont {S.}~\bibnamefont {Ul~Islam}}, \ and\
  \bibinfo {author} {\bibfnamefont {S.~G.}\ \bibnamefont {Ghosh}},\ }\href
  {\doibase 10.1103/PhysRevD.103.084057} {\bibfield  {journal} {\bibinfo
  {journal} {Phys. Rev. D}\ }\textbf {\bibinfo {volume} {103}},\ \bibinfo
  {pages} {084057} (\bibinfo {year} {2021})},\ \Eprint
  {http://arxiv.org/abs/2104.00714} {arXiv:2104.00714 [gr-qc]} \BibitemShut
  {NoStop}%
\bibitem [{\citenamefont {Hakimov}\ and\ \citenamefont
  {Atamurotov}(2016)}]{Hakimov2016a}%
  \BibitemOpen
  \bibfield  {author} {\bibinfo {author} {\bibfnamefont {A.}~\bibnamefont
  {Hakimov}}\ and\ \bibinfo {author} {\bibfnamefont {F.}~\bibnamefont
  {Atamurotov}},\ }\href {\doibase 10.1007/s10509-016-2702-7} {\bibfield
  {journal} {\bibinfo  {journal} {Astrophys.~Space.~Sci.}\ }\textbf {\bibinfo
  {volume} {361}},\ \bibinfo {pages} {112} (\bibinfo {year}
  {2016})}\BibitemShut {NoStop}%
\bibitem [{\citenamefont {Rogers}(2015)}]{Rog:2015a}%
  \BibitemOpen
  \bibfield  {author} {\bibinfo {author} {\bibfnamefont {A.}~\bibnamefont
  {Rogers}},\ }\href {\doibase 10.1093/mnras/stv903} {\bibfield  {journal}
  {\bibinfo  {journal} {Mon.~Not.~R.~Astron.~Soc.}\ }\textbf {\bibinfo {volume}
  {451}},\ \bibinfo {pages} {17} (\bibinfo {year} {2015})}\BibitemShut
  {NoStop}%
\bibitem [{\citenamefont {{Atamurotov}}\ \emph
  {et~al.}(2021{\natexlab{a}})\citenamefont {{Atamurotov}}, \citenamefont
  {{Abdujabbarov}},\ and\ \citenamefont {{Rayimbaev}}}]{Far:2021a}%
  \BibitemOpen
  \bibfield  {author} {\bibinfo {author} {\bibfnamefont {F.}~\bibnamefont
  {{Atamurotov}}}, \bibinfo {author} {\bibfnamefont {A.}~\bibnamefont
  {{Abdujabbarov}}}, \ and\ \bibinfo {author} {\bibfnamefont {J.}~\bibnamefont
  {{Rayimbaev}}},\ }\href {\doibase 10.1140/epjc/s10052-021-08919-x} {\bibfield
   {journal} {\bibinfo  {journal} {Eur.~Phys.~J.~C.}\ }\textbf {\bibinfo
  {volume} {81}},\ \bibinfo {pages} {118} (\bibinfo {year}
  {2021}{\natexlab{a}})}\BibitemShut {NoStop}%
\bibitem [{\citenamefont {Benavides-Gallego}\ \emph {et~al.}(2018)\citenamefont
  {Benavides-Gallego}, \citenamefont {Abdujabbarov},\ and\ \citenamefont
  {Bambi}}]{Car:2018a}%
  \BibitemOpen
  \bibfield  {author} {\bibinfo {author} {\bibfnamefont {C.}~\bibnamefont
  {Benavides-Gallego}}, \bibinfo {author} {\bibfnamefont {A.}~\bibnamefont
  {Abdujabbarov}}, \ and\ \bibinfo {author} {\bibnamefont {Bambi}},\ }\href
  {\doibase 10.1140/epjc/s10052-018-6170-97} {\bibfield  {journal} {\bibinfo
  {journal} {Eur.~Phys.~J.~C.}\ }\textbf {\bibinfo {volume} {78}},\ \bibinfo
  {pages} {694} (\bibinfo {year} {2018})}\BibitemShut {NoStop}%
\bibitem [{\citenamefont {{Atamurotov}}\ \emph
  {et~al.}(2021{\natexlab{b}})\citenamefont {{Atamurotov}}, \citenamefont
  {{Shaymatov}},\ and\ \citenamefont {{Ahmedov}}}]{Atamurotov215}%
  \BibitemOpen
  \bibfield  {author} {\bibinfo {author} {\bibfnamefont {F.}~\bibnamefont
  {{Atamurotov}}}, \bibinfo {author} {\bibfnamefont {S.}~\bibnamefont
  {{Shaymatov}}}, \ and\ \bibinfo {author} {\bibfnamefont {B.}~\bibnamefont
  {{Ahmedov}}},\ }\href {\doibase 10.3390/galaxies9030054} {\bibfield
  {journal} {\bibinfo  {journal} {Galaxies}\ }\textbf {\bibinfo {volume} {9}},\
  \bibinfo {pages} {54} (\bibinfo {year} {2021}{\natexlab{b}})}\BibitemShut
  {NoStop}%
\bibitem [{\citenamefont {{Atamurotov}}\ \emph
  {et~al.}(2021{\natexlab{c}})\citenamefont {{Atamurotov}}, \citenamefont
  {{Shaymatov}}, \citenamefont {{Sheoran}},\ and\ \citenamefont
  {{Siwach}}}]{Atamurotov216}%
  \BibitemOpen
  \bibfield  {author} {\bibinfo {author} {\bibfnamefont {F.}~\bibnamefont
  {{Atamurotov}}}, \bibinfo {author} {\bibfnamefont {S.}~\bibnamefont
  {{Shaymatov}}}, \bibinfo {author} {\bibfnamefont {P.}~\bibnamefont
  {{Sheoran}}}, \ and\ \bibinfo {author} {\bibfnamefont {S.}~\bibnamefont
  {{Siwach}}},\ }\href {\doibase 10.1088/1475-7516/2021/08/045} {\bibfield
  {journal} {\bibinfo  {journal} {J. Cosmol. P. A.}\ }\textbf {\bibinfo
  {volume} {2021}},\ \bibinfo {eid} {045} (\bibinfo {year}
  {2021}{\natexlab{c}})},\ \Eprint {http://arxiv.org/abs/2105.02214}
  {arXiv:2105.02214 [gr-qc]} \BibitemShut {NoStop}%
\bibitem [{\citenamefont {{Atamurotov}}\ \emph
  {et~al.}(2022{\natexlab{b}})\citenamefont {{Atamurotov}}, \citenamefont
  {{Sarikulov}}, \citenamefont {{Abdujabbarov}},\ and\ \citenamefont
  {{Ahmedov}}}]{Atamurotov22epjp}%
  \BibitemOpen
  \bibfield  {author} {\bibinfo {author} {\bibfnamefont {F.}~\bibnamefont
  {{Atamurotov}}}, \bibinfo {author} {\bibfnamefont {F.}~\bibnamefont
  {{Sarikulov}}}, \bibinfo {author} {\bibfnamefont {A.}~\bibnamefont
  {{Abdujabbarov}}}, \ and\ \bibinfo {author} {\bibfnamefont {B.}~\bibnamefont
  {{Ahmedov}}},\ }\href {\doibase 10.1140/epjp/s13360-022-02548-3} {\bibfield
  {journal} {\bibinfo  {journal} {Eur. Phys. J. Plus}\ }\textbf {\bibinfo
  {volume} {137}},\ \bibinfo {eid} {336} (\bibinfo {year}
  {2022}{\natexlab{b}})}\BibitemShut {NoStop}%
\bibitem [{\citenamefont {{Atamurotov}}\ \emph
  {et~al.}(2022{\natexlab{c}})\citenamefont {{Atamurotov}}, \citenamefont
  {{Sarikulov}}, \citenamefont {{Khamidov}},\ and\ \citenamefont
  {{Abdujabbarov}}}]{Atamurotov2022EPJP3}%
  \BibitemOpen
  \bibfield  {author} {\bibinfo {author} {\bibfnamefont {F.}~\bibnamefont
  {{Atamurotov}}}, \bibinfo {author} {\bibfnamefont {F.}~\bibnamefont
  {{Sarikulov}}}, \bibinfo {author} {\bibfnamefont {V.}~\bibnamefont
  {{Khamidov}}}, \ and\ \bibinfo {author} {\bibfnamefont {A.}~\bibnamefont
  {{Abdujabbarov}}},\ }\href {\doibase 10.1140/epjp/s13360-022-02780-x}
  {\bibfield  {journal} {\bibinfo  {journal} {Eur. Phys. J. Plus}\ }\textbf
  {\bibinfo {volume} {137}},\ \bibinfo {eid} {567} (\bibinfo {year}
  {2022}{\natexlab{c}})}\BibitemShut {NoStop}%
\bibitem [{\citenamefont {{Atamurotov}}\ and\ \citenamefont
  {{Ghosh}}(2022)}]{AtamurotovGhosh2022}%
  \BibitemOpen
  \bibfield  {author} {\bibinfo {author} {\bibfnamefont {F.}~\bibnamefont
  {{Atamurotov}}}\ and\ \bibinfo {author} {\bibfnamefont {S.~G.}\ \bibnamefont
  {{Ghosh}}},\ }\href {\doibase 10.1140/epjp/s13360-022-02885-3} {\bibfield
  {journal} {\bibinfo  {journal} {European Physical Journal Plus}\ }\textbf
  {\bibinfo {volume} {137}},\ \bibinfo {eid} {662} (\bibinfo {year}
  {2022})}\BibitemShut {NoStop}%
\bibitem [{\citenamefont {{Atamurotov}}\ \emph
  {et~al.}(2022{\natexlab{d}})\citenamefont {{Atamurotov}}, \citenamefont
  {{Alloqulov}}, \citenamefont {{Abdujabbarov}},\ and\ \citenamefont
  {{Ahmedov}}}]{AtamurotovAllo2022}%
  \BibitemOpen
  \bibfield  {author} {\bibinfo {author} {\bibfnamefont {F.}~\bibnamefont
  {{Atamurotov}}}, \bibinfo {author} {\bibfnamefont {M.}~\bibnamefont
  {{Alloqulov}}}, \bibinfo {author} {\bibfnamefont {A.}~\bibnamefont
  {{Abdujabbarov}}}, \ and\ \bibinfo {author} {\bibfnamefont {B.}~\bibnamefont
  {{Ahmedov}}},\ }\href {\doibase 10.1140/epjp/s13360-022-02846-w} {\bibfield
  {journal} {\bibinfo  {journal} {European Physical Journal Plus}\ }\textbf
  {\bibinfo {volume} {137}},\ \bibinfo {eid} {634} (\bibinfo {year}
  {2022}{\natexlab{d}})}\BibitemShut {NoStop}%
\bibitem [{\citenamefont {Gibbons}\ and\ \citenamefont
  {Werner}(2008)}]{Gibbons:2008rj}%
  \BibitemOpen
  \bibfield  {author} {\bibinfo {author} {\bibfnamefont {G.~W.}\ \bibnamefont
  {Gibbons}}\ and\ \bibinfo {author} {\bibfnamefont {M.~C.}\ \bibnamefont
  {Werner}},\ }\href {\doibase 10.1088/0264-9381/25/23/235009} {\bibfield
  {journal} {\bibinfo  {journal} {Class. Quant. Grav.}\ }\textbf {\bibinfo
  {volume} {25}},\ \bibinfo {pages} {235009} (\bibinfo {year} {2008})},\
  \Eprint {http://arxiv.org/abs/0807.0854} {arXiv:0807.0854 [gr-qc]}
  \BibitemShut {NoStop}%
\bibitem [{\citenamefont {Werner}(2012)}]{Werner_2012}%
  \BibitemOpen
  \bibfield  {author} {\bibinfo {author} {\bibfnamefont {M.~C.}\ \bibnamefont
  {Werner}},\ }\href {\doibase 10.1007/s10714-012-1458-9} {\bibfield  {journal}
  {\bibinfo  {journal} {Gen. Relativ. Gravit.}\ }\textbf {\bibinfo {volume}
  {44}},\ \bibinfo {pages} {3047} (\bibinfo {year} {2012})}\BibitemShut
  {NoStop}%
\bibitem [{\citenamefont {Ishihara}\ \emph {et~al.}(2016)\citenamefont
  {Ishihara}, \citenamefont {Suzuki}, \citenamefont {Ono} \emph
  {et~al.}}]{Ishihara_2016}%
  \BibitemOpen
  \bibfield  {author} {\bibinfo {author} {\bibfnamefont {A.}~\bibnamefont
  {Ishihara}}, \bibinfo {author} {\bibfnamefont {Y.}~\bibnamefont {Suzuki}},
  \bibinfo {author} {\bibfnamefont {T.}~\bibnamefont {Ono}},  \emph {et~al.},\
  }\href {\doibase 10.1103/physrevd.94.084015} {\bibfield  {journal} {\bibinfo
  {journal} {Phys. Rev. D}\ }\textbf {\bibinfo {volume} {94}},\ \bibinfo
  {pages} {084015} (\bibinfo {year} {2016})}\BibitemShut {NoStop}%
\bibitem [{\citenamefont {Ishihara}\ \emph {et~al.}(2017)\citenamefont
  {Ishihara}, \citenamefont {Suzuki}, \citenamefont {Ono},\ and\ \citenamefont
  {Asada}}]{Ishihara:2016sfv}%
  \BibitemOpen
  \bibfield  {author} {\bibinfo {author} {\bibfnamefont {A.}~\bibnamefont
  {Ishihara}}, \bibinfo {author} {\bibfnamefont {Y.}~\bibnamefont {Suzuki}},
  \bibinfo {author} {\bibfnamefont {T.}~\bibnamefont {Ono}}, \ and\ \bibinfo
  {author} {\bibfnamefont {H.}~\bibnamefont {Asada}},\ }\href {\doibase
  10.1103/PhysRevD.95.044017} {\bibfield  {journal} {\bibinfo  {journal} {Phys.
  Rev. D}\ }\textbf {\bibinfo {volume} {95}},\ \bibinfo {pages} {044017}
  (\bibinfo {year} {2017})}\BibitemShut {NoStop}%
\bibitem [{\citenamefont {Ono}\ \emph {et~al.}(2017)\citenamefont {Ono},
  \citenamefont {Ishihara},\ and\ \citenamefont {Asada}}]{Ono:2017pie}%
  \BibitemOpen
  \bibfield  {author} {\bibinfo {author} {\bibfnamefont {T.}~\bibnamefont
  {Ono}}, \bibinfo {author} {\bibfnamefont {A.}~\bibnamefont {Ishihara}}, \
  and\ \bibinfo {author} {\bibfnamefont {H.}~\bibnamefont {Asada}},\ }\href
  {\doibase 10.1103/PhysRevD.96.104037} {\bibfield  {journal} {\bibinfo
  {journal} {Phys. Rev. D}\ }\textbf {\bibinfo {volume} {96}},\ \bibinfo
  {pages} {104037} (\bibinfo {year} {2017})}\BibitemShut {NoStop}%
\bibitem [{\citenamefont {Li}\ and\ \citenamefont
  {\"Ovg\"un}(2020)}]{Li:2020dln}%
  \BibitemOpen
  \bibfield  {author} {\bibinfo {author} {\bibfnamefont {Z.}~\bibnamefont
  {Li}}\ and\ \bibinfo {author} {\bibfnamefont {A.}~\bibnamefont {\"Ovg\"un}},\
  }\href {\doibase 10.1103/PhysRevD.101.024040} {\bibfield  {journal} {\bibinfo
   {journal} {Phys. Rev. D}\ }\textbf {\bibinfo {volume} {101}},\ \bibinfo
  {pages} {024040} (\bibinfo {year} {2020})}\BibitemShut {NoStop}%
\bibitem [{\citenamefont {Li}\ \emph {et~al.}(2020)\citenamefont {Li},
  \citenamefont {Zhang},\ and\ \citenamefont {\"Ovg\"un}}]{Li:2020wvn}%
  \BibitemOpen
  \bibfield  {author} {\bibinfo {author} {\bibfnamefont {Z.}~\bibnamefont
  {Li}}, \bibinfo {author} {\bibfnamefont {G.}~\bibnamefont {Zhang}}, \ and\
  \bibinfo {author} {\bibfnamefont {A.}~\bibnamefont {\"Ovg\"un}},\ }\href
  {\doibase 10.1103/PhysRevD.101.124058} {\bibfield  {journal} {\bibinfo
  {journal} {Phys. Rev. D}\ }\textbf {\bibinfo {volume} {101}},\ \bibinfo
  {pages} {124058} (\bibinfo {year} {2020})}\BibitemShut {NoStop}%
\bibitem [{\citenamefont {Crisnejo}\ and\ \citenamefont
  {Gallo}(2018)}]{Crisnejo:2018uyn}%
  \BibitemOpen
  \bibfield  {author} {\bibinfo {author} {\bibfnamefont {G.}~\bibnamefont
  {Crisnejo}}\ and\ \bibinfo {author} {\bibfnamefont {E.}~\bibnamefont
  {Gallo}},\ }\href {\doibase 10.1103/PhysRevD.97.124016} {\bibfield  {journal}
  {\bibinfo  {journal} {Phys. Rev. D}\ }\textbf {\bibinfo {volume} {97}},\
  \bibinfo {pages} {124016} (\bibinfo {year} {2018})},\ \Eprint
  {http://arxiv.org/abs/1804.05473} {arXiv:1804.05473 [gr-qc]} \BibitemShut
  {NoStop}%
\bibitem [{\citenamefont {\"Ovg\"un}(2018)}]{Ovgun:2018fnk}%
  \BibitemOpen
  \bibfield  {author} {\bibinfo {author} {\bibfnamefont {A.}~\bibnamefont
  {\"Ovg\"un}},\ }\href {\doibase 10.1103/PhysRevD.98.044033} {\bibfield
  {journal} {\bibinfo  {journal} {Phys. Rev. D}\ }\textbf {\bibinfo {volume}
  {98}},\ \bibinfo {pages} {044033} (\bibinfo {year} {2018})},\ \Eprint
  {http://arxiv.org/abs/1805.06296} {arXiv:1805.06296 [gr-qc]} \BibitemShut
  {NoStop}%
\bibitem [{\citenamefont {\"Ovg\"un}(2019{\natexlab{a}})}]{Ovgun:2019wej}%
  \BibitemOpen
  \bibfield  {author} {\bibinfo {author} {\bibfnamefont {A.}~\bibnamefont
  {\"Ovg\"un}},\ }\href {\doibase 10.1103/PhysRevD.99.104075} {\bibfield
  {journal} {\bibinfo  {journal} {Phys. Rev. D}\ }\textbf {\bibinfo {volume}
  {99}},\ \bibinfo {pages} {104075} (\bibinfo {year} {2019}{\natexlab{a}})},\
  \Eprint {http://arxiv.org/abs/1902.04411} {arXiv:1902.04411 [gr-qc]}
  \BibitemShut {NoStop}%
\bibitem [{\citenamefont {\"Ovg\"un}(2019{\natexlab{b}})}]{Ovgun:2018oxk}%
  \BibitemOpen
  \bibfield  {author} {\bibinfo {author} {\bibfnamefont {A.}~\bibnamefont
  {\"Ovg\"un}},\ }\href {\doibase 10.3390/universe5050115} {\bibfield
  {journal} {\bibinfo  {journal} {Universe}\ }\textbf {\bibinfo {volume} {5}},\
  \bibinfo {pages} {115} (\bibinfo {year} {2019}{\natexlab{b}})},\ \Eprint
  {http://arxiv.org/abs/1806.05549} {arXiv:1806.05549 [physics.gen-ph]}
  \BibitemShut {NoStop}%
\bibitem [{\citenamefont {Javed}\ \emph
  {et~al.}(2019{\natexlab{a}})\citenamefont {Javed}, \citenamefont {Abbas},\
  and\ \citenamefont {\"Ovg\"un}}]{Javed:2019kon}%
  \BibitemOpen
  \bibfield  {author} {\bibinfo {author} {\bibfnamefont {W.}~\bibnamefont
  {Javed}}, \bibinfo {author} {\bibfnamefont {J.}~\bibnamefont {Abbas}}, \ and\
  \bibinfo {author} {\bibfnamefont {A.}~\bibnamefont {\"Ovg\"un}},\ }\href
  {\doibase 10.1140/epjc/s10052-019-7208-3} {\bibfield  {journal} {\bibinfo
  {journal} {Eur. Phys. J. C}\ }\textbf {\bibinfo {volume} {79}},\ \bibinfo
  {pages} {694} (\bibinfo {year} {2019}{\natexlab{a}})},\ \Eprint
  {http://arxiv.org/abs/1908.09632} {arXiv:1908.09632 [physics.gen-ph]}
  \BibitemShut {NoStop}%
\bibitem [{\citenamefont {Javed}\ \emph
  {et~al.}(2019{\natexlab{b}})\citenamefont {Javed}, \citenamefont {Abbas},\
  and\ \citenamefont {\"Ovg\"un}}]{Javed:2019rrg}%
  \BibitemOpen
  \bibfield  {author} {\bibinfo {author} {\bibfnamefont {W.}~\bibnamefont
  {Javed}}, \bibinfo {author} {\bibfnamefont {j.}~\bibnamefont {Abbas}}, \ and\
  \bibinfo {author} {\bibfnamefont {A.}~\bibnamefont {\"Ovg\"un}},\ }\href
  {\doibase 10.20944/preprints201906.0101.v1} {\bibfield  {journal} {\bibinfo
  {journal} {Phys. Rev. D}\ }\textbf {\bibinfo {volume} {100}},\ \bibinfo
  {pages} {044052} (\bibinfo {year} {2019}{\natexlab{b}})},\ \Eprint
  {http://arxiv.org/abs/1908.05241} {arXiv:1908.05241 [gr-qc]} \BibitemShut
  {NoStop}%
\bibitem [{\citenamefont {Javed}\ \emph
  {et~al.}(2019{\natexlab{c}})\citenamefont {Javed}, \citenamefont {Babar},\
  and\ \citenamefont {\"Ovg\"un}}]{Javed:2019ynm}%
  \BibitemOpen
  \bibfield  {author} {\bibinfo {author} {\bibfnamefont {W.}~\bibnamefont
  {Javed}}, \bibinfo {author} {\bibfnamefont {R.}~\bibnamefont {Babar}}, \ and\
  \bibinfo {author} {\bibfnamefont {A.}~\bibnamefont {\"Ovg\"un}},\ }\href
  {\doibase 10.1103/PhysRevD.100.104032} {\bibfield  {journal} {\bibinfo
  {journal} {Phys. Rev. D}\ }\textbf {\bibinfo {volume} {100}},\ \bibinfo
  {pages} {104032} (\bibinfo {year} {2019}{\natexlab{c}})},\ \Eprint
  {http://arxiv.org/abs/1910.11697} {arXiv:1910.11697 [gr-qc]} \BibitemShut
  {NoStop}%
\bibitem [{\citenamefont {Javed}\ \emph
  {et~al.}(2020{\natexlab{a}})\citenamefont {Javed}, \citenamefont {Hamza},\
  and\ \citenamefont {\"Ovg\"un}}]{Javed:2020lsg}%
  \BibitemOpen
  \bibfield  {author} {\bibinfo {author} {\bibfnamefont {W.}~\bibnamefont
  {Javed}}, \bibinfo {author} {\bibfnamefont {A.}~\bibnamefont {Hamza}}, \ and\
  \bibinfo {author} {\bibfnamefont {A.}~\bibnamefont {\"Ovg\"un}},\ }\href
  {\doibase 10.20944/preprints201911.0142.v1} {\bibfield  {journal} {\bibinfo
  {journal} {Phys. Rev. D}\ }\textbf {\bibinfo {volume} {101}},\ \bibinfo
  {pages} {103521} (\bibinfo {year} {2020}{\natexlab{a}})},\ \Eprint
  {http://arxiv.org/abs/2005.09464} {arXiv:2005.09464 [gr-qc]} \BibitemShut
  {NoStop}%
\bibitem [{\citenamefont {Javed}\ \emph
  {et~al.}(2019{\natexlab{d}})\citenamefont {Javed}, \citenamefont {Babar},\
  and\ \citenamefont {\"Ovg\"un}}]{Javed:2019qyg}%
  \BibitemOpen
  \bibfield  {author} {\bibinfo {author} {\bibfnamefont {W.}~\bibnamefont
  {Javed}}, \bibinfo {author} {\bibfnamefont {R.}~\bibnamefont {Babar}}, \ and\
  \bibinfo {author} {\bibfnamefont {A.}~\bibnamefont {\"Ovg\"un}},\ }\href
  {\doibase 10.1103/PhysRevD.99.084012} {\bibfield  {journal} {\bibinfo
  {journal} {Phys. Rev. D}\ }\textbf {\bibinfo {volume} {99}},\ \bibinfo
  {pages} {084012} (\bibinfo {year} {2019}{\natexlab{d}})},\ \Eprint
  {http://arxiv.org/abs/1903.11657} {arXiv:1903.11657 [gr-qc]} \BibitemShut
  {NoStop}%
\bibitem [{\citenamefont {\"Ovg\"un}\ \emph {et~al.}(2019)\citenamefont
  {\"Ovg\"un}, \citenamefont {Sakall\i{}},\ and\ \citenamefont
  {Saavedra}}]{Ovgun:2018fte}%
  \BibitemOpen
  \bibfield  {author} {\bibinfo {author} {\bibfnamefont {A.}~\bibnamefont
  {\"Ovg\"un}}, \bibinfo {author} {\bibfnamefont {I.}~\bibnamefont
  {Sakall\i{}}}, \ and\ \bibinfo {author} {\bibfnamefont {J.}~\bibnamefont
  {Saavedra}},\ }\href {\doibase 10.1016/j.aop.2019.167978} {\bibfield
  {journal} {\bibinfo  {journal} {Annals Phys.}\ }\textbf {\bibinfo {volume}
  {411}},\ \bibinfo {pages} {167978} (\bibinfo {year} {2019})},\ \Eprint
  {http://arxiv.org/abs/1806.06453} {arXiv:1806.06453 [gr-qc]} \BibitemShut
  {NoStop}%
\bibitem [{\citenamefont {Javed}\ \emph
  {et~al.}(2020{\natexlab{b}})\citenamefont {Javed}, \citenamefont {Abbas},\
  and\ \citenamefont {\"Ovg\"un}}]{Javed:2019jag}%
  \BibitemOpen
  \bibfield  {author} {\bibinfo {author} {\bibfnamefont {W.}~\bibnamefont
  {Javed}}, \bibinfo {author} {\bibfnamefont {J.}~\bibnamefont {Abbas}}, \ and\
  \bibinfo {author} {\bibfnamefont {A.}~\bibnamefont {\"Ovg\"un}},\ }\href
  {\doibase 10.20944/preprints201906.0124.v1} {\bibfield  {journal} {\bibinfo
  {journal} {Annals Phys.}\ }\textbf {\bibinfo {volume} {418}},\ \bibinfo
  {pages} {168183} (\bibinfo {year} {2020}{\natexlab{b}})},\ \Eprint
  {http://arxiv.org/abs/2007.16027} {arXiv:2007.16027 [gr-qc]} \BibitemShut
  {NoStop}%
\bibitem [{\citenamefont {Pantig}\ and\ \citenamefont
  {Rodulfo}(2020)}]{Pantig:2020odu}%
  \BibitemOpen
  \bibfield  {author} {\bibinfo {author} {\bibfnamefont {R.~C.}\ \bibnamefont
  {Pantig}}\ and\ \bibinfo {author} {\bibfnamefont {E.~T.}\ \bibnamefont
  {Rodulfo}},\ }\href {\doibase 10.1016/j.cjph.2020.06.015} {\bibfield
  {journal} {\bibinfo  {journal} {Chin. J. Phys.}\ }\textbf {\bibinfo {volume}
  {66}},\ \bibinfo {pages} {691} (\bibinfo {year} {2020})}\BibitemShut
  {NoStop}%
\bibitem [{\citenamefont {Pantig}\ \emph {et~al.}(2022)\citenamefont {Pantig},
  \citenamefont {Yu}, \citenamefont {Rodulfo},\ and\ \citenamefont
  {\"Ovg\"un}}]{Pantig2022}%
  \BibitemOpen
  \bibfield  {author} {\bibinfo {author} {\bibfnamefont {R.~C.}\ \bibnamefont
  {Pantig}}, \bibinfo {author} {\bibfnamefont {P.~K.}\ \bibnamefont {Yu}},
  \bibinfo {author} {\bibfnamefont {E.~T.}\ \bibnamefont {Rodulfo}}, \ and\
  \bibinfo {author} {\bibfnamefont {A.}~\bibnamefont {\"Ovg\"un}},\ }\href@noop
  {} {\bibfield  {journal} {\bibinfo  {journal} {Annals of Physics}\ }\textbf
  {\bibinfo {volume} {436}},\ \bibinfo {pages} {168722} (\bibinfo {year}
  {2022})}\BibitemShut {NoStop}%
\bibitem [{\citenamefont {Pantig}\ and\ \citenamefont
  {\"Ovg\"un}(2022)}]{Pantig2022a}%
  \BibitemOpen
  \bibfield  {author} {\bibinfo {author} {\bibfnamefont {R.~C.}\ \bibnamefont
  {Pantig}}\ and\ \bibinfo {author} {\bibfnamefont {A.}~\bibnamefont
  {\"Ovg\"un}},\ }\href {\doibase 10.1140/epjc/s10052-022-10319-8} {\bibfield
  {journal} {\bibinfo  {journal} {Eur. Phys. J. C}\ }\textbf {\bibinfo {volume}
  {82}},\ \bibinfo {pages} {391} (\bibinfo {year} {2022})}\BibitemShut
  {NoStop}%
\bibitem [{\citenamefont {Okyay}\ and\ \citenamefont
  {\"Ovg\"un}(2022)}]{Okyay:2021nnh}%
  \BibitemOpen
  \bibfield  {author} {\bibinfo {author} {\bibfnamefont {M.}~\bibnamefont
  {Okyay}}\ and\ \bibinfo {author} {\bibfnamefont {A.}~\bibnamefont
  {\"Ovg\"un}},\ }\href {\doibase 10.1088/1475-7516/2022/01/009} {\bibfield
  {journal} {\bibinfo  {journal} {JCAP}\ }\textbf {\bibinfo {volume} {01}},\
  \bibinfo {pages} {009} (\bibinfo {year} {2022})}\BibitemShut {NoStop}%
\bibitem [{\citenamefont {Arakida}(2018)}]{Arakida:2017hrm}%
  \BibitemOpen
  \bibfield  {author} {\bibinfo {author} {\bibfnamefont {H.}~\bibnamefont
  {Arakida}},\ }\href {\doibase 10.1007/s10714-018-2368-2} {\bibfield
  {journal} {\bibinfo  {journal} {Gen. Rel. Grav.}\ }\textbf {\bibinfo {volume}
  {50}},\ \bibinfo {pages} {48} (\bibinfo {year} {2018})},\ \Eprint
  {http://arxiv.org/abs/1708.04011} {arXiv:1708.04011 [gr-qc]} \BibitemShut
  {NoStop}%
\bibitem [{\citenamefont {Arakida}(2021)}]{Arakida:2020xil}%
  \BibitemOpen
  \bibfield  {author} {\bibinfo {author} {\bibfnamefont {H.}~\bibnamefont
  {Arakida}},\ }\href {\doibase 10.1088/1475-7516/2021/08/028} {\bibfield
  {journal} {\bibinfo  {journal} {JCAP}\ }\textbf {\bibinfo {volume} {08}},\
  \bibinfo {pages} {028} (\bibinfo {year} {2021})},\ \Eprint
  {http://arxiv.org/abs/2006.13435} {arXiv:2006.13435 [gr-qc]} \BibitemShut
  {NoStop}%
\bibitem [{\citenamefont {Zhang}(2022)}]{Zhang:2021ygh}%
  \BibitemOpen
  \bibfield  {author} {\bibinfo {author} {\bibfnamefont {Z.}~\bibnamefont
  {Zhang}},\ }\href {\doibase 10.1088/1361-6382/ac38d1} {\bibfield  {journal}
  {\bibinfo  {journal} {Class. Quant. Grav.}\ }\textbf {\bibinfo {volume}
  {39}},\ \bibinfo {pages} {015003} (\bibinfo {year} {2022})},\ \Eprint
  {http://arxiv.org/abs/2112.04149} {arXiv:2112.04149 [gr-qc]} \BibitemShut
  {NoStop}%
\bibitem [{\citenamefont {Li}\ and\ \citenamefont {Jia}(2021)}]{Li:2021xhy}%
  \BibitemOpen
  \bibfield  {author} {\bibinfo {author} {\bibfnamefont {Z.}~\bibnamefont
  {Li}}\ and\ \bibinfo {author} {\bibfnamefont {J.}~\bibnamefont {Jia}},\
  }\href {\doibase 10.1103/PhysRevD.104.044061} {\bibfield  {journal} {\bibinfo
   {journal} {Phys. Rev. D}\ }\textbf {\bibinfo {volume} {104}},\ \bibinfo
  {pages} {044061} (\bibinfo {year} {2021})},\ \Eprint
  {http://arxiv.org/abs/2108.05273} {arXiv:2108.05273 [gr-qc]} \BibitemShut
  {NoStop}%
\bibitem [{\citenamefont {D.~Carvalho}\ \emph {et~al.}(2021)\citenamefont
  {D.~Carvalho}, \citenamefont {Alencar}, \citenamefont {Mendes},\ and\
  \citenamefont {Landim}}]{DCarvalho:2021zpf}%
  \BibitemOpen
  \bibfield  {author} {\bibinfo {author} {\bibfnamefont {I.~D.}\ \bibnamefont
  {D.~Carvalho}}, \bibinfo {author} {\bibfnamefont {G.}~\bibnamefont
  {Alencar}}, \bibinfo {author} {\bibfnamefont {W.~M.}\ \bibnamefont {Mendes}},
  \ and\ \bibinfo {author} {\bibfnamefont {R.~R.}\ \bibnamefont {Landim}},\
  }\href {\doibase 10.1209/0295-5075/134/51001} {\bibfield  {journal} {\bibinfo
   {journal} {EPL}\ }\textbf {\bibinfo {volume} {134}},\ \bibinfo {pages}
  {51001} (\bibinfo {year} {2021})},\ \Eprint {http://arxiv.org/abs/2103.03845}
  {arXiv:2103.03845 [gr-qc]} \BibitemShut {NoStop}%
\bibitem [{\citenamefont {Ali}\ and\ \citenamefont
  {Kauhsal}(2022)}]{Ali:2021psk}%
  \BibitemOpen
  \bibfield  {author} {\bibinfo {author} {\bibfnamefont {M.~S.}\ \bibnamefont
  {Ali}}\ and\ \bibinfo {author} {\bibfnamefont {S.}~\bibnamefont {Kauhsal}},\
  }\href {\doibase 10.1103/PhysRevD.105.024062} {\bibfield  {journal} {\bibinfo
   {journal} {Phys. Rev. D}\ }\textbf {\bibinfo {volume} {105}},\ \bibinfo
  {pages} {024062} (\bibinfo {year} {2022})},\ \Eprint
  {http://arxiv.org/abs/2106.08464} {arXiv:2106.08464 [gr-qc]} \BibitemShut
  {NoStop}%
\bibitem [{\citenamefont {Fu}\ \emph {et~al.}(2021)\citenamefont {Fu},
  \citenamefont {Zhao},\ and\ \citenamefont {Liu}}]{Fu:2021akc}%
  \BibitemOpen
  \bibfield  {author} {\bibinfo {author} {\bibfnamefont {Q.-M.}\ \bibnamefont
  {Fu}}, \bibinfo {author} {\bibfnamefont {L.}~\bibnamefont {Zhao}}, \ and\
  \bibinfo {author} {\bibfnamefont {Y.-X.}\ \bibnamefont {Liu}},\ }\href
  {\doibase 10.1103/PhysRevD.104.024033} {\bibfield  {journal} {\bibinfo
  {journal} {Phys. Rev. D}\ }\textbf {\bibinfo {volume} {104}},\ \bibinfo
  {pages} {024033} (\bibinfo {year} {2021})},\ \Eprint
  {http://arxiv.org/abs/2101.08409} {arXiv:2101.08409 [gr-qc]} \BibitemShut
  {NoStop}%
\bibitem [{\citenamefont {Kumar}\ \emph {et~al.}(2019)\citenamefont {Kumar},
  \citenamefont {Ghosh},\ and\ \citenamefont {Wang}}]{Kumar:2019pjp}%
  \BibitemOpen
  \bibfield  {author} {\bibinfo {author} {\bibfnamefont {R.}~\bibnamefont
  {Kumar}}, \bibinfo {author} {\bibfnamefont {S.~G.}\ \bibnamefont {Ghosh}}, \
  and\ \bibinfo {author} {\bibfnamefont {A.}~\bibnamefont {Wang}},\ }\href
  {\doibase 10.1103/PhysRevD.100.124024} {\bibfield  {journal} {\bibinfo
  {journal} {Phys. Rev. D}\ }\textbf {\bibinfo {volume} {100}},\ \bibinfo
  {pages} {124024} (\bibinfo {year} {2019})},\ \Eprint
  {http://arxiv.org/abs/1912.05154} {arXiv:1912.05154 [gr-qc]} \BibitemShut
  {NoStop}%
\bibitem [{\citenamefont {Jusufi}\ \emph {et~al.}(2018)\citenamefont {Jusufi},
  \citenamefont {\"Ovg\"un}, \citenamefont {Saavedra}, \citenamefont
  {V\'asquez},\ and\ \citenamefont {Gonz\'alez}}]{Jusufi:2018jof}%
  \BibitemOpen
  \bibfield  {author} {\bibinfo {author} {\bibfnamefont {K.}~\bibnamefont
  {Jusufi}}, \bibinfo {author} {\bibfnamefont {A.}~\bibnamefont {\"Ovg\"un}},
  \bibinfo {author} {\bibfnamefont {J.}~\bibnamefont {Saavedra}}, \bibinfo
  {author} {\bibfnamefont {Y.}~\bibnamefont {V\'asquez}}, \ and\ \bibinfo
  {author} {\bibfnamefont {P.~A.}\ \bibnamefont {Gonz\'alez}},\ }\href
  {\doibase 10.1103/PhysRevD.97.124024} {\bibfield  {journal} {\bibinfo
  {journal} {Phys. Rev. D}\ }\textbf {\bibinfo {volume} {97}},\ \bibinfo
  {pages} {124024} (\bibinfo {year} {2018})},\ \Eprint
  {http://arxiv.org/abs/1804.00643} {arXiv:1804.00643 [gr-qc]} \BibitemShut
  {NoStop}%
\bibitem [{\citenamefont {{Kumar}}\ \emph
  {et~al.}(2020{\natexlab{b}})\citenamefont {{Kumar}}, \citenamefont
  {{Islam}},\ and\ \citenamefont {{Ghosh}}}]{21}%
  \BibitemOpen
  \bibfield  {author} {\bibinfo {author} {\bibfnamefont {R.}~\bibnamefont
  {{Kumar}}}, \bibinfo {author} {\bibfnamefont {S.~U.}\ \bibnamefont
  {{Islam}}}, \ and\ \bibinfo {author} {\bibfnamefont {S.~G.}\ \bibnamefont
  {{Ghosh}}},\ }\href {\doibase 10.1140/epjc/s10052-020-08606-3} {\bibfield
  {journal} {\bibinfo  {journal} {European Physical Journal C}\ }\textbf
  {\bibinfo {volume} {80}},\ \bibinfo {eid} {1128} (\bibinfo {year}
  {2020}{\natexlab{b}})},\ \Eprint {http://arxiv.org/abs/2004.12970}
  {arXiv:2004.12970 [gr-qc]} \BibitemShut {NoStop}%
\bibitem [{\citenamefont {{Toledo}}\ and\ \citenamefont
  {{Bezerra}}(2019)}]{34}%
  \BibitemOpen
  \bibfield  {author} {\bibinfo {author} {\bibfnamefont {J.~M.}\ \bibnamefont
  {{Toledo}}}\ and\ \bibinfo {author} {\bibfnamefont {V.~B.}\ \bibnamefont
  {{Bezerra}}},\ }\href {\doibase 10.1142/S0218271819500238} {\bibfield
  {journal} {\bibinfo  {journal} {Int. J. Mod. Phys. D}\ }\textbf {\bibinfo
  {volume} {28}},\ \bibinfo {eid} {1950023} (\bibinfo {year}
  {2019})}\BibitemShut {NoStop}%
\bibitem [{\citenamefont {{Sakti}}\ \emph {et~al.}(2021)\citenamefont
  {{Sakti}}, \citenamefont {{Prihadi}}, \citenamefont {{Suroso}},\ and\
  \citenamefont {{Zen}}}]{SC1}%
  \BibitemOpen
  \bibfield  {author} {\bibinfo {author} {\bibfnamefont {M.~F.~A.~R.}\
  \bibnamefont {{Sakti}}}, \bibinfo {author} {\bibfnamefont {H.~L.}\
  \bibnamefont {{Prihadi}}}, \bibinfo {author} {\bibfnamefont {A.}~\bibnamefont
  {{Suroso}}}, \ and\ \bibinfo {author} {\bibfnamefont {F.~P.}\ \bibnamefont
  {{Zen}}},\ }in\ \href {\doibase 10.1088/1742-6596/1949/1/012016} {\emph
  {\bibinfo {booktitle} {Journal of Physics Conference Series}}},\ \bibinfo
  {series} {Journal of Physics Conference Series}, Vol.\ \bibinfo {volume}
  {1949}\ (\bibinfo {year} {2021})\ p.\ \bibinfo {pages} {012016}\BibitemShut
  {NoStop}%
\bibitem [{\citenamefont {{Herscovich}}\ and\ \citenamefont
  {{Richarte}}(2010)}]{SC2}%
  \BibitemOpen
  \bibfield  {author} {\bibinfo {author} {\bibfnamefont {E.}~\bibnamefont
  {{Herscovich}}}\ and\ \bibinfo {author} {\bibfnamefont {M.~G.}\ \bibnamefont
  {{Richarte}}},\ }\href {\doibase 10.1016/j.physletb.2010.04.065} {\bibfield
  {journal} {\bibinfo  {journal} {Phys. Lett. B}\ }\textbf {\bibinfo {volume}
  {689}},\ \bibinfo {pages} {192} (\bibinfo {year} {2010})},\ \Eprint
  {http://arxiv.org/abs/1004.3754} {arXiv:1004.3754 [hep-th]} \BibitemShut
  {NoStop}%
\bibitem [{\citenamefont {Vilenkin}\ and\ \citenamefont
  {Shellard}(2000)}]{Vilenkin:2000jqa}%
  \BibitemOpen
  \bibfield  {author} {\bibinfo {author} {\bibfnamefont {A.}~\bibnamefont
  {Vilenkin}}\ and\ \bibinfo {author} {\bibfnamefont {E.~P.~S.}\ \bibnamefont
  {Shellard}},\ }\href@noop {} {\  (\bibinfo {year} {2000})}\BibitemShut
  {NoStop}%
\bibitem [{\citenamefont {Barriola}\ and\ \citenamefont
  {Vilenkin}(1989)}]{Barriola:1989hx}%
  \BibitemOpen
  \bibfield  {author} {\bibinfo {author} {\bibfnamefont {M.}~\bibnamefont
  {Barriola}}\ and\ \bibinfo {author} {\bibfnamefont {A.}~\bibnamefont
  {Vilenkin}},\ }\href {\doibase 10.1103/PhysRevLett.63.341} {\bibfield
  {journal} {\bibinfo  {journal} {Phys. Rev. Lett.}\ }\textbf {\bibinfo
  {volume} {63}},\ \bibinfo {pages} {341} (\bibinfo {year} {1989})}\BibitemShut
  {NoStop}%
\bibitem [{\citenamefont {{Letelier}}(1979)}]{SC0}%
  \BibitemOpen
  \bibfield  {author} {\bibinfo {author} {\bibfnamefont {P.~S.}\ \bibnamefont
  {{Letelier}}},\ }\href {\doibase 10.1103/PhysRevD.20.1294} {\bibfield
  {journal} {\bibinfo  {journal} {Phys. Rev. D}\ }\textbf {\bibinfo {volume}
  {20}},\ \bibinfo {pages} {1294} (\bibinfo {year} {1979})}\BibitemShut
  {NoStop}%
\bibitem [{\citenamefont {{Tsujikawa}}(2013)}]{Q1}%
  \BibitemOpen
  \bibfield  {author} {\bibinfo {author} {\bibfnamefont {S.}~\bibnamefont
  {{Tsujikawa}}},\ }\href {\doibase 10.1088/0264-9381/30/21/214003} {\bibfield
  {journal} {\bibinfo  {journal} {Classical and Quantum Gravity}\ }\textbf
  {\bibinfo {volume} {30}},\ \bibinfo {eid} {214003} (\bibinfo {year}
  {2013})},\ \Eprint {http://arxiv.org/abs/1304.1961} {arXiv:1304.1961 [gr-qc]}
  \BibitemShut {NoStop}%
\bibitem [{\citenamefont {{Liaqat}}\ and\ \citenamefont
  {{Hussain}}(2022)}]{Liaqat2022}%
  \BibitemOpen
  \bibfield  {author} {\bibinfo {author} {\bibfnamefont {A.}~\bibnamefont
  {{Liaqat}}}\ and\ \bibinfo {author} {\bibfnamefont {I.}~\bibnamefont
  {{Hussain}}},\ }\href {\doibase 10.1088/1674-1137/ac2e66} {\bibfield
  {journal} {\bibinfo  {journal} {Chinese Physics C}\ }\textbf {\bibinfo
  {volume} {46}},\ \bibinfo {eid} {015101} (\bibinfo {year}
  {2022})}\BibitemShut {NoStop}%
\bibitem [{\citenamefont {{Miranda}}\ \emph {et~al.}(2014)\citenamefont
  {{Miranda}}, \citenamefont {{Carneiro}},\ and\ \citenamefont
  {{Pigozzo}}}]{Miranda2014}%
  \BibitemOpen
  \bibfield  {author} {\bibinfo {author} {\bibfnamefont {W.}~\bibnamefont
  {{Miranda}}}, \bibinfo {author} {\bibfnamefont {S.}~\bibnamefont
  {{Carneiro}}}, \ and\ \bibinfo {author} {\bibfnamefont {C.}~\bibnamefont
  {{Pigozzo}}},\ }\href {\doibase 10.1088/1475-7516/2014/07/043} {\bibfield
  {journal} {\bibinfo  {journal} {JCAP}\ }\textbf {\bibinfo {volume} {2014}},\
  \bibinfo {eid} {043} (\bibinfo {year} {2014})},\ \Eprint
  {http://arxiv.org/abs/1405.3673} {arXiv:1405.3673 [astro-ph.CO]} \BibitemShut
  {NoStop}%
\bibitem [{\citenamefont {{Perlick}}\ and\ \citenamefont
  {{Tsupko}}(2022)}]{Perlick2022rev}%
  \BibitemOpen
  \bibfield  {author} {\bibinfo {author} {\bibfnamefont {V.}~\bibnamefont
  {{Perlick}}}\ and\ \bibinfo {author} {\bibfnamefont {O.~Y.}\ \bibnamefont
  {{Tsupko}}},\ }\href {\doibase 10.1016/j.physrep.2021.10.004} {\bibfield
  {journal} {\bibinfo  {journal} {Phys. Rep.}\ }\textbf {\bibinfo {volume}
  {947}},\ \bibinfo {pages} {1} (\bibinfo {year} {2022})},\ \Eprint
  {http://arxiv.org/abs/2105.07101} {arXiv:2105.07101 [gr-qc]} \BibitemShut
  {NoStop}%
\bibitem [{\citenamefont {{Papnoi}}\ and\ \citenamefont
  {{Atamurotov}}(2022)}]{Atamurotov2022papnoi}%
  \BibitemOpen
  \bibfield  {author} {\bibinfo {author} {\bibfnamefont {U.}~\bibnamefont
  {{Papnoi}}}\ and\ \bibinfo {author} {\bibfnamefont {F.}~\bibnamefont
  {{Atamurotov}}},\ }\href {\doibase 10.1016/j.dark.2021.100916} {\bibfield
  {journal} {\bibinfo  {journal} {Physics of the Dark Universe}\ }\textbf
  {\bibinfo {volume} {35}},\ \bibinfo {eid} {100916} (\bibinfo {year}
  {2022})},\ \Eprint {http://arxiv.org/abs/2111.15523} {arXiv:2111.15523
  [gr-qc]} \BibitemShut {NoStop}%
\bibitem [{\citenamefont {{Jusufi}}\ \emph {et~al.}(2020)\citenamefont
  {{Jusufi}}, \citenamefont {{Jamil}},\ and\ \citenamefont
  {{Zhu}}}]{Jusufi2020obs}%
  \BibitemOpen
  \bibfield  {author} {\bibinfo {author} {\bibfnamefont {K.}~\bibnamefont
  {{Jusufi}}}, \bibinfo {author} {\bibfnamefont {M.}~\bibnamefont {{Jamil}}}, \
  and\ \bibinfo {author} {\bibfnamefont {T.}~\bibnamefont {{Zhu}}},\ }\href
  {\doibase 10.1140/epjc/s10052-020-7899-5} {\bibfield  {journal} {\bibinfo
  {journal} {European Physical Journal C}\ }\textbf {\bibinfo {volume} {80}},\
  \bibinfo {eid} {354} (\bibinfo {year} {2020})},\ \Eprint
  {http://arxiv.org/abs/2005.05299} {arXiv:2005.05299 [gr-qc]} \BibitemShut
  {NoStop}%
\bibitem [{\citenamefont {{Akiyama}}\ and\ \citenamefont
  {et~al.}(2022)}]{Akiyama2022sgr}%
  \BibitemOpen
  \bibfield  {author} {\bibinfo {author} {\bibfnamefont {K.}~\bibnamefont
  {{Akiyama}}}\ and\ \bibinfo {author} {\bibnamefont {et~al.}},\ }\href
  {\doibase 10.3847/2041-8213/ac6674} {\bibfield  {journal} {\bibinfo
  {journal} {Astrophys. J. Lett}\ }\textbf {\bibinfo {volume} {930}},\ \bibinfo
  {eid} {L12} (\bibinfo {year} {2022})}\BibitemShut {NoStop}%
\bibitem [{\citenamefont {Sharif}\ and\ \citenamefont
  {Iftikhar}(2015)}]{Sharif:2015oua}%
  \BibitemOpen
  \bibfield  {author} {\bibinfo {author} {\bibfnamefont {M.}~\bibnamefont
  {Sharif}}\ and\ \bibinfo {author} {\bibfnamefont {S.}~\bibnamefont
  {Iftikhar}},\ }\href {\doibase 10.1155/2015/635625} {\bibfield  {journal}
  {\bibinfo  {journal} {Adv. High Energy Phys.}\ }\textbf {\bibinfo {volume}
  {2015}},\ \bibinfo {pages} {635625} (\bibinfo {year} {2015})},\ \bibinfo
  {note} {[Erratum: Adv.High Energy Phys. 2015, 219762 (2015)]}\BibitemShut
  {NoStop}%
\end{thebibliography}%


%

\end{document}